%% file: Manuscript.tex
\documentclass[letterpaper,12pt]{article}

\usepackage{fullpage}  
\usepackage[T1]{fontenc}
\usepackage{authblk}
\usepackage{amsfonts}
\usepackage{amsmath}
\usepackage{amssymb}
\usepackage{amsthm}
\usepackage{graphicx}
\usepackage[colorlinks=true,linkcolor=teal,citecolor=teal,urlcolor=teal,plainpages=false,pdfpagelabels]{hyperref}
\usepackage{color}
\usepackage[dvipsnames]{xcolor}
\usepackage{mathtools}
\usepackage{bbm}
\usepackage[normalem]{ulem}
\usepackage{cite}  
\usepackage{float}
\usepackage{array}
\usepackage{multirow}
\usepackage{longtable}
\usepackage{verbatim}
\usepackage{enumitem}
\usepackage{abstract}
\usepackage{tocloft}
\usepackage[nottoc]{tocbibind}
\usepackage[font=small,labelfont=sc,margin=1cm]{caption}
\usepackage[section]{placeins}

\usepackage{anyfontsize}  
\usepackage{microtype} 

\usepackage{fancyhdr} 

\usepackage{newtxtext}
\usepackage{newtxmath}

\newcommand{\je}[1]{{\textcolor{cyan}{#1}}}

\allowdisplaybreaks

\makeatletter
\g@addto@macro\bfseries{\boldmath}
\makeatother

\newcommand{\bra}[1]{\langle #1|}
\newcommand{\ket}[1]{|#1\rangle}

\newcommand{\e}{\operatorname{e}}

\providecommand{\coloneq}{\mathrel{\mathop:}=}

\makeatletter
\newcommand{\pushright}[1]{\ifmeasuring@#1\else\omit\hfill$\displaystyle#1$\fi\ignorespaces}
\newcommand{\pushleft}[1]{\ifmeasuring@#1\else\omit$\displaystyle#1$\hfill\fi\ignorespaces}
\makeatother

\theoremstyle{definition}

\newtheorem{lemma}{Lemma}

\newtheorem{definition}[lemma]{Definition}

\theoremstyle{plain}

\renewcommand{\qedsymbol}{$\blacksquare$}

\renewcommand{\qedsymbol}{\unskip\nobreak\quad\qedsymbol}
\renewcommand{\qedsymbol}{$\blacksquare$}

\usepackage{physics}

\usepackage{algorithm}
\usepackage{algorithmic}
\floatname{algorithm}{Algorithm}
\renewcommand{\algorithmicrequire}{\textbf{Input: }}
\renewcommand{\algorithmicensure}{\textbf{Output: }}
\newcommand{\algorithmicparam}{\textbf{Parameters: }}
\newcommand{\algorithmicinit}{\textbf{Initialize: }}
\newcommand{\algorithmichyperparam}{\textbf{Hyperparameters:}}

\definecolor{hollywoodcerise}{rgb}{0.96, 0.0, 0.63}

\newcommand{\AW}[1]{\textcolor{hollywoodcerise}{[AW: #1]}}
\definecolor{laurascolor}{rgb}{0.82, 0.0, 1.0}

\definecolor{electricblue}{RGB}{89,203,232}
\newcommand{\DR}[1]{\textcolor{electricblue}{[DR: #1]}}

\newcommand{\SK}[1]{{\color{orange} [SK: #1]}}

\begin{document}

\renewcommand\Affilfont{\small}

\title{\textbf{Quantum reinforcement learning of classical rare dynamics: Enhancement by intrinsic Fourier features}}

%
%

\author[1,2]{Alissa Wilms}
\author[3,4]{Laura Ohff}
\author[7]{Andrea Skolik}
\author[1,5,6]{\\Jens Eisert}
\author[1,8,9]{Sumeet~Khatri}
\author[1]{David A. Reiss}


\affil[1]{Dahlem Center for Complex Quantum Systems, Freie Universität Berlin, 14195 Berlin, Germany}
\affil[2]{Porsche Digital GmbH, 71636 Ludwigsburg, Germany}
\affil[3]{University of Bamberg,
Bamberg, Germany}
\affil[4]{Formerly at Porsche Digital GmbH, 71636 Ludwigsburg, Germany}
\affil[5]{Fraunhofer Heinrich Hertz Institute, 10587 Berlin, Germany}
\affil[6]{Helmholtz-Zentrum Berlin für Materialien und Energie, 14109 Berlin, Germany}
\affil[7]{QuiX Quantum GmbH, Heilbronner Str. 150, 70191 Stuttgart, Germany}
\affil[8]{Department of Computer Science, Virginia Tech, Blacksburg, VA 24061, USA}
\affil[9]{Virginia Tech Center for Quantum Information Science and Engineering, Blacksburg, VA 24061, USA}

\date{}


\maketitle              
\begin{abstract}
    Rare events are essential for understanding the behavior of non-equilibrium and industrial systems. It is of ongoing interest to develop methods for effectively searching for rare events.
    With the advent of quantum computing and its potential advantages
    over classical computing for applications like sampling certain probability distributions, the question arises whether quantum computers could also provide an advantage or inspire new methods for sampling the statistics of rare events. In this work, we propose a quantum reinforcement learning (QRL) method for studying rare dynamics, and we 
    investigate their benefits over classical approaches based on neural networks. 
    As a proof-of-concept example, we demonstrate that our QRL agents can learn and generate the rare dynamics of random walks, and we are able to explain this success as well as the different contributing factors to it via the intrinsic Fourier features of the parametrized quantum circuit. Furthermore, we show better learning behavior with fewer parameters compared to classical approaches. This is the first investigation of QRL applied to generating rare events and suggests that QRL is a promising method to study their dynamics and statistics. 
    

\end{abstract}

\tableofcontents

\section{Introduction}
Among the many discussed potential applications of quantum computers, especially in the presumed \emph{noisy intermediate-scale quantum} (NISQ) era~\cite{preskill2018quantum} and beyond, machine learning has attracted significant attention in recent years~\cite{biamonte2017quantum, dunjko2018machine, schuld2021machine}.
Much research has been devoted to understanding possible advantages that quantum computers may have in machine learning tasks compared to classical computers~\cite{cerezo2022challenges,gyurik2022separations}. 
The notion of a ``quantum advantage'' itself can be defined in different ways:
It may be seen in terms of sample complexity, computational complexity or generalization.
One definition that is commonly referred to is a strict 
and provable separation between the performance of quantum and classical computers in terms of notions of computational complexity.
While results of this type are 
highly desirable --- examples of this kind including the results of
Refs.~\cite{Jerbi, huang2021information, sweke2021quantum, liu2021rigorous} --- it increasingly seems to turn out that such results alone may not make a decisive case in the field. Quantum algorithms are known to perform well on a number of highly structured problems~\cite{AshleyAlgorithms,aaronson2022structurespeedups}. For machine learning problems, it is often not so clear to what extent the demands of such high levels of structure can be met. It has been argued that quantum advantages in this sense may be 
only provide a partial answer~\cite{PRXQuantum.3.030101}.
What is more, some expected quantum
advantages have later been ``dequantized'' via the discovery of 
substantially improved classical algorithms~\cite{tang2022dequantizing, cotler2021revisiting, sweke2023potential}.
These considerations invite the line of thought that 
quantum advantages in this sense may not quite be the end of the story, and that it is advisable to comprehensively explore regimes in which rigorous results and heuristic insights come together. It is the main point of this work to suggest a line of thought within this mindset in probabilistic modeling, combining rigorous and heuristic approaches, to make a case that this reasoning may be worth considering more.

A natural starting point here are sampling problems which seem to be particularly suited for quantum architectures.
Sampling problems have long been 
known to be
a source of a provable quantum advantage~\cite{boixo2018supremacy,2022arXiv220604079H}, with prominent examples being (Gaussian) boson sampling~\cite{aaronson2013bosonsampling,kruse2019GaussianBosonSampling} as well as sampling from the Born distribution of universal random quantum circuits~\cite{boixo2018supremacy,boixo2017sampling,arute2019quantum}. In this work, we take steps of considering
the intersection of sampling and quantum machine learning algorithms. This intersection is at the essence of generative learning~\cite{silver2016AlphaGo,goodfellow2014GAN,silver2018AlphaGo,goodfellow2020GAN,cao2023RLgenerativeAI,franceschelli2024RLgenerativeAI,foster2023generativedeeplearning_book}. Specifically, we consider the practically motivated sampling problem of \textit{rare-trajectory sampling} and the potential advantages of machine learning with quantum computers for this 
set of examples.
The study of rare dynamics is essential for understanding the non-equilibrium behavior of physical, chemical, biological, economic, and other systems. In particular, sampling from, learning and generating rare trajectories for first-passage-time problems is an important task in the finance and insurance industries, and for understanding rare events such as crashes of financial markets and earthquakes~\cite{mcneil2015quantitative, l2009importance, bousquet2021extreme, embrechts2013modelling}. In the engineering industry, surrogate models (also known as digital twins) are utilized in simulations of car crashes~\cite{HAY2022433} or airplane parts, requiring computational tools to simulate rare dynamic behavior or outliers~\cite{aircrafrtrareevents, Kracker_Dhanasekaran_Schumacher_Garcke_2023}.

Classical methods for rare-trajectory sampling include Monte-Carlo simulation (which is known to be sample inefficient~\cite{beck2015rare}), importance sampling~\cite{beck2015rare}, and more specific techniques like backtracking of rare events and exponential tilting of probability distributions~\cite{aguilar2023unified}. For example, the latter (alternatively called ``exponential change of measure'' or ``exponential twisting'') has been applied to the analysis of molecular dynamics~\cite{2018MolPh.116.3104W}. Still, these methods are to an extent limited, 
since, \textit{e.g.}, determining the optimal exponential tilting is equivalent to solving the original problem~\cite{touchette2012basic}, 
and poor choices of exponential tilting can render it ineffective.
For importance sampling in general an analogous problem occurs~\cite{touchette2012basic}. 
Similarly, backtracking as a recursion technique is computationally inefficient in large sample spaces, thus poorly scalable and only suitable for problems with structure~\cite{levitin2012introduction}.  

Consequently, machine learning methods have been developed for instances of rare-trajectory sampling~\cite{strahan2023predicting, RaoMl, asghar2024efficient}. However, conventional supervised learning models  (\textit{e.g.}, those based on neural networks) are limited in capturing the rare dynamics of complex systems, because neural networks depend on data samples and samples of rare trajectories are, by definition, statistically infrequent. In contrast to conventional supervised learning, Rose \textit{et al.}~\cite{rose2021reinforcement} have successfully applied \emph{reinforcement learning} (RL) techniques to this problem. They have considered the problem of a one-dimensional random walker, and the ``rare events'' were rare trajectories, \textit{e.g.}, those that return to the initial position after a fixed number of time steps, known as ``\emph{random walk bridges}'' (RWBs). They have defined the parameterized policy of the RL agents by classical \emph{neural networks} (NNs). 
Their work is an example of an RL-based generative model in which the goal of the RL agents is to learn a probability distribution such that they can generate samples from the true, unknown distribution; see, \textit{e.g.}, Refs.~\cite{cao2023RLgenerativeAI,franceschelli2024RLgenerativeAI,foster2023generativedeeplearning_book} for an overview of this thriving field of research.


Turning to quantum computing inspired approaches, in recent years, RL applications based on variational quantum algorithms in various forms and flavors have been emerging~\cite{jerbi2021variational,lockwood2020reinforcement,benedetti2019parameterized}. Such quantum RL applications model the RL agent using a \textit{parameterized quantum circuit} (PQC), a reasonable quantum analogue of a neural network. There have been numerical indications on possible separations between the performance of this model and NN-based RL, for example, in terms of the number of required learning parameters~\cite{skolik2022quantum}. 
Despite estimates of the resource requirements in quantum machine learning models~\cite{Jerbi_2023}, the effects of many choices within the PQC architecture---such as the number of qubits and data-uploading layers, the kind of data uploading, etc.---on the performance of different quantum RL methods remain largely unstudied, with Ref.~\cite{skolik2022quantum} being a notable exception. Furthermore, the application of quantum reinforcement learning to more practical problems remains 
largely unexplored, and the dependence of its success on the specific problem characteristics has yet to be fully understood. 

In the light of these points, the central questions addressed in this work are the following.
\begin{enumerate}
    \item How effectively does quantum RL perform in the task of sampling rare trajectories of a one-dimensional random walk?
    \item Can quantum RL possibly achieve an advantage over NN-based RL methods for this task, and if so, in what sense?
    \item What specific characteristics of this use case influence the model’s success or failure? 
\end{enumerate}  

\begin{figure}
    \begin{center}
        \includegraphics[width=15cm]{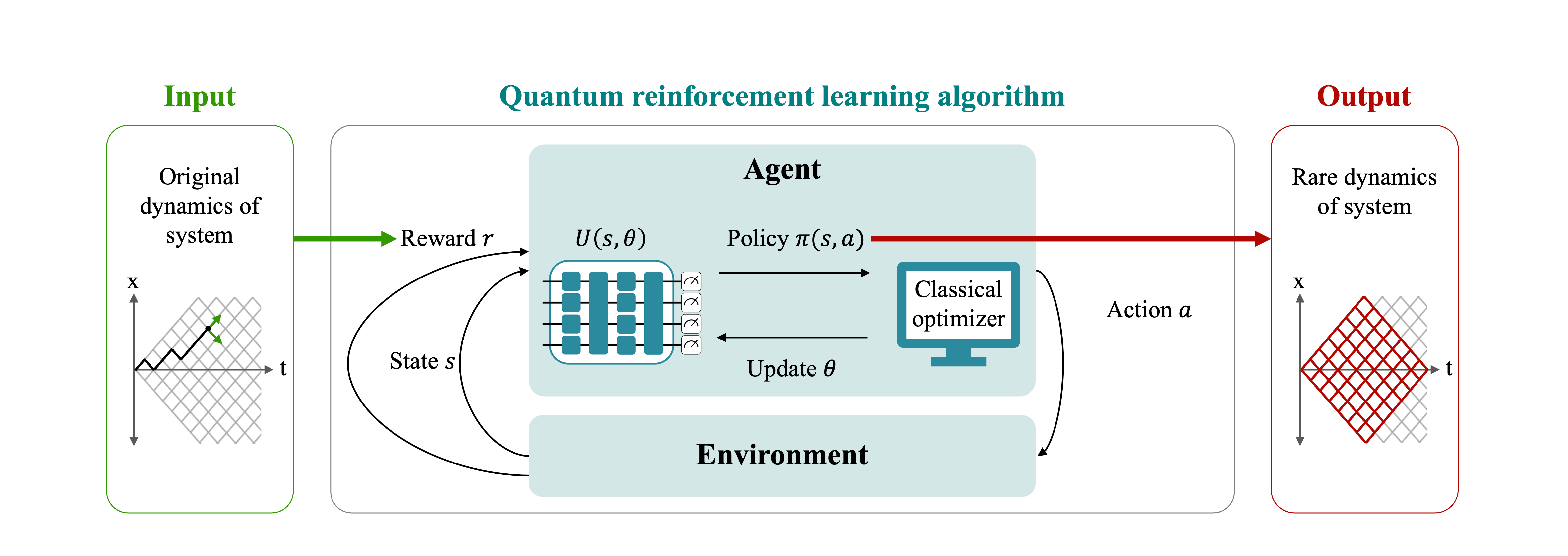}
        \caption{Schematic depiction of our quantum reinforcement learning model. The input is illustrated using trajectories from a random walk. The output consists of samples of rare trajectories, which we take to be random walk bridges.}
        \label{QRLFramework}
    \end{center}
\end{figure}

\paragraph*{Our contributions.} First, we introduce two models within the domain of \emph{quantum reinforcement learning}  (QRL) for the purpose of rare-trajectory sampling: policy-gradient and actor-critic QRL. The general QRL framework is depicted in Fig.~\ref{QRLFramework}. Here, the RL agent (see blue box labeled ``agent'') generates rare trajectories of the random walk (see red box ``output'') according to a policy $\pi(s, a)$ parameterized by the unitary transformation $U(s, \theta)$ of a PQC with variational parameters $\theta$. This policy denotes probabilities to perform specific actions $a$ at specific states $s$ of the environment (depicted by the blue box at the bottom). After performing an action, the agent receives the reward, signaling how close the policy is to the reweighted dynamics. The parameters $\theta$ of the PQC are then optimized via a classical optimizer. After successful learning, the policy maximizes the reward and in the end the optimized policy gives rise to the rare dynamics of interest.

We then demonstrate that our QRL models perform sampling rare trajectories of a one-dimensional random walk better than NN-based RL methods, even with just one and two qubits only; see Section~\ref{sec:1- and 2-qubit policy gradient PQCs}, where we demonstrate that our QRL models can learn the rare trajectories of the random walk, and Section~\ref{sec:Comparison to classical neural networks} for the comparison to classical NNs. 
It is hence important to stress that here, no asymptotic quantum advantage is being investigated, but rather a separation in the performance depending on the number of degrees of freedom made use of.
It is rather surprising that our model can perform well with just one and two qubits, and, therefore, we provide an explanation of this success with an analysis of PQC-based policies via computing and fitting their truncated Fourier series in Section~\ref{sec:1-qubit PQC}. This analysis shows that, similar to Fourier encodings in classical NNs mapping input coordinates to a set of trigonometric functions~\cite{tancik2020fourier}, the data encoding of the PQCs maps the input data to a Fourier series (in this case the input data are the possible states of the random walk). This mapping is intrinsic to the PQC and enhances the performance of QRL for rare-trajectory sampling. Thus, we name this phenomenon ``enhancement by intrinsic Fourier features''. We further use the equivalence between Fourier encodings and one layer sinusoidal activation function~\cite{benbarka2021seeing} in classical NNs and observe that this increases the performance of the classical NNs, similar to the PQC. The application area of sinusoidal activation functions, low-dimensional domains with high frequency features, aligns well with our problem of interest and may provide insights into identifying suitable areas of applications for PQCs.
We also examine the non-trivial dependence of the performance on the number of data-uploading layers of the PQCs, increasing performance first and later depreciating performance, and the (beneficial) effects of different parameters and architectures of the PQCs in Section~\ref{sec:Comparison of data re-uploading and parameter variation}.

Naturally the question arises what benefits, if any, can be gained by using additional qubits and data-uploading layers. More complex variants of the problem might need a scaling of the PQC. One direction to think of is a higher dimensional random walk, where the additional coordinates are encoded in gates acting on additional qubits or a problem size scaling in the length of the trajectory; the latter is investigated in Section~\ref{sec:Scaling the number of time steps}. Another direction would be to apply a parallel encoding with more than two qubits, but to keep the dimension of the encoded data the same, as can be seen in Section~\ref{sec:Scaling the number of time steps}, where we present results for an eight-qubit PQC. This might be useful for trajectories which are more constrained, e.g., to specific dynamical observables. In turn, more complex constraints would lead to more complex optimal policies, for which the PQCs would need a higher expressivity to yield good approximations. Both higher dimensional random walks and more complex optimal policies are relevant for many practical problems of interest. 

Nevertheless, even for these more complex variants of the problem, it might be the case that polynomial many Fourier features might be sufficient for a good enough approximation of the data set, which is precisely what we find in Section~\ref{sec:1-qubit PQC} for small instances of the problem. If so, identifying this polynomially-sized subset can still be tricky and depends on the structure of the target functions as well as the data distribution. Algorithmic approaches for identifying subsets of Fourier features are often based on sparsity of the target function in the frequency domain, and this sparsity affects the approximation accuracy. If such a subset would be sufficient and identifiable, an effective dequantization (as presented in Section~\ref{sec:1-qubit PQC}) would be possible, e.g., by using quantum-inspired RL via Fourier neural networks. Without being able to identify such a polynomially-sized subset, sampling methods inspired by random Fourier features~\cite{landman2022VQCRFF,sweke2023potential} have been explored, which essentially choose polynomially-sized subsets at random. It would be interesting to apply this method to our problem in future work.

\section{Theory and methodology}
\label{sec:theory_methodology}

In this section, we review the considered stochastic 
processes as well as the basics and methods of reinforcement learning and variational quantum algorithms necessary for this article. We elaborate all methods along the paradigmatic stochastic process of a random walk.


    

\subsection{Random walk} \label{sec:Stochastic model}

\paragraph*{Original dynamics.} 
To capture the classical dynamics,
we consider a one-dimensional random walk process (e.g., see Ref.~\cite[Section~9.7]{Stewart09_book} for a pedagogical overview), in which the position $x_t$ at time $t\in\mathbb{N}_0\coloneqq\{0,1,2,\dotsc\}$ is given by an integer, \textit{i.e.}, $x_t \in \mathbb{Z}$. Starting from the initial position $x_0\in\mathbb{Z}$, a random step of size $\Delta x_t\in\mathbb{Z}$ is taken at each time step. Consequently, 
\begin{equation}
    x_t  = x_0 + \sum_{j=1}^t \Delta x_{j}.
\end{equation}
Without loss of generality, we choose $x_0 = 0$. Additionally, we assume the steps $\Delta x_{j}$ to be independent and identically distributed random variables taking values $\pm 1$ with probabilities $1/2\pm\epsilon$ for some $\epsilon\in [0, 1/2]$.\footnote{Note that the methods presented in this work can also be applied to the case of variable step sizes.} Thus, the random walk constitutes a Markov chain, whose transition probabilities are given by
\begin{equation}\label{eq:walker_transition_probabilities}
    P(x_{t + 1}|x_{t}) = 
    \begin{cases}
    \frac{1}{2}\pm\epsilon & \text{for } x_{t + 1} = x_{t} \pm 1, \\
    0 & \text{else}.
    \end{cases}
\end{equation}

\begin{figure}[h]
    \begin{center}
        \includegraphics[width=0.2\textwidth]{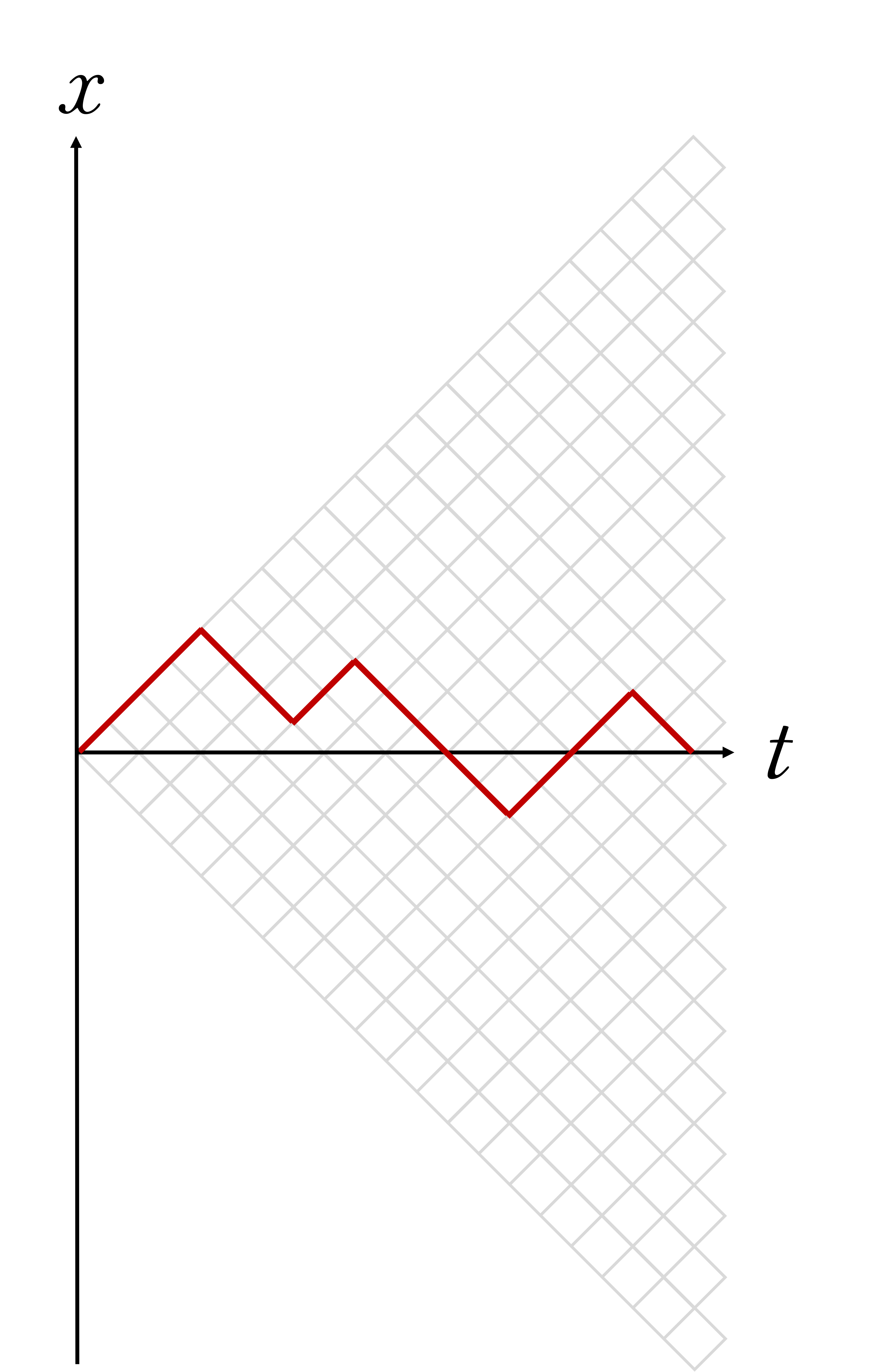}\qquad
        \includegraphics[width=0.2\textwidth]{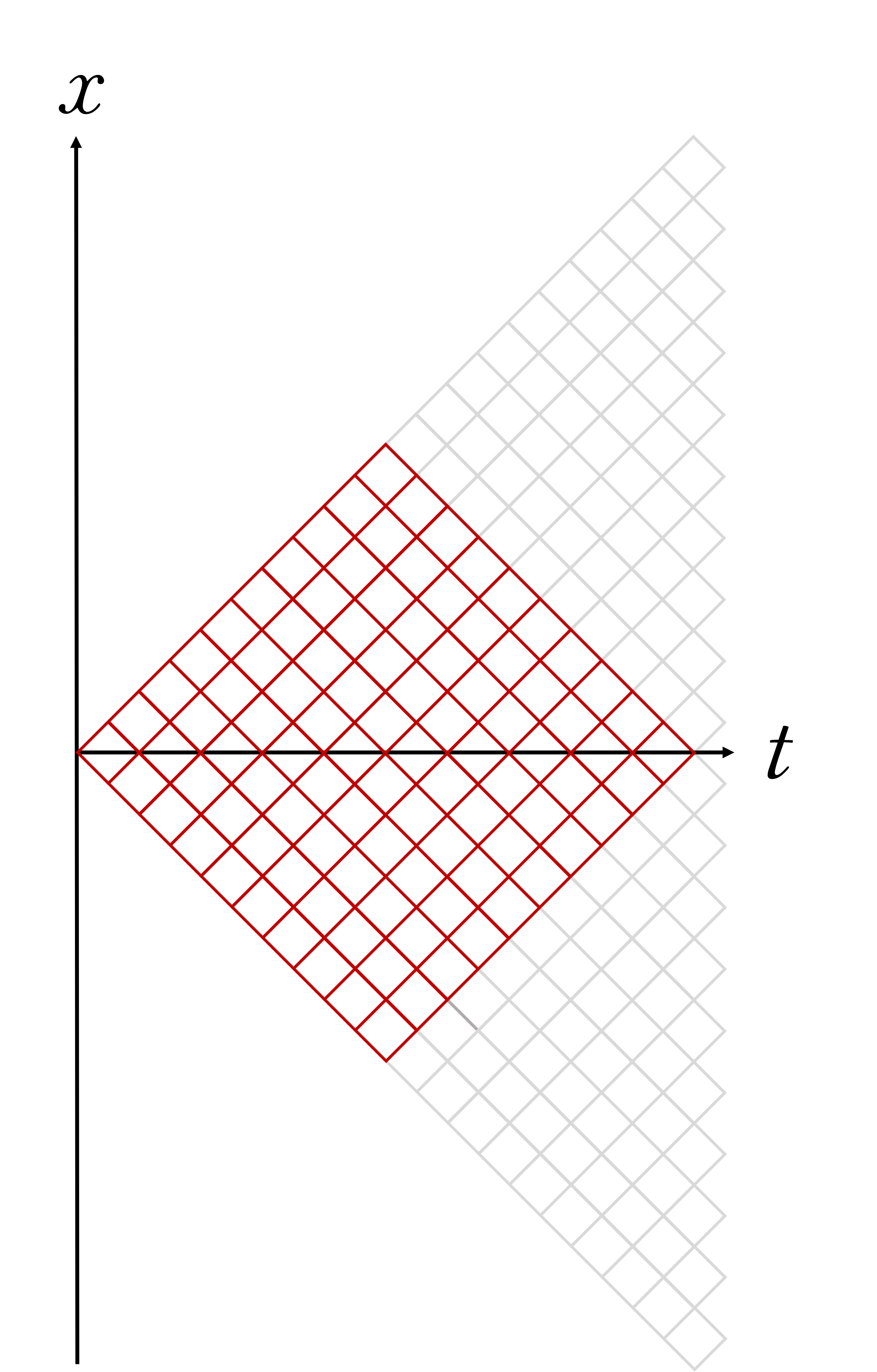}
         \caption{Left: example of a trajectory (red line) created by a random walk. Right: trajectory ensemble of random walk bridges (red lines). (The gray lines show all possible trajectories.)}
        \label{fig:trajectory}
    \end{center}
\end{figure}

The central object of our investigation is a \textit{trajectory} $\omega_{0}^T\coloneq(x_0,x_1,x_2,\dotsc,x_T)$, which is a  sequence of states of the random walk from time $t=0$ to $T$. An example of a trajectory is shown in Fig.~\ref{fig:trajectory}. The number $N$ of possible trajectories scales exponentially with time $T$ (in the special case considered here: $N = 2^T$ ). The probability $P\!(\omega_0^T )$ to obtain a particular trajectory is given by
\begin{align}
    P\!\left(\omega_0^T \right) = \left(\frac{1}{2}+\epsilon\right)^{n_+}\left(\frac{1}{2}-\epsilon\right)^{n_-},
    \label{eq-original_dynamics}
\end{align}
where $n_{\pm}$ is the number of steps $\pm 1$ taken in the trajectory, with $n_{+} + n_{-} = T$. Note that for $\epsilon = 0$, Eq.\ \eqref{eq-original_dynamics} simplifies to $P\!\left(\omega_0^T \right) = 1/2^T$, so all $2^T$ trajectories $\omega_0^T$ are equally likely. Following Ref.~\cite{rose2021reinforcement}, we call the trajectory distribution $P\!(\omega_0^T )$ the \textit{original dynamics} of the random walk.

The probability distribution $P(x_T = x)$ of the endpoint $x_T$ of a trajectory is the binomial distribution,
\begin{equation}
\label{eq-walker_endpoint_distribution}
    P(x_T = x) = \binom{T}{(T + x)/2}\left(\frac{1}{2} + \epsilon\right)^{(T + x)/2}\left(\frac{1}{2} - \epsilon\right)^{(T - x)/2},
\end{equation} 
since $n_\pm$ is determined by $T$ and $x$, where $x\in\{-T ,-T + 2, \dotsc, +T\}$ (for other values of $x$ the probability is zero). In the special case of $\epsilon = 0$, this reduces to $P(x_T = x) = \binom{T}{(T + x)/2} \frac{1}{2^T}$. We are especially interested in the end point $x_T=0$.

\begin{definition}[Random walk bridge~\cite{Benichou}]
    Let $T$ be an even number of time steps. A \textit{random walk bridge} (RWB) is any trajectory $\omega_{0}^T=(x_0,x_1,\dotsc,x_T)$ of the random walk such that $x_0=x_T=0$.
\end{definition}

The trajectory ensemble of all random walk bridges of a certain length $T$ is illustrated in the right part of Fig.~\ref{fig:trajectory}. From Eq.\ \eqref{eq-walker_endpoint_distribution} it follows that the probability to obtain such an RWB is equal to
\begin{equation}
\label{eq:P(x_T_=_0)}
    P(x_T = 0)=\frac{T!}{{(T/2!)}^2} \left(\frac{1}{4}-\epsilon^2\right)^{T/2},
\end{equation} 
for all $T\in\mathbb{N}_0$ even. Observe that RWB trajectories are rare, in the sense that their probability decreases exponentially as $T$ increases. In particular, it holds that
\begin{equation}\label{eq:probRWBlargeT}
    \lim_{T\to\infty} \left[ -\frac{1}{T}\ln P(x_T = 0) \right] =-\ln\!\left(2\sqrt{\frac{1}{4}-\epsilon^2}\right)\eqqcolon I(\epsilon),
\end{equation}
which implies that $\Pr[x_T=0]\approx \e^{-TI(\epsilon)}$ for large $T$.

\paragraph*{Reweighted dynamics.} Our goal is essentially to learn a \textit{reweighted trajectory distribution} $P_W\!(\omega_0^T )$ for RWB trajectories via reinforcement learning. Following Ref.~\cite{rose2021reinforcement}, the  distribution\footnote{In Ref.~\cite{rose2021reinforcement}, this is referred to as ``reweighted trajectory ensemble'.} is defined as
\begin{equation}
\label{eq:reweighted_probs}
    P_W\!\left(\omega_0^T \right) \coloneq \frac{W\!\left(\omega_0^T \right) P\!\left(\omega_0^T \right) }{\mathbb{E}_{\omega_0^T\sim P}\!\left[W\!\left(\omega_0^T \right) \right]},
\end{equation}
where $W(\omega_0^T)\coloneq \prod_{t=1}^T W(x_t, x_{t-1}, t)$ with $W(x_t, x_{t-1}, t)\geq 0$ is the \textit{weight function}, and $\mathbb{E}_{\omega_0^T\sim P}[\cdot]$ denotes the expectation value for a trajectory observable with respect to the original dynamics $P$ defined in 
Eq.\ \eqref{eq-original_dynamics}. In the case considered here, we take the weight function to be
\begin{equation}\label{eq:RWB_weight}
    W(x_t, x_{t-1}, t) = e^{-s x_t^2 \delta_{t,T}},
\end{equation}
with $s>0$, which yields a total weight of $1$ for RWB trajectories and between $0$ and $1$ for  trajectories with $x_T \neq 0$. 

In principle the reweighted trajectory distribution $P_W$ can be calculated analytically. In Ref.~\cite[App.~A]{rose2021reinforcement}, Rose \textit{et al.} show that it can be expressed as a time-dependent Markov chain, \textit{i.e.},
\begin{equation}
    \label{eq:reweighted_transition_probabilities}
    P_W\!\left(\omega_0^T \right) =\prod_{t=1}^T P_W(x_t|x_{t-1},t-1),
\end{equation} 
where
\begin{align}
    P_W(x_t|x_{t-1}, t-1) &= \frac{g(x_t, t)}{g(x_{t-1}, t-1)} W(x_t, x_{t-1}, t) P(x_t|x_{t-1}), 
    \label{optimalPW}\\
    g(x_t, t) &= \mathbb{E}_{x_{t+1} \sim P} \big[ W(x_{t+1}, x_t, t+1) g(x_{t+1}, t+1)\big], 
    \label{eq:g_recursive}
\end{align}  
with boundary condition $g(x, T) = 1$ for all $x$, and $P(x_t|x_{t-1})$ the transition probability of Eq.\  \eqref{eq:walker_transition_probabilities}.
In order to determine $P_W$ one has to compute $g(x_t, t)$ according to Eq.~\eqref{eq:g_recursive}, \textit{e.g.}, via an iterative or recursive algorithm. 

\begin{figure}[h]
    \centering
\includegraphics[width=0.45\textwidth]{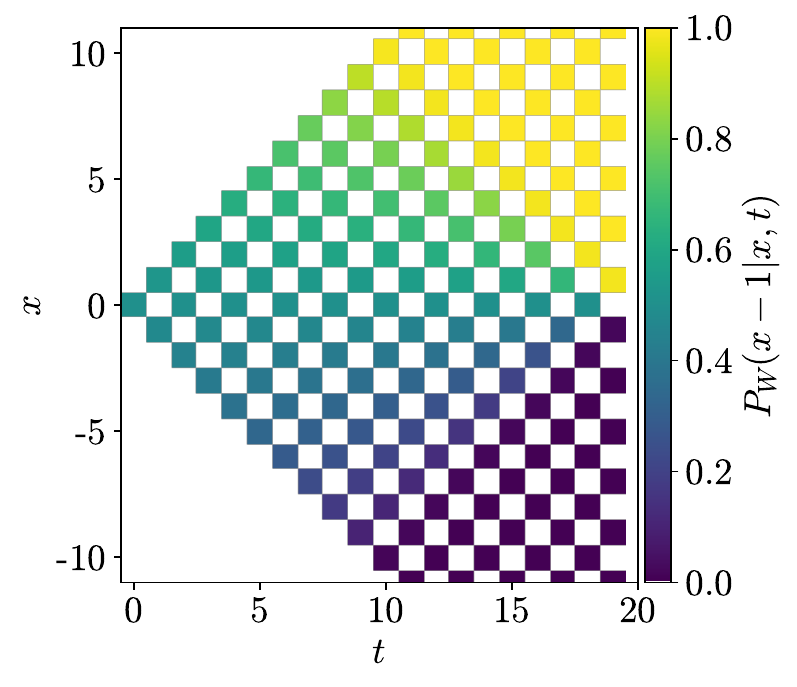}
    \caption{Reweighted transition probabilities $P_W(x - 1 | x, t)$ for $T = 20$ and weights according to Eq.\  \eqref{eq:RWB_weight} with $s=1$, calculated via Eqs.~\eqref{optimalPW} and \eqref{eq:g_recursive}. The coloring of each cell at $x, t$ indicates the probability to one step down.}
    \label{fig:P_W}
\end{figure}

As an example, in Fig.~\ref{fig:P_W} we show a plot of the exact reweighted distribution computed in this way. 
As expected, at the upper (lower) right edge of the red rhombus depicted in Fig.~\ref{fig:trajectory} the probabilities for one step down are close to $1$ ($0$) to ensure that the resulting trajectories are an RWB to a high probability.

Still, the computational complexity to solve Eqs.~\eqref{optimalPW} and \eqref{eq:g_recursive}  can quickly become 
infeasibly large, \textit{e.g.}, for increasing values of $T$, a case which is the main aim of this line of research.\footnote{Another case of interest are random walks in more than one dimension (in contrast to the one-dimensional case considered here), for which the computational complexity increases exponentially with the number of dimensions.} 
Additionally, it is likely that the algorithm is numerically unstable due to extremely small or large values involved. Therefore we aim at an alternative method to determine the reweighted transition probabilities (approximately), which is presented in the following.

\subsection{Markov decision process formulation}
\label{sec:RL}

We now outline how to formulate the random walk process from the previous subsection as a Markov decision process as a theoretical formulation of RL. Here, a \emph{Markov decision process}  (MDP)~\cite{Put14_book,sutton2018reinforcement} models a sequence of interactions of an agent with the environment, with the goal to select actions that maximize its cumulative rewards from the environment. 

Every MDP is defined by the tuple $(\mathcal{S},\mathcal{A},\{ \mathcal{T}_a \}_{a\in\mathcal{A}},r)$, which consists of the following elements:
\begin{itemize}
    \item 
    A set $\mathcal{S}$ (the ``state space'') of possible states $s$ of the environment.
    
    \item 
    A set $\mathcal{A}$ consisting of the  possible actions $a$ of the agent.
    
    \item 
    A set $\{ \mathcal{T}_a \}_{a\in\mathcal{A}}$ of transition probabilities $\mathcal{T}_a$, which give the probability for the environment to transition from any state $s \in\mathcal{S}$ to any other state,
    given that the action of the agent was $a\in\mathcal{A}$. 
    Note that the transition depends only on the current state $s$ and action $a$ (and not past ones), which is what makes the decision process Markovian.
    
    \item 
    A function $r:\mathcal{S}\times\mathcal{A}\times\mathcal{S} \times \mathbb{N} \to\mathbb{R}$, $(s_{t},a_{t-1},s_{t - 1}, t) \mapsto r(s_{t},a_{t-1},s_{t - 1}, t)$, which quantifies the reward that the agent receives at every (discrete) time step $t\in\mathbb{N}$, as a function of the previous state $s_{t-1}$, the action $a_{t-1}$, and the current state $s_t$. 
\end{itemize}
The interaction of the agent with the environment is modeled by its \textit{policy} $\pi$, 
in the way that $\pi (a, s, t)$ is the probability that the agent takes action $a\in\mathcal{A}$, given that the state of the environment is $s\in\mathcal{S}$ at time $t \in \mathbb{N}_0$. The goal of the agent is to find a policy such that its cumulative rewards are maximized. How the agent updates its policy is specified by the particular reinforcement learning method applied, which we discuss in the next subsection. But first, in the following, we formulate the random walk process as a Markov decision process, abstractly defined above. 


\begin{definition}[Random walk process as an MDP~\cite{rose2021reinforcement}]
    Consider the random walk defined in Section~\ref{sec:Stochastic model} for a given total time $T\in \mathbb{N}$. We model this process as an MDP in the following manner: Let the state space $\mathcal{S}$ consist of the possible positions $x$ of the random walker, i.e., $\mathcal{S}=\{-T,-T+1,\dotsc,+ T\}$. 
    The action set is $\mathcal{A}=\{+1,-1\}$, where ``$+1$'' denotes the action ``go one step up'' and ``$-1$'' the action ``go one step down''. The transition probabilities $\mathcal{T}_{\pm 1}$ are defined as
\begin{equation}
\label{eq-walker_action_transitions}
        \mathcal{T}_{\pm 1}(x \pm 1;x)=1,
    \end{equation}
    with all others equal to zero. Let $\Omega_T\coloneqq\{\omega_0^T=(x_0,x_1,\dotsc,x_T):x_t\in\mathcal{S},x_t-x_{t-1}\in\mathcal{A}\}$ with $T\in\mathbb{N}$ be the set of all valid trajectories of the random walk. Note that the valid trajectories are defined such that the agent only moves one step up or down in every time step. 
\end{definition}

\paragraph*{Random walker policy.} The policy of the random walker at time $t\in\mathbb{N}$ is given by a parameterized probability distribution $\pi_{\theta}$, where $\theta$ denotes the parameters that are to be optimized to obtain a policy maximizing the rewards. Then the parameterized transition probabilities of the random walker at time $t$ are given by
\begin{align}
    P_{\theta}(x_t|x_{t-1},t - 1)&= \mathcal{T}_{+1}(x_{t};x_{t-1})\pi_{\theta}(+1, x_{t-1},t-1,)+\mathcal{T}_{-1}(x_{t};x_{t-1})\pi_{\theta}(-1, x_{t-1},t-1)
    \nonumber\\
    &=\left\{\begin{array}{l l} \pi_{\theta}(\pm 1, x_{t-1},t-1) & \text{if }x_t=x_{t-1} \pm 1, \\[1ex] 0 & \text{else},    \end{array}   \right. \label{eq-random_walker_policy_transition}
\end{align}
where we have made use of \eqref{eq-walker_action_transitions}. This implies that, for every trajectory $\omega_0^T\in\Omega_T$,
\begin{equation}
    P_{\theta}\!\left(\omega_0^T \right)= \prod_{t=1}^T \pi_{\theta}(a_{t - 1}, x_{t - 1}, t - 1),
    \label{eq:parameterized_dynamics}
\end{equation}
where $a_{t - 1} \coloneq x_t - x_{t - 1}$ and choosing $x_0 = 0$ with unit probability.

\paragraph*{Random walker reward.} Since in the case considered here the transitions following an action, described by $\mathcal{T}_{\pm 1}$, are deterministic, we can use the shorthand $r(x_{t}, x_{t - 1}, t) \coloneq r(x_{t},a_{t-1},x_{t - 1}, t)$ with $a_{t - 1} = x_t - x_{t - 1}$ for the reward function $r$. In order to pursue our goal of approximating the reweighted dynamics $P_W$ in Eq.\ \eqref{eq:reweighted_probs} to be able to generate \emph{random walk bridges} 
(RWBs) with a high(er) probability, we choose the reward function $r$ to be 
\begin{align}
    r(x_{t},x_{t- 1}, t)=\ln W(x_t,x_{t-1},t)-\ln\frac{\pi_{\theta}(a_{t - 1}, x_{t - 1}, t - 1)}{P(x_t|x_{t-1})}.
    \label{eq:reward_definition}
\end{align}
In this case, the expectation value of the return $R$, defined as the sum of rewards, 
\begin{align}
R\!\left(\omega_0^T \right)&\coloneq \sum_{t=1}^T r(x_{t},x_{t- 1}, t),
\label{eq:return_definition}
\end{align}
for a trajectory $\omega_0^T\in\Omega_T$ is given by the \emph{Kullback-Leibler} (KL) divergence $D(P_{\theta}\Vert P_W)$ of the parameterized dynamics $P_{\theta}\!(\omega_0^T )$ in Eq.\ \eqref{eq:parameterized_dynamics} to the reweighted dynamics $P_W\!(\omega_0^T)$ in Eq.\ \eqref{eq:reweighted_probs}, 
\begin{align}
    \mathbb{E}_{\omega_0^T\sim P_{\theta}}\!\left[  R \!\left(\omega_0^T \right) \right] &=- D(P_{\theta}\Vert P_W) + 
    \mathbb{E}_{\omega_0^T\sim P_{W}}\!\left[  R \!\left(\omega_0^T \right) \right]
    \label{eq:KLdivergence}
\end{align}
up to the second term, which is constant w.r.t.\ the optimization parameters $\theta$. Here, we have used the definition\footnote{This formula holds as long as $\{x:Q'(x)>0\}\subseteq\{x:Q(x)>0\}$.}
\begin{equation}
    D(Q'\Vert Q)\coloneqq \sum_{x\in\mathcal{X}}Q'(x)\ln\left(\frac{Q'(x)}{Q(x)}\right)=\mathbb{E}_{x\sim Q'}\!\left[\frac{Q'(x)}{Q(x)}\right]
\end{equation}
of the KL divergence of two probability distributions $Q',Q:\mathcal{X}\to[0,1]$ over a finite alphabet $\mathcal{X}$.

From Eq.\ \eqref{eq:KLdivergence} we can see that maximizing the expected return for trajectories by adjusting the parameters $\theta$ of the policy is equivalent to minimizing the KL divergence $D(P_{\theta}\Vert P_W)$. In this sense, the agent aims to find a policy that is ``close'' to $P_W$, quantified by the KL divergence.

Another way to interpret the reward function is provided by the equality
\begin{align}
R\!\left(\omega_0^T \right)
    &=-sx_T^2-\sum_{t=1}^T \ln\frac{\pi_{\theta}(a_{t - 1}, x_{t - 1}, t - 1)}{P(x_t|x_{t-1})},\label{eq:return_RWB}
\end{align}
which follows from the choice of weight function in Eq.~\eqref{eq:RWB_weight}. Thus maximizing the return is a trade-off between achieving a high probability to generate RWBs via the policy (represented by the first term) and a high exploration of different trajectories/RWBs generated by the policy (second term). The parameter $s$ controls this trade-off, where large values favor the former and small ones the latter.

\subsection{Reinforcement-learning methods}\label{rlmethods}

We use two different 
methods of \emph{reinforcement learning} (RL) to find the policy that heuristically maximizes the return: the policy-gradient and the actor-critic framework. 
The \textit{policy-gradient} framework guarantees convergence to local optima, as it updates the parameters $\theta$ of the policy using (approximations of) the gradient of the expected return w.r.t.\  $\theta$. In our case this amounts to computing the gradient of the KL divergence $D(P_{\theta}\Vert P_W)$, as follows from Eq.\ \eqref{eq:KLdivergence}.  The gradient can be written as~\cite{rose2021reinforcement}
\begin{equation}\label{eq:gradientKL}
    \nabla_\theta D(P_{\theta}\Vert P_W) = - \mathbb{E}_{a_t \sim \pi_{\theta}}\! \left[\sum_{t=1}^T R\!\left(\omega_{t-1}^T \right) \nabla_\theta \ln \pi_{\theta}(a_{t - 1}, x_{t - 1}, t - 1) \right].
\end{equation}
Since in most cases it is not feasible to calculate this gradient exactly, we estimate it based on Monte-Carlo sampling with samples of $N$ trajectories generated according to $\pi_\theta$, \textit{i.e.}, we approximate the gradient by its empirical estimate as
\begin{align}
   \nabla_\theta D(P_{\theta}\Vert P_W) \approx -  \frac{1}{N} \sum_{i=1}^N \left\lbrack \sum_{t=1}^T R\!\left( \!\left(\omega^T_{t-1} \right)^{(i)} \right) \nabla_\theta \ln \pi_{\theta}\!\left(a_{t - 1}^{(i)}, x_{t - 1}^{(i)}, t - 1 \right) \right\rbrack. 
   \label{eq:gradientKL_estimate}
\end{align} 
Here the index $i$ labels the $N$ individual trajectories (called ``episodes'' in the language of RL) of the sample (called ``batch''). In our algorithm, we refer to the KL divergence as a ``loss function'' $L_P$ (see App.~\ref{pseudocode}). The parameters $\theta$ of the policy $\pi_\theta$ are then updated according to the formula 
\begin{align}
   \theta_{n+1} = \theta_n + \alpha_n \frac{1}{N} \sum_{i=1}^N \left\lbrack \sum_{t=1}^T R\!\left( \!\left(\omega^T_{t-1} \right)^{(i)} \right) \nabla_\theta \ln \pi_{\theta}\!\left(a_{t - 1}^{(i)}, x_{t - 1}^{(i)}, t - 1 \right) \right\rbrack,
   \label{update}
\end{align} 
where $\alpha_n$ is the learning rate (which is an adjustable hyperparameter) and $n$ denotes the batch.
%
\begin{figure}[h]
    \centering
\includegraphics[width=0.45\textwidth]{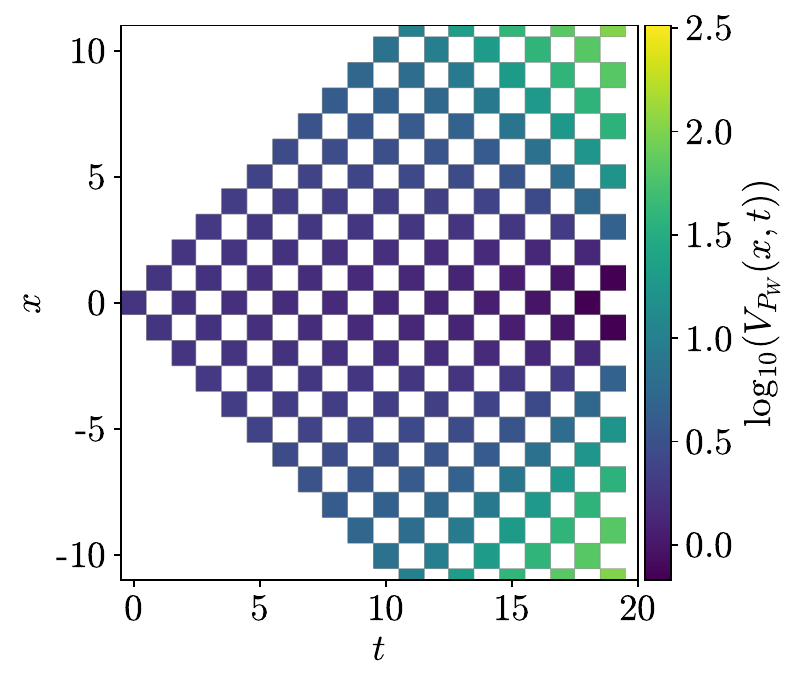}
    \caption{Exemplary result for the value function $V_{P_W}$ of the reweighted dynamics $P_W$ for $T = 20$ and weights according to Eq.\  \eqref{eq:RWB_weight} with and $s=1$, calculated via Eq.\ \eqref{eq:Bellman}. The logarithm $\log(-V_P)$ is taken for better visualization only. 
    } 
    \label{fig:value_functions}
\end{figure}
The \textit{actor-critic} framework is the other method of reinforcement learning that we employ in the same way as in Ref.~\cite{rose2021reinforcement}. In this framework, the so-called ``actor'' (in the policy-gradient framework referred to as the ``agent'') decides on an action based on the policy, and the so-called ``critic'' evaluates the action based on a value function, which captures how valuable the new state after the action is with respect to the overall goal. Thus, using a  value function is a less greedy and more long-term oriented approach. The value function $V_\pi$ of a state $s$ at time $t$ is defined\footnote{(The general definition of the value function also includes a discounting factor $0 \leq \gamma \leq 1$, but $\gamma < 1$ is not necessary here. Thus we set $\gamma = 1$ and arrive at the definition in the main text.)} as the expected sum of future rewards given the agent starts from state $s$ at time $t$ and follows the policy $\pi$, \textit{i.e.},
\begin{equation}
    V_{\pi} (s, t)= \mathbb{E}_\pi\!\left[ R\!\left(\omega_t^T | s_t = s \right) \right].
    \label{eq:value_function_def}
\end{equation}
This definition implies that the value function satisfies the Bellman equation 
as a fixed point equation (switching notation to the specific case considered here)
\begin{equation}
    V_{\pi_\theta} (x_t, t) = \mathbb{E}_{x_{t + 1} \sim \pi_\theta}\left[
 V_{\pi_\theta} (x_{t+1}, t + 1) + r(x_{t+1}, x_t, t + 1)\right], \label{eq:Bellman}
\end{equation}
where, as before, $r(x_{t+1}, x_t, t + 1)$ describes the reward of the transition/action. Thus the Bellman equation describes how the value function of the current state and time is related to that after the next time step. 
Due to the normalization $\sum_{a_t} \pi_{\theta}(a_{t}, x_{t}, t) = 1$ of the policy $\pi_\theta$, the value function can be used for a modification of the gradient of the KL divergence in Eq.\ \eqref{eq:gradientKL}, \textit{i.e.}, 
\begin{equation}
     \nabla_\theta D(P_\theta\Vert P_W) = -\mathbb{E}_{\pi_{\theta}}\!\left[\sum_{t=0}^{T - 1} \big(R(\omega^T_{t})- V_{\pi_{\theta}}(x_t,t) \big) \nabla_\theta \ln \pi_{\theta}(a_{t}, x_{t }, t ) \right],
\end{equation}
where the value function $V_{\pi_{\theta}}$ is used as a baseline for the future return $R(\omega^T_{t})$. For most problems of interest it is impossible to calculate the value function via the Bellmann Eq.\ \eqref{eq:Bellman} \textit{exactly}. Similarly to the challenge of computing the reweighted transition probabilities $P_W(x_t|x_{t-1}, t-1)$ via Eqs.~\eqref{optimalPW} and \eqref{eq:g_recursive} exactly, typical reasons in real-world applications are far time horizons $T$ and high-dimensional state spaces (coined ``the curse of dimensionality'') such that the computational complexity scales exponentially with their dimension. Thus, we introduce an approximate value function $V_\psi(x,t)$ with parameters $\psi \in \mathbb{R}^{d_V}$, whose closeness to $V_{\pi_\theta}$ is quantified via the ``mean squared deviation'' loss function $L_V(\psi)$,
\begin{equation}
L_V(\psi)= \mathbb{E}_{x_{t} \sim \pi_{\theta}}\!\left[\frac{1}{2}\sum_{t=0}^{T-1} {\big(V_\psi(x_t,t) -V_{\pi_\theta}(x_t,t) \big)}^2\right].
\end{equation}
Calculating its gradient, we get as an approximation
\begin{equation}
    \nabla_\psi L_V(\psi)\approx - \mathbb{E}_{x_{t} \sim \pi_{\theta}}\!\left[ \sum_{t=0}^{T-1} \delta_{\rm TD}(x_{t + 1}, x_{t}, t + 1)\nabla_\psi V_\psi(x_{t}, t) \right],
    \label{eq:AC_loss_function_gradient}
\end{equation}
with the temporal difference error 
\begin{equation}
   \delta_{\rm TD}(x_{t+1}, x_{t}, t + 1)=V_\psi(x_{t + 1}, t + 1) + r(x_{t+1},x_{t} , t + 1) - V_\psi(x_{t}, t).
   \label{eq:temporal_difference_error}
\end{equation}
This error evaluates how close the difference of the values of the next and the current state are to the reward. Using this error, the gradient of the KL divergence can be estimated as  
\begin{equation}
   \nabla_\theta D(P_\theta\Vert P_W) \approx - \mathbb{E}_{ \pi_{\theta}}\!\left[ \sum_{t=0}^{T - 1} \delta_{\text{TD}}(x_{t+1}, x_{t}, t + 1) \nabla_\theta \ln \pi_\theta(a_{t}, x_{t }, t ) \right]. \label{eq:KLvaluefunction}
\end{equation}
 
The difference between this gradient of the KL divergence in the actor-critic and that in the policy-gradient approach (see  Eq.~\eqref{eq:gradientKL}) lies in the substitution of the sum of returns with a sum of temporal difference errors, which include the value function. Analogously to the parameter updates in the policy-gradient framework, see Eqs. \eqref{eq:gradientKL}, \eqref{eq:gradientKL_estimate}, and \eqref{update}, the parameters of the actor and the critic are updated via empirical estimates of the KL divergence and loss function gradient in Eqs. \eqref{eq:KLvaluefunction} and \eqref{eq:AC_loss_function_gradient}, respectively, each with batch size and learning rate, all of which can be chosen independently.
To summarize the actor-critic framework: The updates of the parameters $\theta$ of the ``actor's'' policy $\pi_\theta$ are not only governed by the rewards and the current parameter values, but also by the ``critic's'' value function $V_{\psi}$; in turn the actor's policy $\pi_\theta$ governs the choice of actions and thus the trajectories generated, which not only influence the updates of $\theta$ via the rewards, but also that of the parameters $\psi$ of the ``critic's'' value function $V_{\psi}$ -- resulting in two ``intertwined control loops'' instead of one in the policy-gradient case.
he key innovation of this work is that we calculate the policy and the value function for reinforcement learning of classical rare dynamics with a parameterized quantum circuit, as explained below in Section~\ref{sec:quantum_RL}.

\subsection{Quantum policy model} \label{sec:quantum_RL}

\paragraph*{PQC policy and value function.} The idea of hybrid 
quantum-classical algorithms \cite{McClean_2016} is to embed quantum algorithms for specific computational tasks inside a classical algorithm. In this work, we embed a quantum algorithm to estimate the policy (or value function) inside the RL algorithms explained in Section~\ref{rlmethods}. Specifically, we use an instance of a variational quantum algorithm~\cite{VQA}, where certain measurements of the quantum state at the output of a \emph{parameterized quantum circuit} (PQC)  correspond to the target function of the computational task. A PQC is defined by a unitary operation $U(s,\theta)$ applied to a fixed $n$-qubit state vector $\ket{0^{\otimes n}}$, such that its output is $|\psi_{s,\theta}\rangle = U(s,\theta)|0^{\otimes n}\rangle$. The parameter $s \in \mathbb{R}^d$ represents the data input from the classical algorithm for the respective computational task. The vector $\theta= (\phi, \lambda, \omega)$ of trainable parameters is updated classically, and consists of the input scaling parameters $\lambda \in \mathbb{R}^{|\lambda|}$, the variational parameters $\phi \in {[0,2\pi)}^{|\phi|}$, which are the rotation angles of the gates in the PQC, and scalar weights $\omega_{a,i} \in \mathbb{R}$ whose meaning will be discussed below. The expectation value $\left<O_a\right>_{s,\theta}\coloneqq\bra{\psi_{s,\theta}}O_a\ket{\psi_{s,\theta}}$ of an observable $O_a$ of the PQC with respect to an action $a$ of the RL agent encodes the values of the target function. In our case, the target function is the policy.

We implement a quantum version of the softmax-RL algorithm called softmax-PQC \cite{jerbi2021variational}, for which the policy is given by
\begin{equation}
\pi_{\theta}(a, s) = \frac{e^{{\beta\langle O_a\rangle}_{s,\theta}}}{\sum\limits_{a'} e^{{\beta\langle O_{a'}\rangle}_{s,\theta}}},
\label{eq:softmax-PQC_policy}
\end{equation}
where the input data $s = (x, t)$ (different from the state $s$ used in the previous subsections) and the observable $O_a$ is restricted to a weighted sum of given Hermitian operators $H_{a,i}$, i.e., $O_a = \sum_i \omega_{a,i} H_{a,i}$, with scalar weights $\omega_{a,i} \in \mathbb{R}$ that are also called the output scaling parameters. One can think of $\beta$ as a parameter that is similar to an inverse temperature, as the policy $\pi_{\theta}(a, s)$ according to Eq.~\eqref{eq:softmax-PQC_policy} above behaves like the probability that a quantum system at inverse temperature $\beta = 1 / k_{\rm B} T$ ($k_{\rm B}$ is the Boltzmann constant, $T$ the temperature) is in a quantum state with ``energy'' $\langle O_a \rangle_{s,\theta}$. The gradient of the policy embedded in the classical gradient of the KL divergence is calculated according to Ref.~\cite{jerbi2021variational}
\begin{equation}
    \nabla_\theta\log\pi_\theta(a, s) = \beta\left(\nabla_\theta \langle O_a \rangle_{s,\theta} -\sum_{a'}\pi_\theta(a', s)\nabla_\theta \langle O_{a'} \rangle_{s,\theta}\right).
\end{equation} 
In the actor-critic case, we use two separate PQCs, one for calculating the policy of the actor and one for calculating the value function of the critic. Thus, the actor-critic PQC has twice as many trainable parameters compared to the policy gradient PQC, both parameters optimized separately and denoted by the superscripts ``A'' and ``C'', respectively. We implement the critic PQC as a function approximator for the value function, defined as 
\begin{equation}
    V_{\psi}(s) = V_{\theta^{\rm C}}(s) = {\langle O\rangle}_{s,\theta^{\rm C}}, 
    \label{eq:PQC_value_function}
\end{equation}
where $s = (x, t)$, i.e., the input data $s$ consist of the random walk position $x$ and time $t$, and analogously to the observable $O_a$ above $O = \sum_i \omega_{i}^{\rm C} H_{i}$, which will be specified in the next paragraph. 

\begin{figure}[h]
    \centering
    \includegraphics[width=0.9\textwidth]{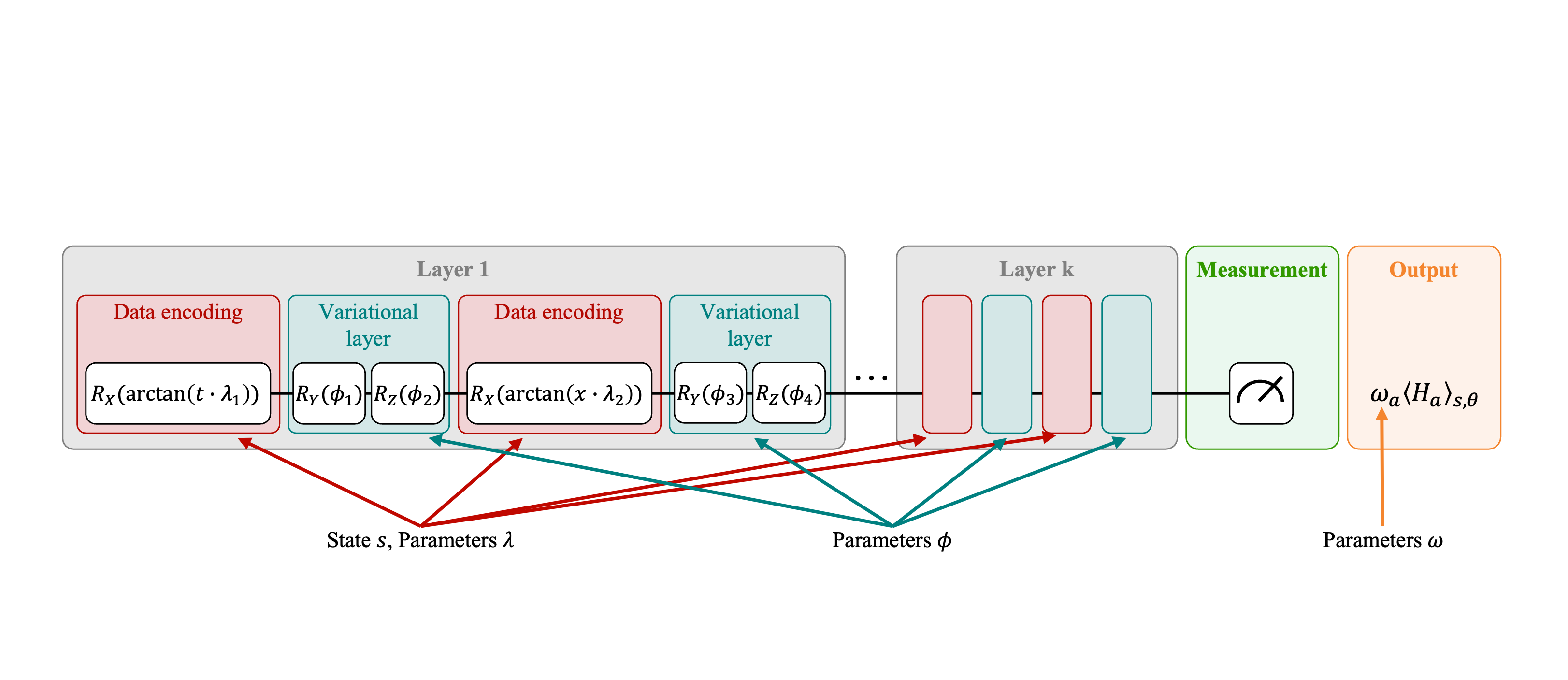}
    \caption{One-qubit parameterized quantum circuit of a hybrid quantum-classical algorithm for reinforcement learning of rare dynamics with data re-uploading. }
    \label{Circuit1Q}
\end{figure}

\begin{figure}[h]
    \centering
    \includegraphics[width=0.9\textwidth]{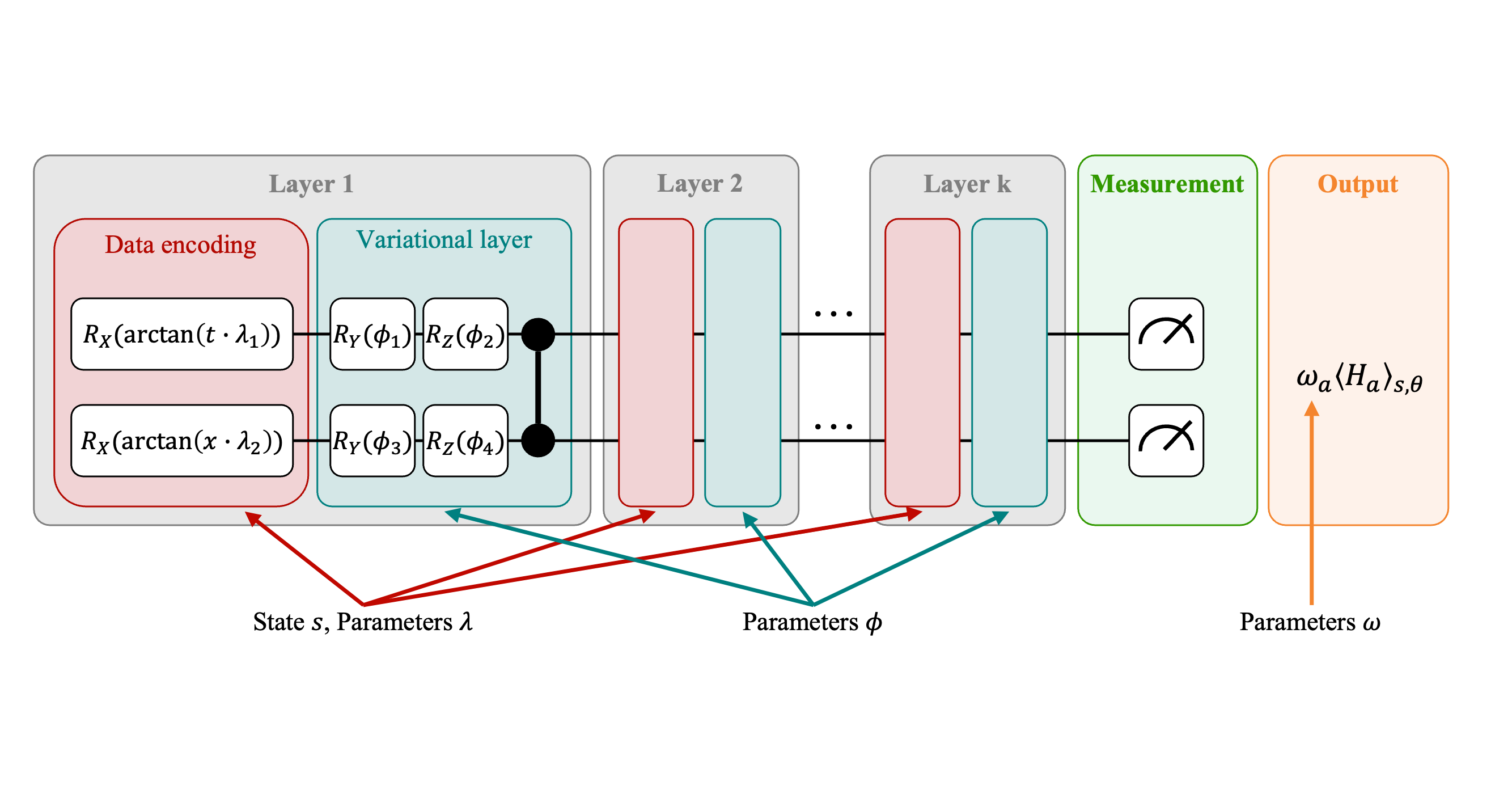}
    \caption{Two-qubit parameterized quantum circuit similar to Fig.~\ref{Circuit1Q}. The two-qubit gate is the controlled-$Z$ gate.}
    \label{CircuitReUploading}
\end{figure}

\paragraph*{Data encoding and readout.} Our data-encoding PQCs are shown in Fig.~\ref{Circuit1Q} and Fig.~\ref{CircuitReUploading}. Based on data-encoding strategies presented in Ref.~\cite{schuld2020circuit}, we encode the position $x$ into the angle $x'$ of the $R_X(x')$ gate. ($R_{X/Y/Z}(\alpha)$ denotes a rotation by an angle $\alpha$ around the $X/Y/Z$-axis of the Bloch sphere of a qubit.) Analogous to Ref.~\cite{skolik2022quantum}, the input $x'$ is related to the original input $x$ and the input scaling parameter $\lambda_x$ via 
$x' = \arctan (x \cdot \lambda_{x})$. The time step $t$ is encoded into the angle $t'$ of the second $R_X(t')$ gate (either on the second qubit or after a variational layer) via $t'= \arctan (t \cdot \lambda_{t})$. 
Although this strategy is specifically designed for continuous state spaces, we use it for the discrete state space of the random walk. This is because, compared to other encoding strategies, this type of encoding is arbitrarily scalable in $t$ and $x$ without needing to increase the number of qubits. The repetition of data-encoding layers (gray boxes labeled ``layer 1'' up to ``layer k'') is referred to as data re-uploading~\cite{perez2020data}. Without data re-uploading, a PQC is only able to model sinusoidal functions of the input with a single frequency; thus, data re-uploading is used to increase the expressivity of the circuit~\cite{schuld2020circuit}. With data re-uploading, already a one-qubit PQC, as shown in Fig.~\ref{Circuit1Q}, can be a universal quantum classifier~\cite{perez2020data}. 

As shown in Fig.~\ref{Circuit1Q} and Fig.~\ref{CircuitReUploading}, the first data-encoding layer (red box) is followed by a variational layer (blue box) consisting of $R_Y$ and $R_Z$ gates, as well as an entangling gate (to be specific, a controlled-$Z$ gate; not used in the one-qubit case). Data re-uploading consists of exact repetitions of these basic layers and is followed by a projective measurement (green box) of the observables $O_a$ to determine their expectation values. For the two actions $a = \pm 1$ to go one step up/down, we choose the observables to be $O_1 = \omega Z$ in the one-qubit case and $O_1 = \omega Z_0 Z_1$ in two-qubit case. In both cases, we let $O_{-1}=-O_{1}$.  
The probabilities to go up/down are obtained from the expectation values of their respective observables $O_{\pm 1}$ via the softmax function of Eq.~\eqref{eq:softmax-PQC_policy}. Analogously, the observable $O$ for the value function is chosen to be $O = \omega^{\rm C} Z$ in the one-qubit case and $O = \omega^{\rm C} Z_0 Z_1$.  
The variational parameters $\phi$ are updated classically, via Eq.~\eqref{update}. Each class of circuit parameters $\lambda$, $\phi$, and $\omega$ is optimized by a separate optimizer, each with its own learning rate $\alpha_{\lambda}$, $\alpha_{\phi}$, and $\alpha_{\omega}$. These learning rates as well as the inverse-temperature parameter $\beta$ are chosen via a hyperparameter search, whose details we present in Section~\ref{sec:numerical_results}.

\section{Results}\label{sec:numerical_results}

In this section, we present our numerical results for the quantum \emph{policy-gradient} (PG) and quantum \emph{actor-critic} (AC) \emph{reinforcement learning} (RL) approaches to find rare trajectories 
for the one-dimensional random walker, as defined in Section~\ref{rlmethods} and Section~\ref{sec:quantum_RL}. In order to perform the reinforcement learning and simulate the parameterized quantum circuits (PQCs), we used the Python library TensorFlow Quantum \cite{broughton2020tensorflow}. The full Python code is publicly available on GitHub \cite{quantum_rl_plot}. 
To optimize the performance of the RL agents, which are considered sensitive to hyperparameters, we conducted a heuristic hyperparameter search. This involved testing various values for each learning rate parameter---$\alpha_\lambda$, $\alpha_\phi$, $\alpha_\omega$ in the PG approach and $\alpha^{\rm A}_\lambda$, $\alpha^{\rm A}_\phi$, $\alpha^{\rm A}_\omega$, $\alpha^{\rm C}_\lambda$, $\alpha^{\rm C}_\phi$, $\alpha^{\rm C}_\omega$ in the AC approach (the superscript ``A'' refers to the actor and the superscript ``C'' refers to the critic)---and the inverse temperature parameter $\beta$. Specifically, we assessed the values  0.01, 0.05, 0.1, 0.3, 0.5, 0.7, 0.9, and 1 for each parameter and selected the combination of values that achieved an optimal balance between learning speed, convergence properties and learning outcome.  The hyperparameter values used in the actor-critic and policy gradient methods, as presented in the pseudocode (Appendix~\ref{pseudocode}), are listed for each simulation in Appendix~\ref{app:hyperparameters}.

\subsection{One- and two-qubit PQC-based policy-gradient reinforcement learning}
\label{sec:1- and 2-qubit policy gradient PQCs}

\begin{figure}[h!]
    \centering
    \includegraphics[width=8cm]{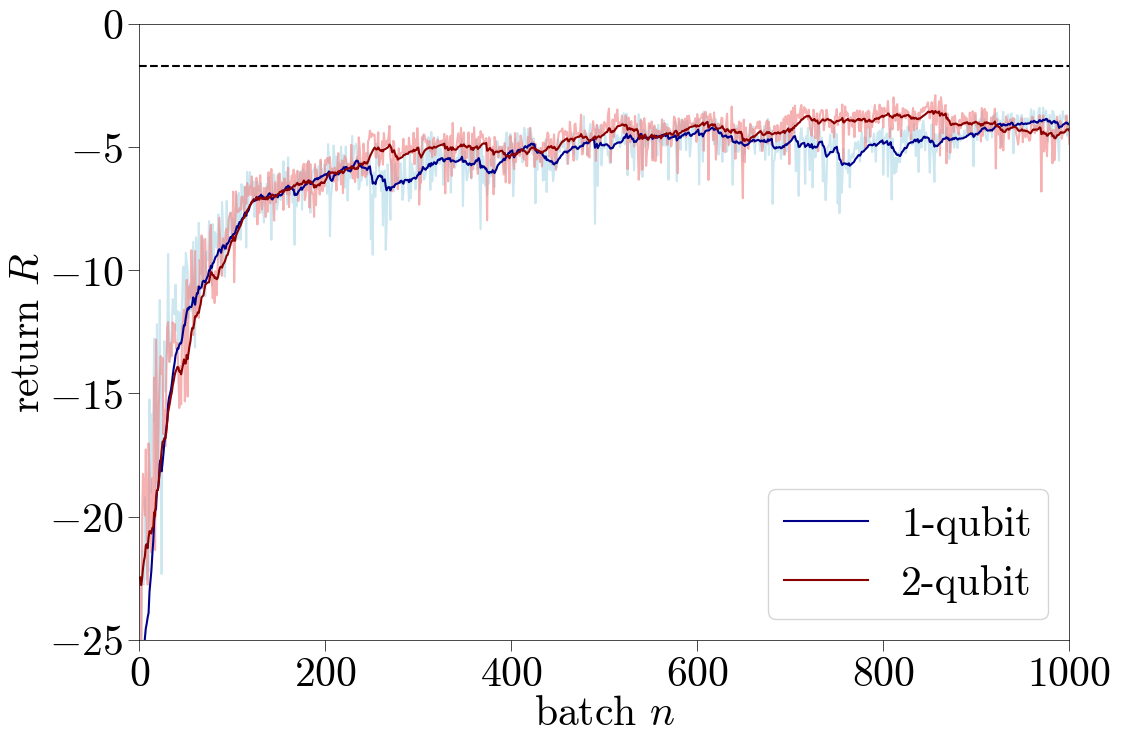}
    \includegraphics[width=8cm]{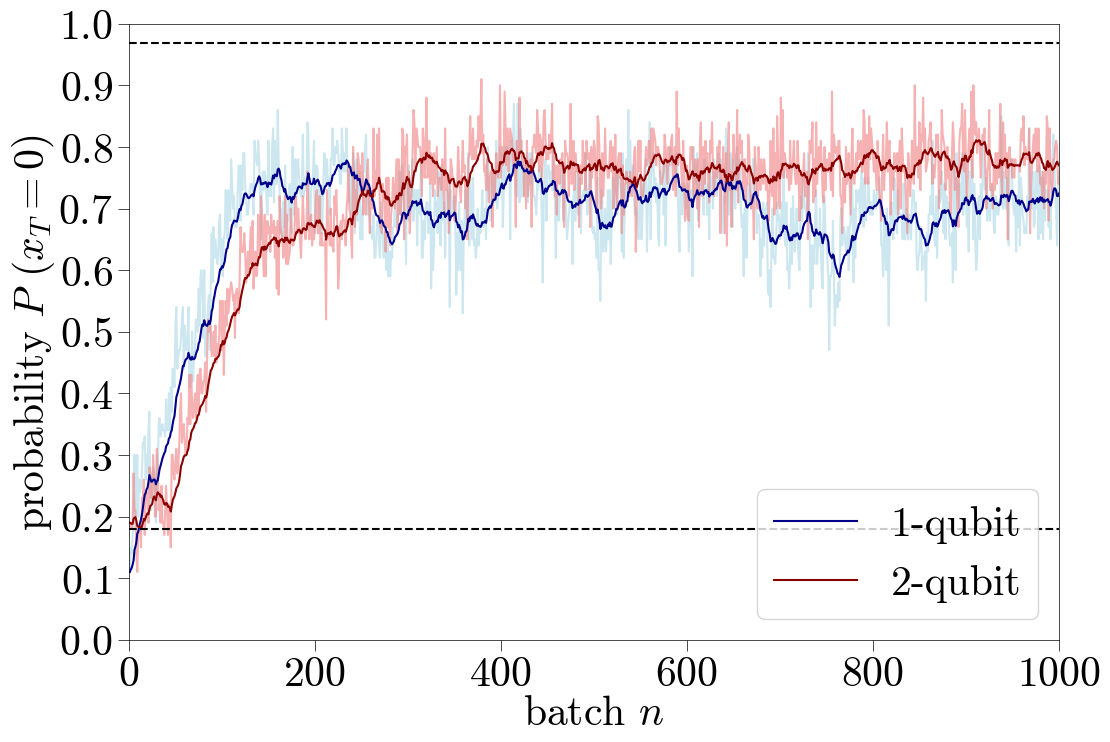}
    \caption{Performance of agents using one- (blue) and two-qubit (red) PQCs with three data re-uploading layers for policy-gradient reinforcement learning, trained to generate \emph{random walk bridges} (RWBs) with trajectory length $T = 20$. The lighter lines represent the average of 10 agents, each trained using the same hyperparameters. The darker lines represent the exponential moving average. Left: Return $R$ as a function of batches $n$, as given in Eq.~\eqref{eq:return_RWB}, each consisting of 10 
    episodes (trajectories). Right: Probability of generating a RWB with endpoint $x_T = 0$ per batch. The hyperparameters  of the agents used in these plots can be found in Tab.~\ref{paramsFig6} of App.~\ref{app:hyperparameters}. 
    }
    \label{Fig6}
\end{figure}

\begin{figure}[h!]
    \centering
    \includegraphics[width=8cm]{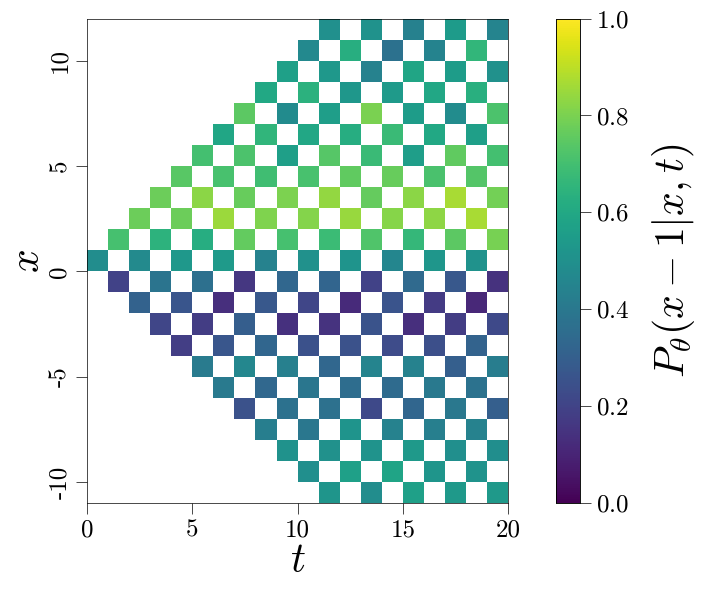}
    \includegraphics[width=7cm]{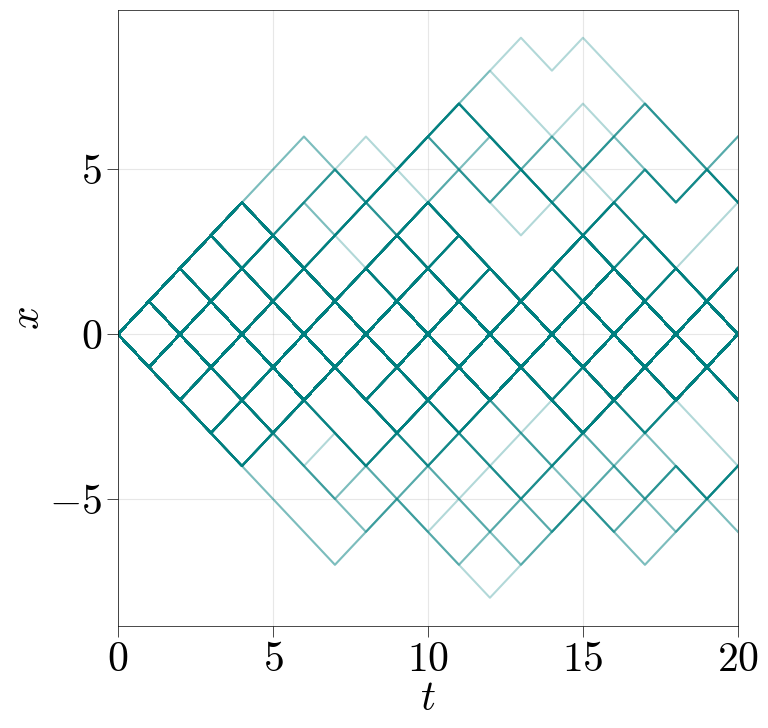}
    \caption{Results of agents using two-qubit PQCs with three data re-uploading layers for policy-gradient reinforcement learning trained to generate \emph{random walk bridges} (RWBs) with trajectory length $T = 20$. Left: Final policy averaged over 10 trained agents, where the coloring indicates the probability that the agent jumps one step down from its current state. Right: Sampled 1000 trajectories from the final policy of one randomly selected agent. 
    The thicker the line, the more often the transition indicated by the line occurred. The hyperparameters of the agent used in these plots can be found in App.\ \ref{app:hyperparameters}, Tab. \ref{paramsFig6}, right column. 
    \label{Fig7}}
\end{figure}

First, we assess whether the one- and two-qubit PQC-based policy-gradient RL agents are able to learn to generate \emph{random walk bridges} (RWBs). Our specific results are shown in Figs.~\ref{Fig6} and \ref{Fig7}. Both plots in Fig.~\ref{Fig6} show results averaged over 10 independently trained agents, where the lightly colored lines indicate these average results and the darkly colored lines indicate the trend of these averages, in the form of an \emph{exponential moving average} (EMA) with a smoothing factor $\alpha = 0.1$. (Calculating the EMA of $x_t$ gives the smoothed value $S_t$ via $S_t=\alpha x_t +(1-\alpha)S_{t-1}$ \cite{NIST_Handbook}.) The stochastic fluctuations arise 
from the initial random sampling of variational parameters and the Monte-Carlo gradient sampling, see Section~\ref{sec:RL} for further details. Averaging the results of 10 agents and calculating the EMA reduces the appearance of these fluctuations and increases the representativity of the results.

In Fig.~\ref{Fig6} (left), we plot the agents' return (as defined in Eq.~\eqref{eq:return_RWB}) as a function of the number of batches. We can see that the agents achieve a high degree of learning, as indicated by an increasing return that converges to a value close to the optimal return. The dashed line indicates the expected return for the optimal policy $P_W$. This is approximated by generating $N = 10^5$ trajectories according to the probabilities $P_W(x_t|x_{t-1}, t-1)$, computing the return for each trajectory, as if an agent acted according to the policy $P_W$, and taking the mean of the results. The number $N$ is chosen sufficiently large to ensure convergence. The difference between this expected return and the actual return is the KL divergence $D(P_{\theta}\Vert P_W)$, which quantifies the proximity of the agents' policy $P_\theta$ to $P_W$. 
Successful learning is also characterized by a growing likelihood of generating a RWB as the batch number increases. In Fig.~\ref{Fig6} (right), we plot this probability of generating a RWB, which we estimate by taking the fraction of the rare trajectories obtained during training. For every agent and for every batch $n$, consisting of 10 episodes, we count the number of RWBs obtained within this batch. We then compute the average of these values across the 10 agents. 
These probabilities should be compared to the probability $P(x_T = 0)=T! \left(\frac{T}{2}! \right)^{-2} 2^{-T}  \approx 0.18$ (for $T = 20$) of generating an RWB via the original random walk dynamics $P$ (lower black line in the right side of Fig.~\ref{Fig6}). Thus, the algorithm achieves a significantly better result in generating an RWB, compared to $P$. The probability of generating an RWB via $P_W$ is 0.97 (represented by the top black line in the right side of Fig.~\ref{Fig6}; also approximated by generating $N = 10^5$ trajectories).


Both plots in Fig.~\ref{Fig6} show that the one- and the two-qubit PQCs (each with three data-uploading layers) successfully learn to generate RWBs, though with small differences in their learning behavior, as can be seen in the right plot showing the probabilities of generating an RWB. 
While the one-qubit PQCs learn faster, the two-qubit PQCs achieve higher probabilities in the long term.
These differences in learning behavior might be explainable by the different PQC architecture and 
thus different "return landscapes" analogous to "loss landscapes" (\textit{i.e.}, graphs/hypersurfaces of loss functions as functions of the parameters to be optimized) in other optimization tasks. Both faster initial learning and less accuracy after training suggest a less complicated return landscape. Still, an analysis of the return landscapes is beyond the scope of our work, as well as an analysis of the effects of scaling the number of data-uploading layers on the return landscapes, compared to that of scaling the number of qubits onto which data-uploading gates are applied. Both of these open questions are deferred to future work.
It is to be noted though that the differences (averaged with an EMA, see darker lines in Fig.\  \ref{Fig6}) between one- and two-qubit agents are small fractions of the overall results and approximately within the statistical fluctuations (see lighter lines in Fig.\  \ref{Fig6}). 

The left plot of Fig.~\ref{Fig7} visualizes the dynamics generated by the average final policy of 10 agents after training using a two-qubit PQC with policy-gradient RL. 
We observe that around the $x=0$ axis, the probability is roughly $50\%$, while for larger values of $x$, there is a high probability to ``jump'' back to the $x=0$ axis. The high values of $x>8$ or $x<-8$ are explored very little, in comparison to the policy of $P_W$, see Fig.\  \ref{fig:P_W}.
From this, we can conclude that the agent learned an essential piece of information about the problem: along the origin, the walker has more freedom to move farther away from the origin in order to traverse more RWBs; on the other hand, for larger $x$, the walker has less freedom to move away from the origin.
The right-hand side of Fig.~\ref{Fig7} shows 1000 sample trajectories generated by the agent using the trained policy. The percentage of approximately $81.5\%$ seems consistent with the right-hand side of Fig.~\ref{Fig6}. 

\FloatBarrier
\subsection{Details of one- and two-qubit PQC-based policies via Fourier series} 
\label{sec:1-qubit PQC}

As explained in Subsection \ref{sec:quantum_RL}, the parameterized dynamics $P_\theta$ derives from the softmax policy in Eq.~\eqref{eq:softmax-PQC_policy}, with the expectation value ${\langle O_a\rangle}_{s,\theta}$ as quantum input from PQCs.
The numerical results of the last section (cf. Figs.~\ref{Fig6} and \ref{Fig7}) indicate that $P_\theta$ for PQCs already with only \textit{few} qubits and \textit{few} data-uploading layers can approximate the optimal policy $P_W$ well. (In the following, $P_\theta$ is called ``policy'' for the sake of brevity.)

The aim of this subsection is to understand the reasons underlying this  success from yet another perspective: in terms of a Fourier analysis of the expectation value ${\langle O_a\rangle}_{s,\theta}$. To this end, we
\begin{enumerate}
    \item[(i)]
    express ${\langle O_a\rangle}_{s,\theta}$ in terms of a truncated Fourier series with the input state $s = (x, t)$ as variable,

    \item[(ii)]
     determine which of its Fourier coefficients can be non-zero and are not identical to others, 

     \item[(iii)]
     fit $P_\theta$ to $P_W$ with these Fourier coefficients as fitting parameters, using 
     a classical optimization procedure, and

     \item[(iv)]
     investigate the results of the previous steps.
\end{enumerate}
We note that this approach essentially constructs a ``classical surrogate'' of the PQC, as presented in Ref.~\cite{schreiber2023surrogates}, but applied within the context of reinforcement learning. The central idea of this approach is that for few data-uploading layers few enough Fourier coefficients are non-zero such that the optimization problem of fitting independent Fourier coefficients should be simpler than that of fitting all variational parameters $\theta$, \textit{i.e.}, formulated in the language of the theory of PQCs: The policy $\pi_\theta$ and the ``optimization landscape'' is simplified to increase the \textit{trainability}, potentially at the expense of \textit{expressivity}. In the following we show that this is indeed the case such that a classical optimization procedure can solve the optimization problem sufficiently well.

The results in this subsection will support the plausibility that the softmax policy $P_\theta$ of Eq.~\eqref{eq:softmax-PQC_policy} of PQCs with only \textit{few} qubits and \textit{few} data-uploading layers can approximate the optimal policy $P_W$ effectively. Furthermore, the results will illuminate the reasons for some of the results in the subsequent subsections. Most prominently: why RL agents using PQCs with two qubits and one data-uploading layer perform like the original dynamics \textit{without} reweighting and thus exhibit poor performance; 
and why with increasing numbers of layers the performance first ameliorates, but then deteriorates. The latter effect will be further discussed in Sec.~\ref{sec:Comparison of data re-uploading and parameter variation}. In summary, the Fourier analysis of this section explains the mentioned results via the expressivity of the softmax policy $P_\theta$, depending on the hyperparameters of the PQCs (\textit{e.g.}, the number of data-uploading layers). 

\paragraph*{Step (i).} The first step of our Fourier analysis outlined above (as well as the second) was motivated by the results of Ref.~\cite{schuld2020circuit}.
In App.~\ref{app:Fourier_analysis}, we show for our data-encoding and data-uploading strategies in Fig.~\ref{Circuit1Q} that ${\langle O_a\rangle}_{s,\theta}$ with variational parameters $\theta= (\phi, \lambda, \omega)$ can be expressed in terms of a truncated Fourier series, \textit{i.e.},
\begin{align}
   {\langle O_a\rangle}_{s,\theta} = \sum_{n_x = - N_x}^{N_x} \sum_{n_t = - N_t}^{N_t} c_{n_x n_t} e^{i (n_x x + n_t t)}, 
   \label{eq:truncated_Fourier_series}
\end{align}
(for the moment omitting input scaling) with Fourier coefficients $c_{n_x n_t}$ depending on the observable $O_a$, the structure of the PQC, and the variational parameters $\theta$ only (and not on the state $s = (x, t)$). (Input scaling with trainable parameters $\lambda$ would lead to adaptive frequencies and thus more expressivity, see Ref. \cite{schuld2020circuit} and the references therein, but also to a more complicated form than Eq.\ \eqref{eq:truncated_Fourier_series}, see App.\ \ref{app:Fourier_analysis}, Eq.\ \eqref{eq:ungrouped_truncated_Fourier_series_multi-index_adaptive_freqs}.) In the sum above, $N_x$ and $N_t$ denote the finite maximum frequencies for the variations in $x$ and $t$, respectively, both $N_x$ and $N_t$ increase linearly with the number of data-uploading layers. In order to ensure that ${\langle O_a\rangle}_{s,\theta}$ is real, it must hold that $c_{n_x,n_t} = c_{-n_x,-n_t}^*$. This result is similar to that of Schuld \textit{et al.} in Ref.~\cite{schuld2020circuit}: They show that for certain (other) data-encoding and data-uploading strategies, the expectation value ${\langle O_a\rangle}_{s,\theta}$ (without input scaling) is given by a multivariate partial Fourier series like Eq.\ \eqref{eq:truncated_Fourier_series}.


\paragraph*{Step (ii).} The next step of our analysis as stated above is to determine which of the Fourier coefficients $c_{n_x n_t}$ can be non-zero and are not identical to others. This is done in two different ways as explained in App.\ \ref{app:Fourier_analysis}. The results suggest that, except for the cases \textit{without} data-reuploading, all $c_{n_x n_t}$ are general non-zero complex numbers and not identical to each other except the relation $c_{n_x,n_t} = c_{-n_x,-n_t}^*$; again, see App.~\ref{app:Fourier_analysis}. 

\paragraph*{Step (iii).} 
The third step of our analysis is to fit the parameterized policy $P_\theta$ of Eq.\ \eqref{eq:softmax-PQC_policy}  
to the optimal policy $P_W$. 
To this end, both the fitting parameters and the loss function to be minimized by the fits are chosen 
as explained in App.\ \ref{app:Fourier_analysis}. 
In all cases the state variables $x$ and $t$ are replaced by the scaled variables $x' = \arctan (x \cdot \lambda_{x})$ and $t'= \arctan (t \cdot \lambda_{t})$, respectively, and the fitting parameters include only two input scaling parameters, $\lambda_x$ and $\lambda_t$. In contrast to the other parts of this work where $\lambda_x$ and $\lambda_t$ can be trained independently for each encoding gate $R_X(x_t')$, this is a simplification. This follows the central idea of this subsection as stated above; it is a disadvantage in terms of the expressivity, but can be an advantage in terms of trainability and avoiding overfitting. In this context also note that our implementation of a classical fitting algorithm neither guarantees to find a global nor a (good) local optimum.

\begin{figure}[h]
    \centering
\includegraphics[width=0.46\textwidth]{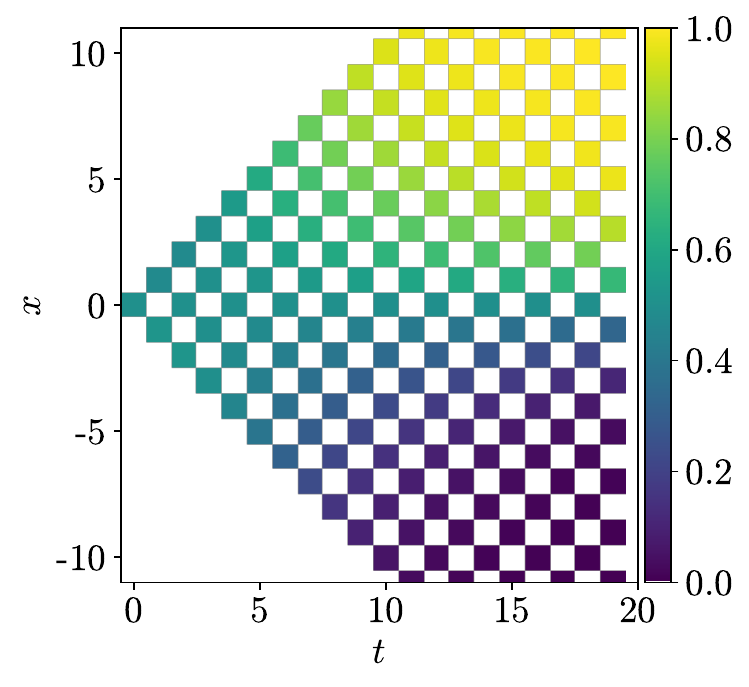}
\includegraphics[width=0.49\textwidth]{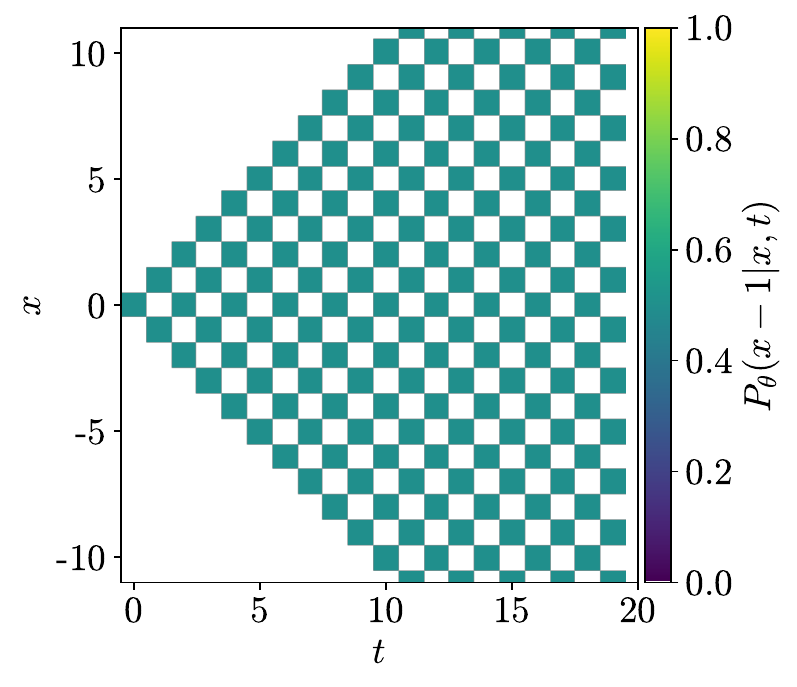}
    \caption{Plots of the best fits  for the policy $P_\theta$ with quantum input from the 1- and the 2-qubits PQC in Figs. \ref{Circuit1Q} and \ref{CircuitReUploading}, respectively, each with 1 data-uploading layer. These results are based on the Fourier analysis presented in Section\  \ref{sec:1-qubit PQC}. In both cases $P_\theta$ is fitted as function of the modulus and phase of its Fourier coefficients in Eq.\ \eqref{eq:truncated_Fourier_series} and its input and output scaling parameters (for more details see main text). Left: best fit for 1-qubit PQC; right: best fit for 2-qubits PQC. The values for the hyperparameters $T$ and $s$ used for these plots can be found in App.\ \ref{app:hyperparameters}, Tab. \ref{paramsFig6}.}
\label{fig:policy_fits_1_qubit_1_layer}
\end{figure}

\begin{figure}[h]
    \centering
\includegraphics[width=0.5\textwidth]{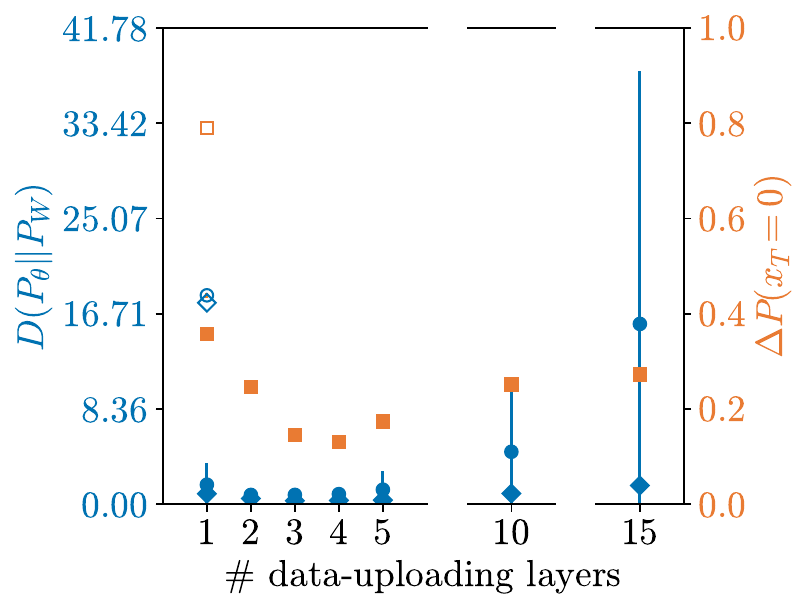}
    \caption{
    Plot of the main numerical results for the Fourier analysis and fits of the policy $P_\theta$, as presented in Section\  \ref{sec:1-qubit PQC}, for the fitting procedure which yields the best results. All results are obtained from 100 fits with randomly initialized fitting parameters. The vertical axis indicates the number of data-uploading layers of the utilized PQCs. The red diamond symbols, circles, and bars indicate the results for the \emph{Kullback-Leibler} (KL) divergence $D(P_\theta\Vert P_W)$ of $P_\theta$ fitted to $P_W$, while the green squares indicate the results for the difference $\Delta P(x_T = 0)$ of the probability to generate a rare trajectory via $P_W$ and via $P_\theta$ with the least KL divergence. In the case of 1 data-uploading layer, the filled symbols denote the results for the 1-qubit PQC and the empty one that for the 2-qubits PQC. For more data-uploading layers, the results apply to both 1- and 2-qubits PQCs. The diamond symbols denote the minimum, the circles the mean and the bars the standard deviation of $D(P_\theta\Vert P_W)$. For the numerical values used in this plot see App.\ \ref{app:Fourier_analysis}, Tab.  \ref{tab:overview_further_num_results_step_2_ii}.
    }
\label{fig:plot_table_results_few_qubits_cases}
\end{figure}

\paragraph*{Step (iv).} Finally, we analyze the results of the three steps explained above. All results are obtained for randomly initialized fitting parameters. Tables with all results of step (iii) for the fitted policy $P_\theta$ in terms of loss after fitting (``loss'' in short), 
\emph{Kullback-Leibler} (KL) divergence $D(P_\theta\Vert P_W)$ to $P_W$, and probability $P(x_T = 0)$ to generate a random walk bridge can be found in App.\ \ref{app:Fourier_analysis}. Exemplary plots of the best found fits of $P_\theta$ for 1 qubit and 2 qubits with 1 data-uploading layer are shown in Fig.\  \ref{fig:policy_fits_1_qubit_1_layer}; the best found fit for 3 data-uploading layers is so close to that for 1 qubit with 1 data-uploading layer such that we do not display it. (Here and in the following ``best'' is used in the sense that the loss of the respective fit is minimal compared to all other plots with the same set of (hyper-)parameters.) 
These plots are supplemented with Fig.\  \ref{fig:plot_table_results_few_qubits_cases}, which shows the \emph{mean squared error} (MSE) of $P_\theta$ to $P_W$
and $P(x_T = 0)$ of the best found fits, both with respect to different numbers of data-uploading layers. 

The results for more than 1 data-uploading layer are both valid for the corresponding 1-qubit and 2-qubits PQCs, because as discussed above the results of step (ii) suggest that all Fourier coefficients of all considered 1-qubit and 2-qubits PQCs with 2 and more layers are general non-zero complex numbers and thus the results of step (iii) (with Fourier coefficients as fitting parameters) are valid for both.\footnote{As a consequence, from this viewpoint one cannot explain the differences between 1- and 2-qubit cases shown in Fig.~\ref{Fig6}. Since the frequencies and number of Fourier coefficients are equal in both cases, we expect the differences to result from the fact that in detail the Fourier coefficients depend differently on the variational parameters $\theta$. An investigation of this hypothesis is beyond the scope of our work and thus is left as future work.}

One can infer from the plots in Fig.\  \ref{fig:policy_fits_1_qubit_1_layer}, 
the MSE,
and $P(x_T = 0)$ of the best found fit that the results for 1 qubit are already quite good for 1 data uploading layer, while that for 2 qubits and 1 layer are quite bad. The latter can probably be explained by the numerical evidence that in this case the truncated Fourier series is almost as restricted as it can be (it has only one cosine amplitude as fitting parameter, see App.\ \ref{app:Fourier_analysis}).

Furthermore, when increasing the number of data uploading layers, these results improve up to 3 data-uploading layers and then deteriorate, while the standard deviation $\sigma$(MSE) of the MSE
of all fits for each set of (hyper-)parameters tends to increase throughout. This suggests that the results first improve, because the number 
of available fitting parameters increases, 
but when the fitting parameters become too many, it becomes harder to find the global minimum or even (good) local minima with the fitting algorithm. This seems to cause 
$\sigma$(MSE)
still to increase and the other results to deteriorate. 

Now let us compare the best found fit of $P_\theta$ for 3 data-uploading layers, visually indistinguishable from the left plot in Fig.\  \ref{fig:policy_fits_1_qubit_1_layer}, to the corresponding exemplary result in the left part of Fig.\  \ref{Fig7} obtained by quantum reinforcement learning. It is obvious that both results are overall similar, but the RL results have features of higher frequencies appearing more irregularly. The reason for these features might be adaptive frequencies/more expressivity due to independently trainable input scaling parameters $\lambda_x$ and $\lambda_t$ for each encoding gate $R_X(x_t')$, as mentioned above in steps (i) and (iii) (see also Eq.\ \eqref{eq:ungrouped_truncated_Fourier_series_multi-index_adaptive_freqs} in App.\ \ref{app:Fourier_analysis}). This might also cause such features in analogous plots in subsequent subsections.





A disadvantage of the loss function of the best fits is that it requires knowing $P_W$, which can be computed for small enough values of $T$, but not for large values. This can be remedied by other choices of the loss function, as discussed in App.~\ref{app:Fourier_analysis}. One such loss function is the Kullback-Leibler divergence to $P_W$ used to update the variational parameters of the \emph{reinforcement learning} (RL) agents, see Eqs.~\eqref{eq:gradientKL_estimate} and \eqref{update}. 

In summary, the four-step Fourier analysis of this subsection suggests that softmax policies $P_\theta$ with quantum input from 1-qubit and 2-qubits PQCs can approximate the optimal policy $P_W$ well \textit{in principle} (except for the case of the 2-qubits PQC with 1 data-uploading layer, which will be checked for RL in Section\  \ref{sec:Comparison of data re-uploading and parameter variation}). 
Now the central question is whether the PQCs (aside from those in the last subsection) are trainable enough so that the RL agents find such good approximations \textit{in practice}. (Note that the ``loss landscape'' defined by the loss function and the chosen fitting parameters might differ from that traversed by the RL agent). This will be answered in the next subsections.


\FloatBarrier
\subsection{Results for actor-critic PQC-based quantum reinforcement learning}
\label{sec:Comparison of models for policy gradient and actor-critic reinforcement learning}

\begin{figure}[h!]
    \centering
    \includegraphics[width=8cm]{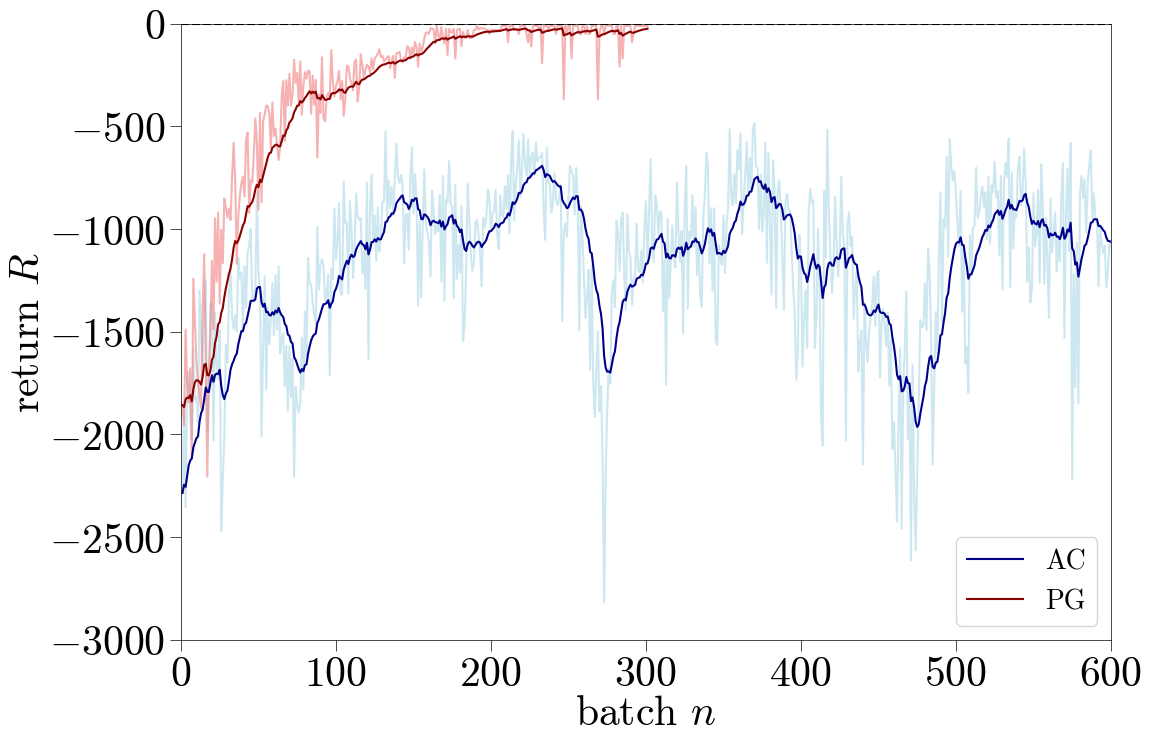}
    \includegraphics[width=8cm]{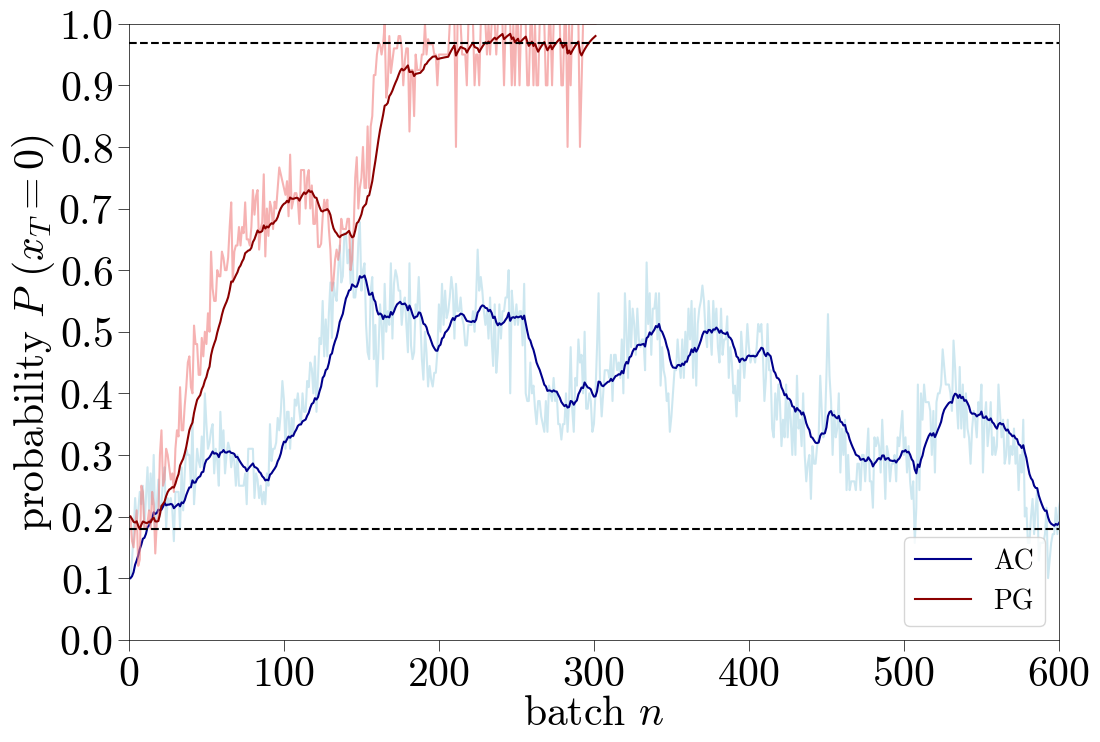}
    \caption{Comparison of the quantum policy-gradient (PG, red lines) and quantum actor-critic (AC, blue lines) reinforcement-learning models trained to generate \emph{random walk bridges} (RWBs) with trajectory length $T=20$ and 3 re-uploading layers. The lighter lines
represent the average of 10 agents, each trained using the same hyperparameters. The darker lines
represent the exponential moving average.
    Left: Return $R$ as a function of batches $n$, as given in Eq.~\eqref{eq:return_RWB}, each consisting of 10 episodes (trajectories). Right: Probability of
generating an RWB with endpoint $x_T = 0$ per batch.  The hyperparameters  of the agents used in these plots can be found in Appendix~\ref{app:hyperparameters}, Table~\ref{paramsFig8}.} \label{fig:modelcomparison}%
\end{figure}

In order to compare the performance of the different RL models with each other and their classical counterpart of Ref.~\cite{rose2021reinforcement}, we trained 10 AC and PG RL agents, as described in Section~\ref{sec:RL} and Section~\ref{sec:quantum_RL}. Both models use a two-qubit PQC with three data-uploading layers. All agents were trained until they had generated 100 rare trajectories. As shown in Fig.~\ref{fig:modelcomparison}, both policy-gradient RL agents learn such that they achieve a return close to zero in the end, compared to the actor-critic which shows clear learning behavior, but fails to get a return close to the PG RL agents. This learning behavior is similar to the results in Ref.~\cite{rose2021reinforcement} regarding RL without PQCs. In both scenarios, the AC agents exhibit slower convergence towards achieving higher returns, yet they maintain a substantial degree of exploration. This is evidenced by the persistent fluctuations in both the high- and low-frequency components of the return curve. The behavior of the PG agents is reversed, indicating a higher exploitation: They show a fast convergence to a high return and a lower exploration, indicated by less fluctuations of the return curves. Since similar effects were observed in Ref.~\cite{rose2021reinforcement} it seems reasonable to conclude that this difference is not caused by the quantum parts of the hybrid quantum-classical algorithms, but by the different RL models. One of the reasons is that during training information about the reward structure propagates more slowly in the AC model than in the PG model \cite{rose2021reinforcement}. This is due to the different structure of the gradient of the KL divergence, as seen in Eqs.~(\ref{eq:KLdivergence}) and (\ref{eq:KLvaluefunction}), since the AC also takes the TD error into account. 

\begin{figure}[h!]
    \centering
    \includegraphics[width=0.3\textwidth]{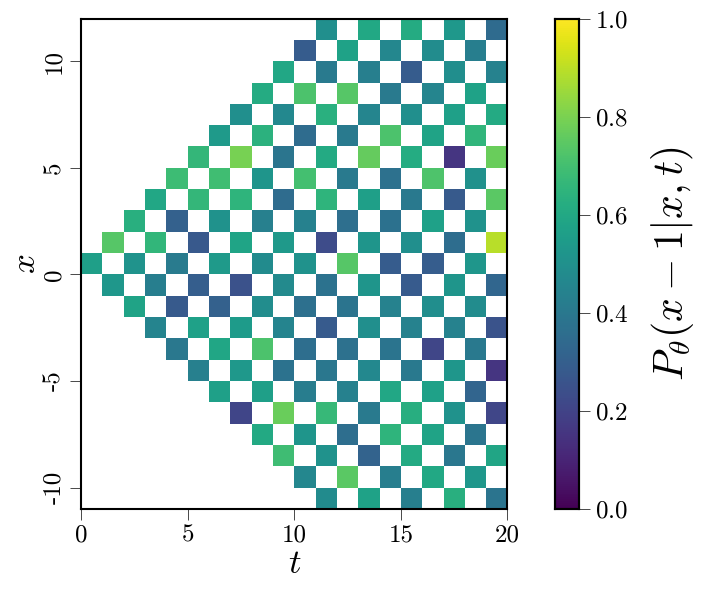}
    \includegraphics[width=0.3\textwidth]{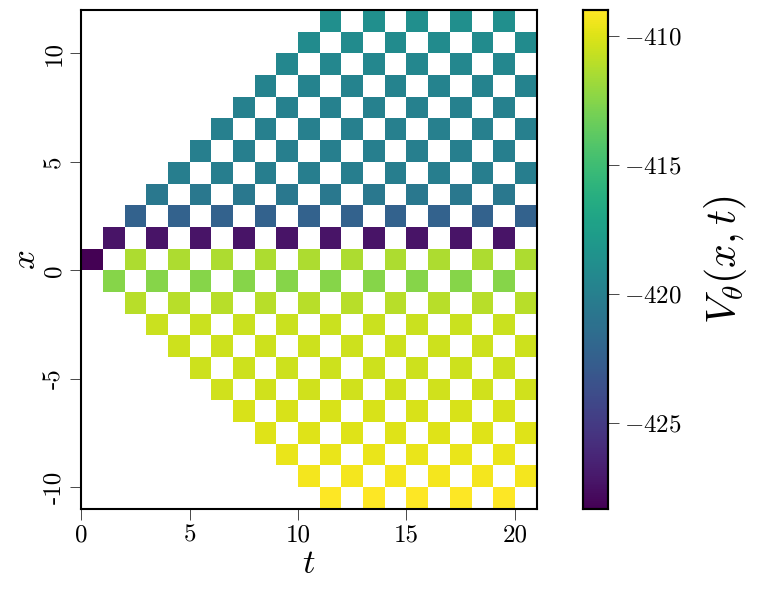}
    \includegraphics[width=0.25\textwidth]{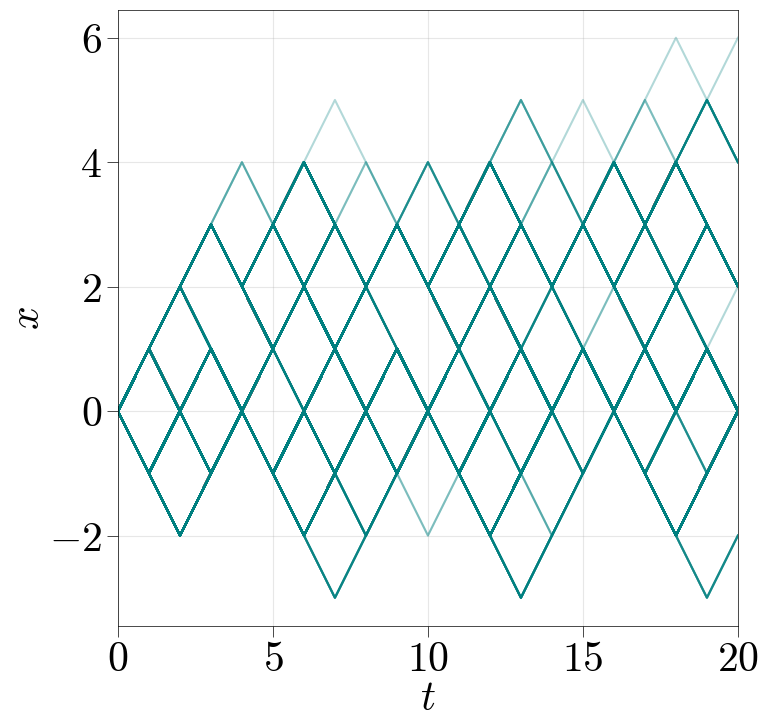}
    \caption{Results of agents using two-qubit PQCs with three data re-uploading layers for
actor-critic reinforcement learning to generate random walk bridges
(RWBs) with trajectory length $T = 20$. Left: Final policy averaged over 10 trained agents,
where the coloring indicates the probability that the agent jumps one step down from its
current state. Middle: Final value function $V_{{\theta^{\rm C}}}(x,t)$ averaged over 10 trained agents. Right: Sampled 1000 trajectories from the final policy of one randomly selected agent.  The thicker the line, the more often the transition indicated by the line occurred. 
    The hyperparameters of the agent used in these plots can be found in Appendix~\ref{app:hyperparameters}, Table~\ref{paramsFig8}.}
    \label{actorCriticPlots}
\end{figure}

In Fig.~\ref{actorCriticPlots}, the policy (left) and the approximation of the value function (middle) are computed from the average of 10 agents. The generated trajectories (right of Fig.~\ref{actorCriticPlots}) are visualized for one randomly selected AC agent after training. The obtained policy has a very different structure from that of the optimal $P_W$, as shown in Fig.~\ref{fig:P_W}. It is also different from the policy of the PG agent, with respect to a notable $x$-axis mirror-symmetric structure, probably because the overall optimisation also takes the value function into account. The value function, in contrast, shows a clear mirror symmetric structure, which favors little deviation of the trajectories from the origin. For comparison, the optimal value function is shown in Fig.\  \ref{fig:value_functions}. The visualized trajectories of a single agent do \textit{not} necessarily reflect the  exploratory behavior of the average agent.

\FloatBarrier
\subsection{Effects of data re-uploading and variation of parameter in PQCs}
\label{sec:Comparison of data re-uploading and parameter variation}

\begin{figure}[h]
    \centering
    \includegraphics[width=8cm]{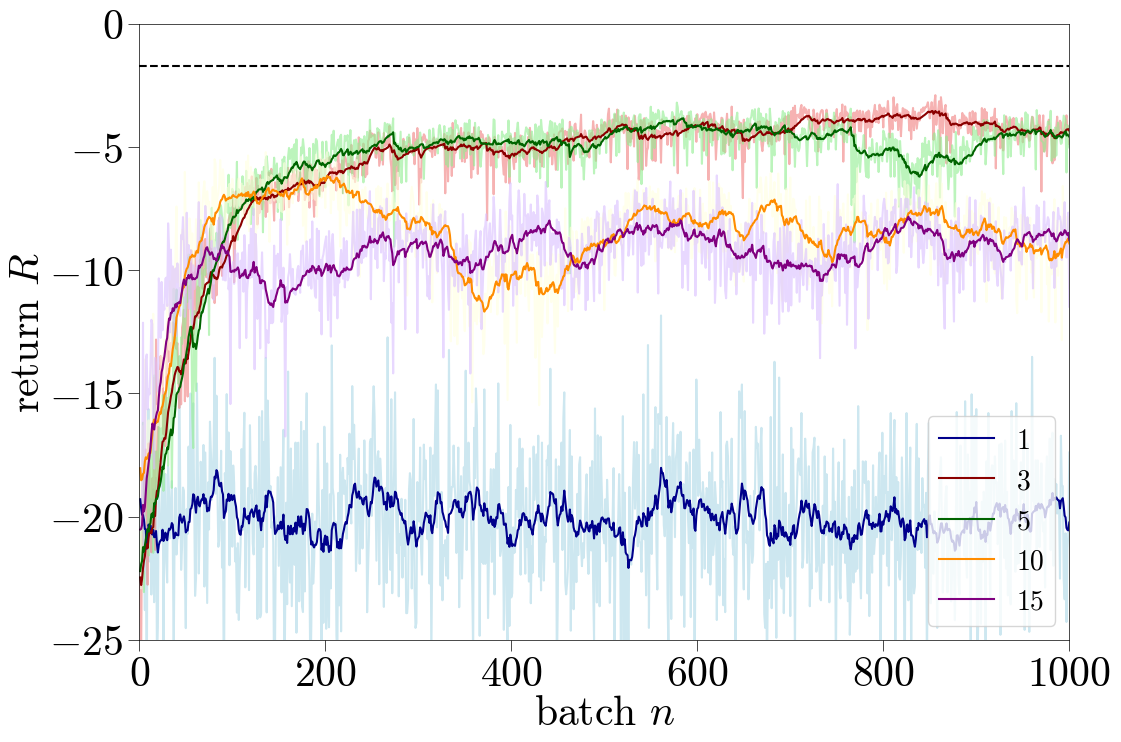}
    \includegraphics[width=8cm]{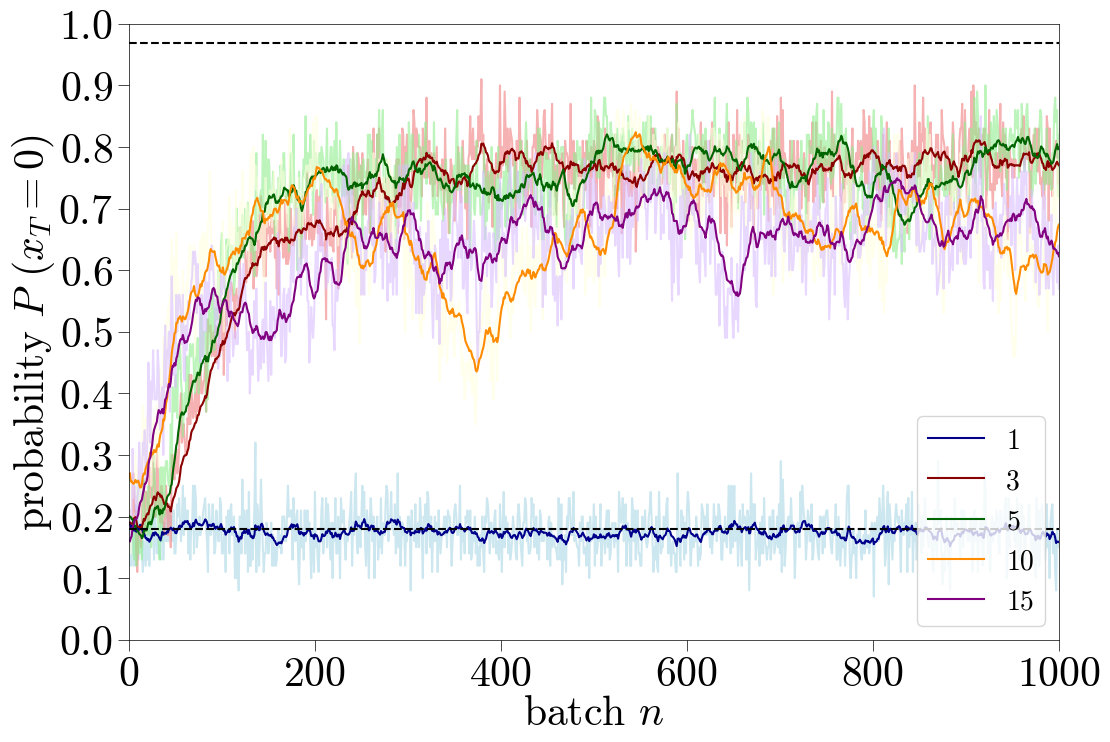}
    \caption{Comparison of varying numbers of data-uploading layers (as indicated in the legend) in the two-qubit PQCs of the policy-gradient reinforcement-learning approach, trained to generate \emph{random walk bridges} (RWBs) with trajectory length $T = 20$. 
    The lighter lines
represent the average of 10 agents, each trained using the same hyperparameters. The darker lines
represent the exponential moving average.
    Left: Return $R$ as a function of batches $n$, as given in Eq.~\eqref{eq:return_RWB}, each consisting of 10 episodes (trajectories). Right: Probability of
generating a random walk bridge with endpoint $x_T = 0$ per batch.  The hyperparameters of the agents used in these plots can be found in Table~\ref{paramsFig11} of Appendix~\ref{app:hyperparameters}.}
    \label{fig:layers1}
\end{figure}

We have investigated the effect of data re-uploading on the learning and exploration behavior of the quantum policy-gradient RL agents. To this end, we trained RL agents using two-qubit PQCs with varying numbers of data-uploading layers to generate RWBs of length $T = 20$, as shown in Fig.~\ref{fig:layers1}. In the left part of Fig.~\ref{fig:layers1}, PG agents using two-qubit PQCs with one data-uploading layer (red line) are not able to learn, 
most likely because four of the six in general non-zero and independent Fourier coefficients are zero for the specific two-qubit PQC with one data-uploading layer here. Moreover, the remaining two Fourier coefficients are equal to each other, as shown in Section~\ref{sec:1-qubit PQC}. A clear improvement in the maximal return and the rate to reach it can be observed when the number of data re-uploading layers of the PQCs for policy-gradient RL is increased from one to three and five, while a further increase reverses this trend. This result is in line with the empirical results of Skolik \textit{et al.}~\cite{skolik2022quantum}, where additional layers first improve the convergence of the agent, and then slow it, too. The latter seems surprising and does not fit with the observations of Ref.~\cite{jerbi2021variational} that, in general, an increase in the depth of the PQC used for reinforcement learning improves the performance of the agent. As shown in Ref.~\cite{schuld2020circuit}, the expressivity for data re-uploading circuits increases with the addition of more layers. The observed effect in our simulations could be attributed to the often non-convex loss landscape of shallow underparametrized PQCs (which do not exhibit barren plateaus), which is characterized by numerous local minima, and thus optimization with gradient-based methods is harder in such loss landscapes. In Refs.~\cite{Anschuetz_2022,you2021exponentiallylocalminimaquantum}, this is shown for quantum neural networks. Our results suggest that the decrease in trainability with increasing circuit depth, which may increase again at greater depths, could also be a phenomenon observed in data re-uploading circuits. We leave this investigation to further studies. The results obtained from fitting the truncated Fourier series of the policy $P_\theta$ in Sec.~\ref{sec:1-qubit PQC} point in a similar direction, see Fig.\  \ref{fig:plot_table_results_few_qubits_cases}: When increasing the number of data-uploading layers, the KL divergence $D(P_\theta\Vert P_W)$ (both its minimum and its mean for independent fits) decreases up to three layers, but increases for more layers (we have investigated up to 15). In comparison, Ref.~\cite{jerbi2021variational} increase the number of layers only up to 10, 
and thus might not reach the regime in their work where performance decreases again. 

\begin{figure}
    \centering
    \includegraphics[width=15cm]{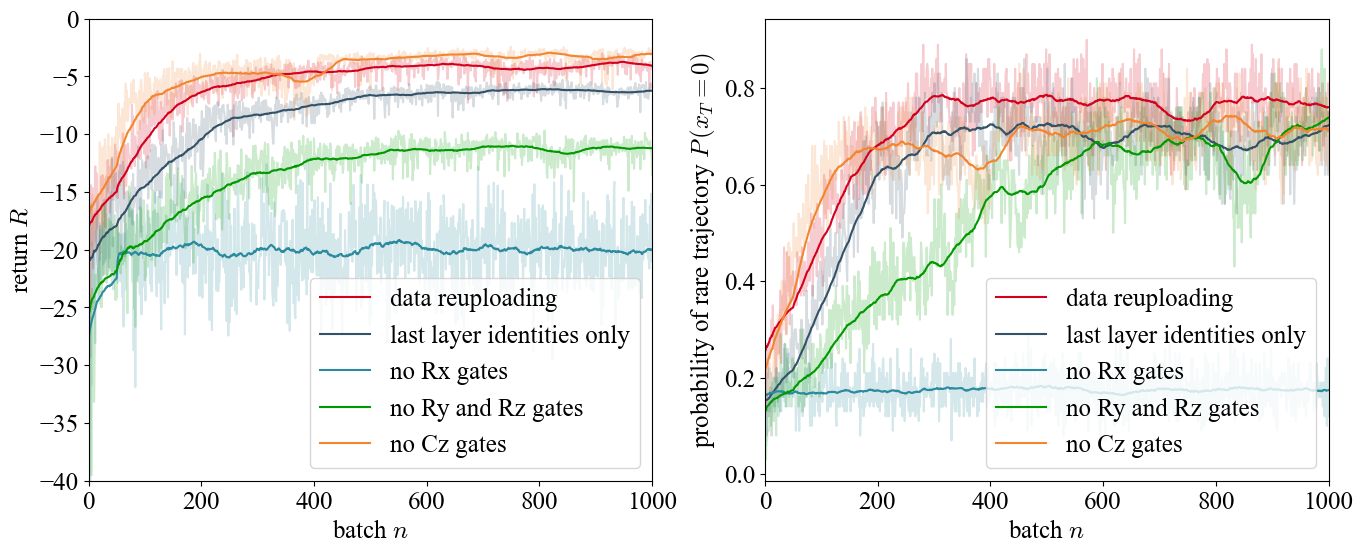}
    \caption{Comparison of deactivating specific gates of the two-qubit data-uploading PQCs of the
policy-gradient reinforcement-learning approach, trained to generate random
walk bridges (RWBs) with trajectory length $T = 20$. The PQC includes input scaling, parameterized by $\lambda$. The lighter lines
represent the average of 10 agents, each trained using the same hyperparameters. The darker lines
represent the exponential moving average. Left: Return $R$ as a function of batches $n$, as given in Eq.~\eqref{eq:return_RWB}, each consisting of 10 episodes (trajectories). Right: Probability of
generating a random walk bridge with endpoint $x_T = 0$ per batch. The hyperparameters of the agents used in these plots can be found in Table~\ref{paramsFig13} of Appendix~\ref{app:hyperparameters}.}
    \label{dataReUploading}
\end{figure}

\begin{figure}
    \centering
    \includegraphics[width=15cm]{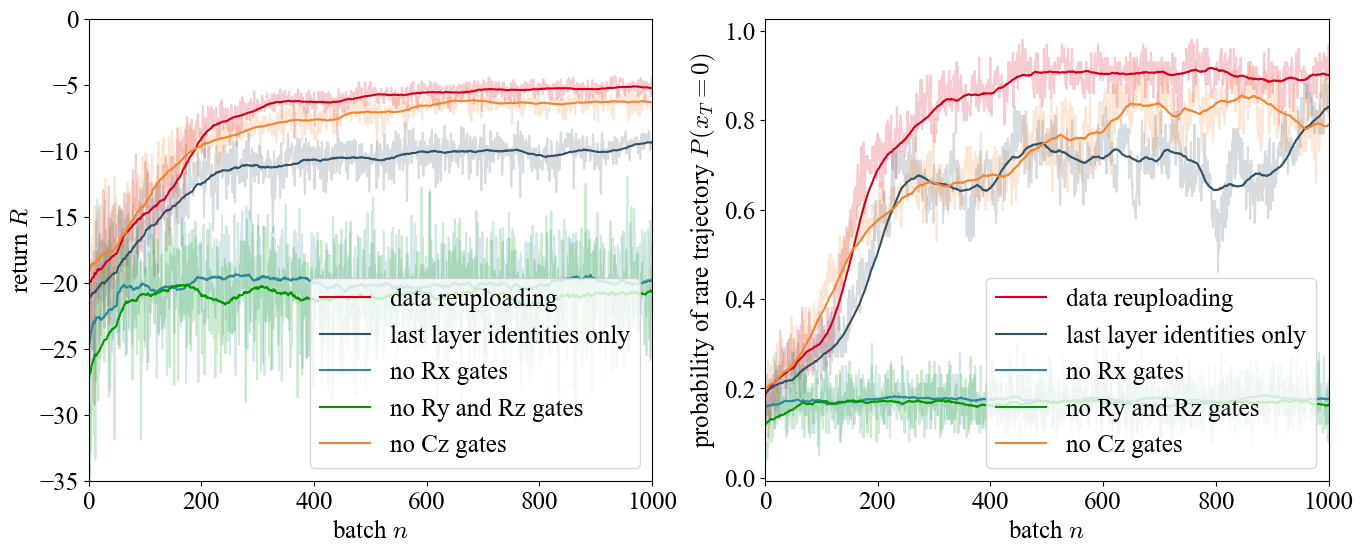}
    \caption{Comparison deactivating specific gates in the PQCs of reinforcement-learning agents (the trajectory length $T=20$). The PQC includes no input scaling parameterized by $\lambda$. The lighter lines
represent the average of 10 agents, each trained using the same hyperparameters. The darker lines
represent the exponential moving average. Left: Return $R$ as a function of batches $n$, as given in Eq.~\eqref{eq:return_RWB}, each consisting of 10 episodes (trajectories). Right: Probability of
generating a random walk bridge with endpoint $x_T = 0$ per batch. The hyperparameters of the agents used in these plots can be found in Tab. \ref{paramsFig13} of App.\ \ref{app:hyperparameters}.}
    \label{dataReUploading2}
\end{figure}

In order to further investigate which types of gates, layers or scaling parameters are relevant for the ability to generate RWBs, we removed certain parts of the policy gradient PQCs, i.e., replaced them with identity gates. The results can be seen in Fig.~\ref{dataReUploading} and Fig.~\ref{dataReUploading2}. We see that removing the entangling gate (controlled-$Z$ gate, $C_Z$) has a small positive effect on the rate of convergence and maximal value of the return and a small negative one on the maximal probability to generate RWBs. One reason for this behavior could be that only the first two entangling $C_Z$ gates in a circuit with $k = 3$ layers have an effect on the expectation value of the observable $Z_1 Z_2$ and thus on the policy, because the last $C_Z$ gate commutes with $Z_1 Z_2$ and hence cancels with its Hermitian conjugate in the expectation value (this is true 
for general $k$). 
The small effect on the performance of the agent implies that no entangling gate layer is necessary for the problem under investigation, which is further supported by the good numerical results of the one-qubit PQCs. In general, quantum circuits resulting in an $n$-qubit state without entanglement have an easy hardware implementation and are compatible with the NISQ era (further discussed in Section~\ref{sec:Outlook and discussion}). Such states are separable into $n$ tensor products of one-qubit states and thus can in general be simulated efficiently with classical computers. Nevertheless, the numerical results of the quantum RL approaches presented here are ``better'' than that of RL approaches which choose the specific parameterization of policy and value function in terms of classical neural networks, as we present below in Section~\ref{sec:Comparison to classical neural networks}. This indicates that the better ability to fit the reweighted dynamics $P_W$ is determined by the specific structure of the PQCs considered here. Motivated by this, we investigated the structure further by deactivating other parts of the PQC.

In both Fig.~\ref{dataReUploading} and Fig.~\ref{dataReUploading2}, it can be seen that the rotation gates have the largest influence on the learning rate and ability of the agent. Without these, the agent cannot obtain the optimal return. As expected, the largest effect is in the absence of the $R_{X}(\alpha)$ gates (blue line). This agent was not able to learn, since this layer encodes the state of the random walker. The $R_{Y}(\alpha)$ and $R_{Z}(\alpha)$ gates play a crucial role with the parameters $\phi$ and thus for learning the parametrized dynamics $P_\theta$. It can be seen in Fig.~\ref{dataReUploading} that the agent is able to learn at a slower rate.

Figure~\ref{dataReUploading2} shows the results of similar ``deactivation'' of specific gates, but with the input scaling parameters set to one, i.e., $\lambda=1$. It can be seen clearly that in comparison to Fig.~\ref{dataReUploading}, the agent is not able to learn without the $R_{Y/Z}(\alpha)$ gates. This hints at the input parameters $\lambda$ having enough influence on the agent to still be able to learn successfully, even without $R_{Y/Z}(\alpha)$ gates. It can be seen clearly in Fig.~\ref{dataReUploading2} that the effect of an absence of the input parameters $\lambda$ is that it slows down the learning behavior independent of the data re-uploading, although the effect is the least visible in the reference circuit. Aided by the Fourier mapping we can see the effect of an ``adaptive frequency matching'' \cite{schuld2020circuit} for further details see Sec.~\ref{sec:1-qubit PQC}. Incidentally, an analogous result is observed in Ref.~\cite{jerbi2021variational}, but in a different context. This suggests that, in general for RL with PQCs, training the input scaling parameter $\lambda$ can be beneficial.

\FloatBarrier
    \subsection{Quantum reinforcement learning compared to classical neural networks}
\label{sec:Comparison to classical neural networks}

In this section, we compare the performance of \emph{reinforcement-learning} (RL) agents whose policy (and value function, in the actor-critic case) are parameterized by either \emph{parameterized quantum circuits} (PQCs), as discussed in the previous sections, or \emph{neural networks} (NNs) with varying number of neurons in the hidden layers and different activation functions. We have implemented the NNs with the Python version of the software library Tensorflow\footnote{https://www.tensorflow.org/} and used a fully connected NN architecture. For each NN, the input layer consists of two neurons, one for the time $t$ and the other for the position $x_t$ of the walker. The output layer consists of two neurons (one for each action) for the policy-NN and only one neuron for the critic-NN. Both the policy-NN and the critic-NN/actor-NN have two hidden layers. For a fair comparison to the number of parameters in the PQC architectures, the parameter count of the NNs is way below current standards (e.g. 256 or 512 neurons per hidden layer). The NN parameters are updated and the policy is computed as described in Section~\ref{sec:Comparison of data re-uploading and parameter variation} 
\begin{figure}[h!]
    \centering
    \includegraphics[width=7.5cm]{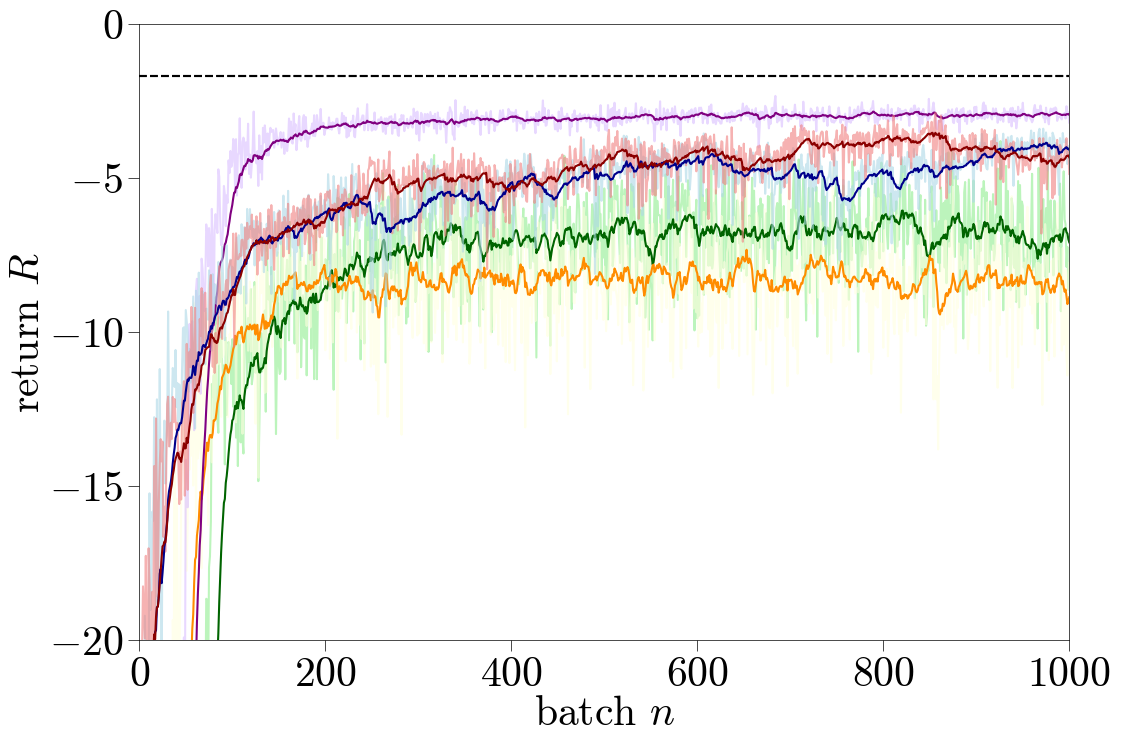} 
    \includegraphics[width=7.5cm]{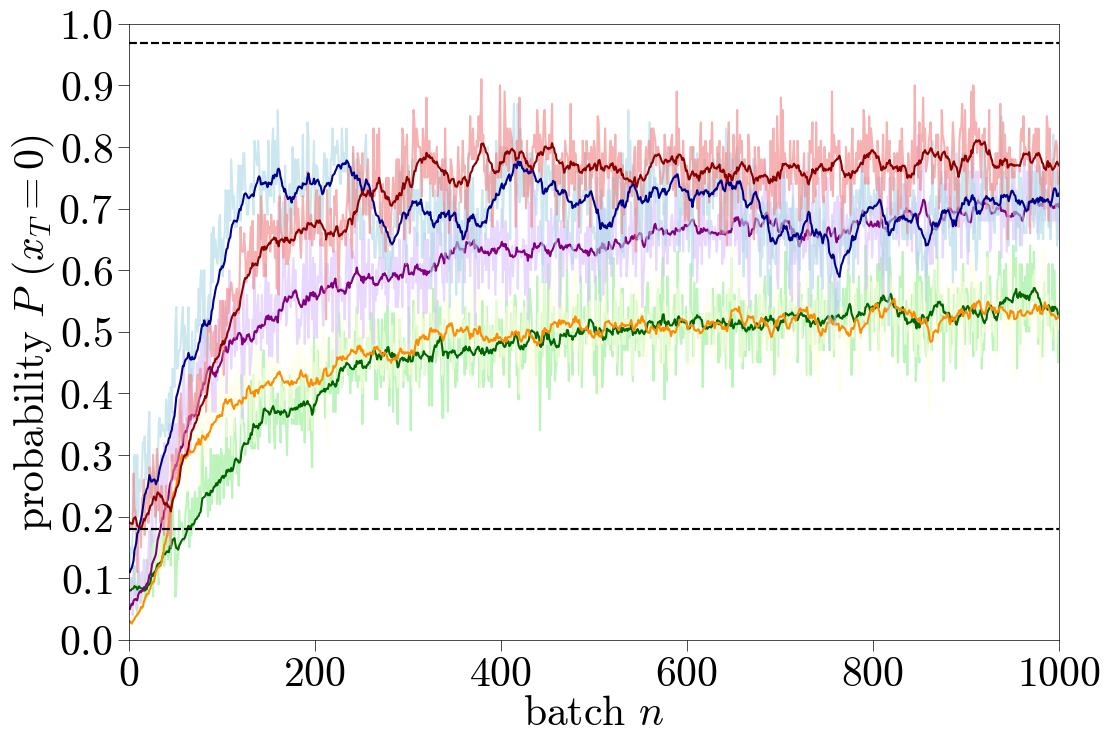} 
    %
    \caption{Comparison between classical policy-gradient neural-network agents (NN) with ReLU activation function and one- and two-qubit policy-gradient PQC agents, trained
to generate \emph{random walk bridges}  (RWBs) with trajectory length $T = 20$. Plot legend: 2-qubit PQC (20 params)(red), 1-qubit PQC (20 params)(blue), NN (18 params)(green), NN (42 params)(orange), and NN (57 params)(purple). The lighter lines represent the average of 10 agents, each trained using the same hyperparameters. The darker lines represent the exponential moving average. Left: Return $R$ as a function of batches $n$, as given in Eq.~\eqref{eq:return_RWB}, each consisting of 10 episodes (trajectories). Right: Probability of
generating a random walk bridge with endpoint $x_T = 0$ per batch. The hyperparameters of the agents used in these plots can be found in Table~\ref{paramsFig15} of Appendix~\ref{app:hyperparameters}.}
    \label{ReluNN}
\end{figure}

\begin{figure}[h!]
    \centering
    \includegraphics[width=7.5cm]{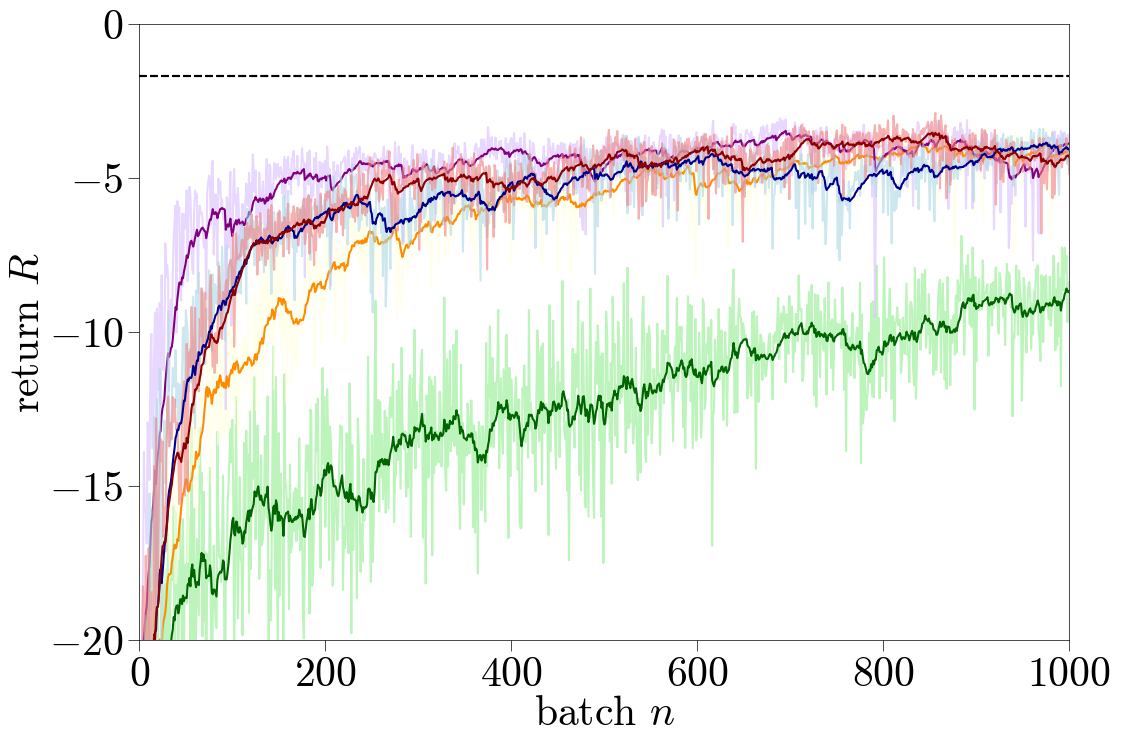}
     \includegraphics[width=7.5cm]{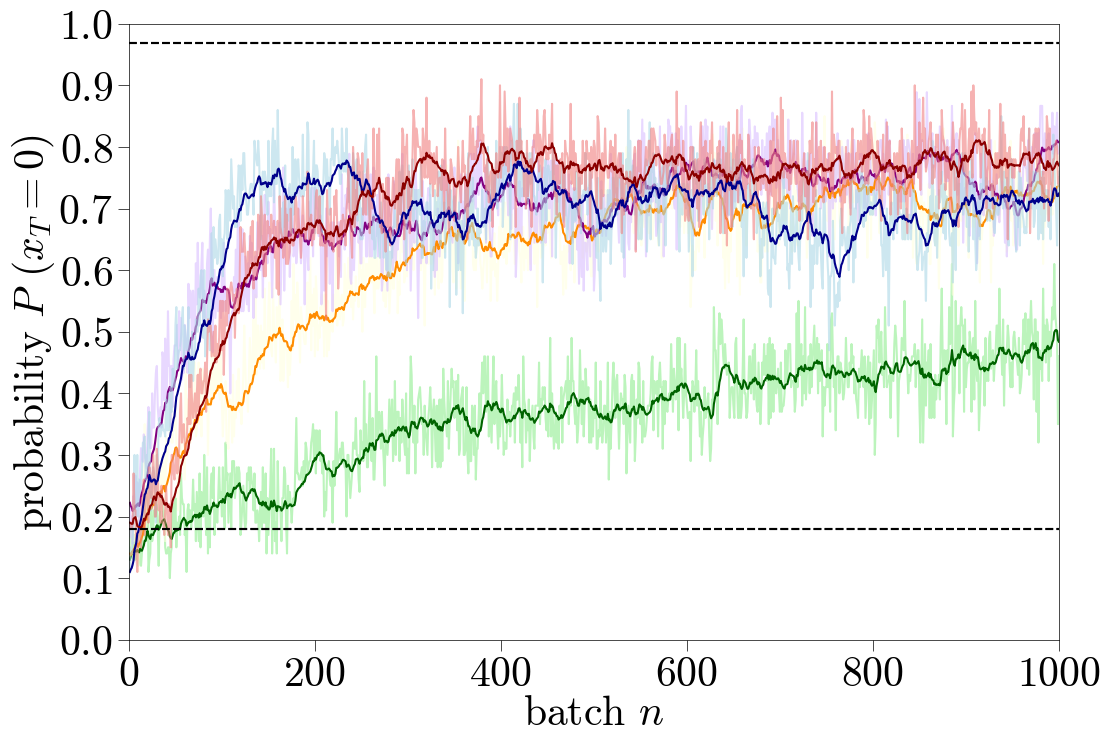}
    \caption{Comparison of policy-gradient neural-network agents (NN) with sinusoidal activation function and one- and two-qubit policy-gradient PQC agents, trained
to generate \emph{random walk bridges}  (RWBs) with trajectory length $T = 20$. Plot legend: 2-qubit PQC (20 params)(red), 1-qubit PQC (20 params)(blue), NN (18 params)(green), NN (42 params)(orange), and NN (57 params)(purple). The lighter lines represent the average of 10 agents, each trained using the same hyperparameters. The darker lines represent the exponential moving average. Left: Average return $R$ per batch $n$.  Right: Probability of generating a rare trajectory $P(x_T = 0)$. The hyperparameters of the agents used in these plots can be found in Tab. \ref{paramsFig15} of App.\  \ref{app:hyperparameters}. }
    \label{SinNN}
\end{figure}

In Fig.~\ref{ReluNN} we compare the NNs to one-qubit (bottom) and two-qubit (top) PQC. The NN a ReLU (rectified linear unit) activation function given by $f(x)= \max\{x,0\}$, where $x$ represents the input data~\cite{aggarwal2018neural}. This activation function is used in machine learning to introduce non-linearity, and it is commonly used due to its efficiency. It can be be seen that one- and two-qubit PQCs (red and blue lines) outperform the NN with a comparable number of parameters (18 parameters, green line) and twice the number of parameters (42 parameters, orange line), both in terms of the return convergence and in the probability to generate a rare trajectory. The NN with 57 parameters (purple lines) outperforms the PQCs in terms of rate of convergence and final received the return, but not in the probability of generating a rare trajectory.  This indicates that an NN with a ReLU activation function needs at least three times the number of parameters compared to a PQC for similar performance. 
While ReLU activation functions are the most widely utilized in neural networks, we also trained the networks using a periodic activation function to ensure an unbiased comparison. Figure~\ref{SinNN} shows this comparison between NNs with sinusoidal activation function with varying parameters and a one-qubit (blue line) and two-qubit (red line) PQC, respectively. There is a clear difference in the achieved return between the two-qubit PQC (red line) and the NN with approximately the same number of parameters (18, green line). Similar to the NNs with ReLU activation function, the NNs with double or triple the parameters achieve results approximately as good as or better than the two-qubit PQC, in terms of return and probability of generating a rare trajectory. The effect of the activation function can be seen much more clearly for small numbers of parameters. In this case, the sinusoidal-NN with double the parameter count (42 parameters) demonstrates comparable performance to the PQC in terms of achieved return and the probability of generating a rare trajectory. In contrast, the ReLU-NN with the same 42 parameters performs worse on both metrics. The plots in the bottom row show that the one-qubit PQC (red line) achieves results similar to that of the NN with approximately the same number of parameters, both in terms of final return and probability of generating a rare trajectory. The notable difference lies in the rate of convergence: the one-qubit PQC needs approximately 200 batches, whereas the NN needs around 1000 batches to reach a similar result. \\
In brief, the ReLU and sinusoidal activation functions result in markedly different neural network performance. The ReLU activation function produces an unbounded positive output, is computationally efficient, and typically maintains meaningful gradients during training. In contrast, the sinusoidal activation function has a bounded output ranging from -1 to 1, is more computationally intensive, and may exhibit oscillating gradients. Benbarka \textit{et al.}~\cite{benbarka2021seeing} have shown that a Fourier encoding of the input data is structurally equivalent to one layer of hidden neurons with a sinusoidal activation function, but with a fixed weight matrix and a fixed bias vector. In detail, the Fourier encoding $f$ of the input data $\bold{x}\in \mathbb{R}^{d_{\rm in}}$ is given by 
\begin{equation}
f(\bold{x})= \begin{pmatrix}
\cos(\bold{B} · \bold{x})  \\
\sin(\bold{B} · \bold{x})  
\end{pmatrix} 
\end{equation}
with the  matrix 
$\bold{B} \in \mathbb{R}^{m\times d_{\rm in}}$, whose elements can be interpreted as discrete frequencies \cite{benbarka2021seeing}. This is similar to the angle encoding of the PQC
\begin{equation}
R_x(\bold{x'}) |0\rangle =  \exp(-i \sigma_x \bold{x'} /2) |0\rangle = \begin{pmatrix}
\cos(\bold{x'}/2) \\
-i \sin(\bold{x'}/2)
\end{pmatrix}.
\end{equation}
except the discrete frequencies in the former case and 
the factor $-i$, which can be absorbed into the subsequent rotation $R_y$. Thus the Fourier encoding, and the one layer training with sinusoidal activation in classical NNs is similar to the method of angle encoding in PQCs, which motivated us to compare their performance empirically. This can also motivate surrogate modelling of PQCs, which will be further discussed in the outlook. 

Given that a sinusoidal activation function appears to be a good fit for this data, it 
is prudent to examine existing studies on the use of periodic activation functions in classical neural networks, particularly regarding the types of inputs for which these functions are effective, and to assess whether these findings can be applied to this project. Typically, Fourier encodings are used in low dimensional domains (domains where the input is $x\in\mathbb{R}^d$ with d being small) to learn high-frequency functions \cite{tancik2020fourier}. Classical NNs exhibit a spectral bias favoring the learning of low-frequency functions, a tendency that becomes particularly pronounced in low-dimensional domains. This results in the NNs to first learn the global phase or overall structure instead of rapid variations. Analytical and numerical explanations for this behavior have been investigated by Tancik \textit{et al.}~\cite{tancik2020fourier}. Fourier encodings on the input data allow tuning the frequencies that are learned by the NN, including but not limited to the high frequency components. This improves training speed and generalization of the NN. In our case the input data is low-dimensional
(2d) and the optimal policy, $P_W$, has a high-frequency variation. In conclusion, quantum angular encodings, classical Fourier encoding as well as NNs with sinusoidal activation functions seem to be a good problem-architecture fit.

 We found that increasing the number of parameters in the NN and PQC shows varying performance of the agents in the policy-gradient case; see Appendix~\ref{sec:Comparison of data re-uploading and parameter variation}. We witness that the rate of convergence improves when increasing the number of neurons in the hidden layers in the policy-gradient NNs. For the challenge of generating rare trajectories, on the one hand an increase in parameters of the NNs is beneficial, whereas on the other hand an increase in data re-uploading in PQCs (and with that an increase in parameters) is not beneficial. In Ref.~\cite{skolik2022quantum} a similar behavior of the NNs is seen. This hints into the direction that the effects of scaling the number of parameters can be different for classical and quantum learning algorithms, depend on the problem and cannot be extrapolated trivially.

\FloatBarrier

\subsection{Results for scaling up the qubit count and problem size}
\label{sec:Scaling the number of time steps}
In this section, we study the aspect of scaling in two different variants. First, we study the effect of increasing the number $T$ of time steps and thus the state space. As $T$ increases, the trajectories become more complex and the probability of obtaining an RWB decreases exponentially; see the statement directly below Eq.~\eqref{eq:probRWBlargeT}. Second, we keep the state space fixed and scale up the number of qubits used to solve the task.

\begin{figure}[h!]
    \centering\includegraphics[width=15cm]{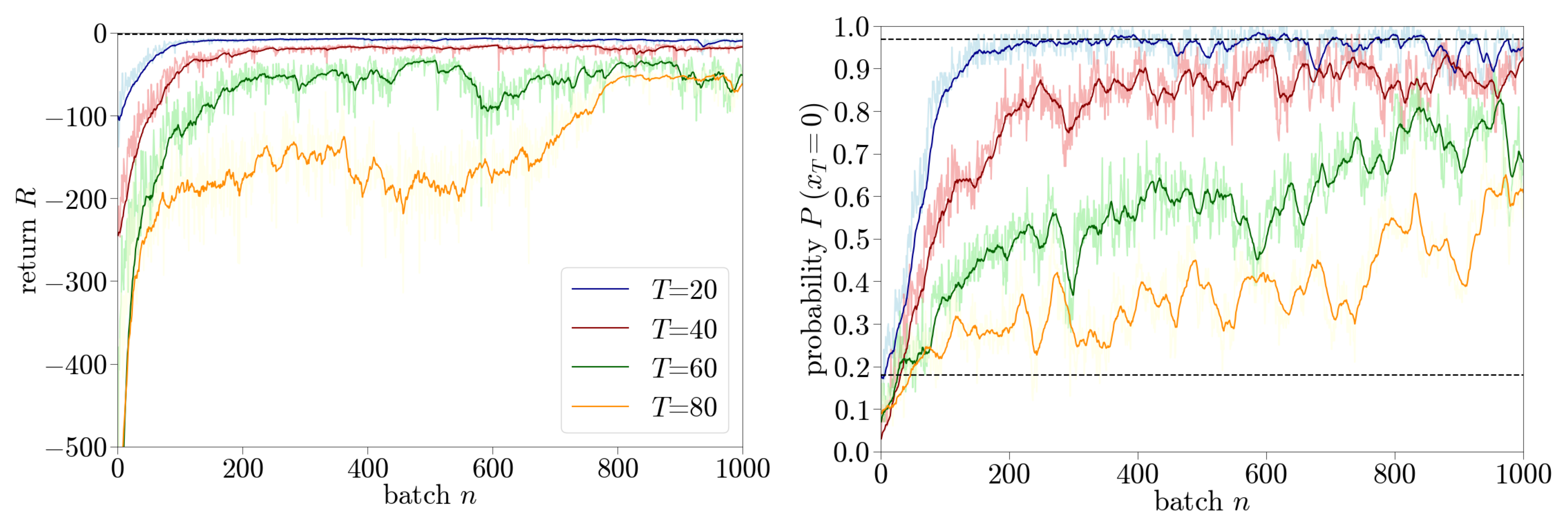}
    \caption{Effects of scaling the trajectory length $T$ on the performance of two-qubit PQC-based policy-gradient RL agents with three data-uploading layers, trained to generate \emph{random walk bridges} (RWBs). Blue: $T=20$. Red: $T=40$. Green: $T=60$. Yellow: $T=80$. The lighter lines represent the average of 10 agents, each trained using the same hyperparameters. The darker lines represent the exponential moving average. Left: Average return $R$ per batch $n$.  Right: Probability of generating a rare trajectory $P(x_T = 0)$. The hyperparameters of the agents used in these plots can be found in Tab. \ref{paramsFig23} of App.\ \ref{app:hyperparameters}.}
    \label{scaling1}
\end{figure}

Figure~\ref{scaling1} shows the return $R$ as a function of batches $n$ of two-qubit policy-gradient RL agents with three data-uploading layers for different trajectory lengths $T$. As $T$ increases the convergence rate of the return slows. Nonetheless, for large $T$ the PQC successfully learns and exhibits consistent converging behavior.
In combination with the good learning behavior of the PQC-based policy-gradient RL agents (improved compared to that of NN-based ones, see Section~\ref{sec:Comparison to classical neural networks}), it seems plausible that the PQC-based approach can cope with the increase in state space size. Since real-world RL applications often operate with large state spaces, our results suggest that quantum reinforcement learning (and our method) might be applicable in practical use cases.

\begin{figure}[h!]
    \centering
    \includegraphics[width=15cm]{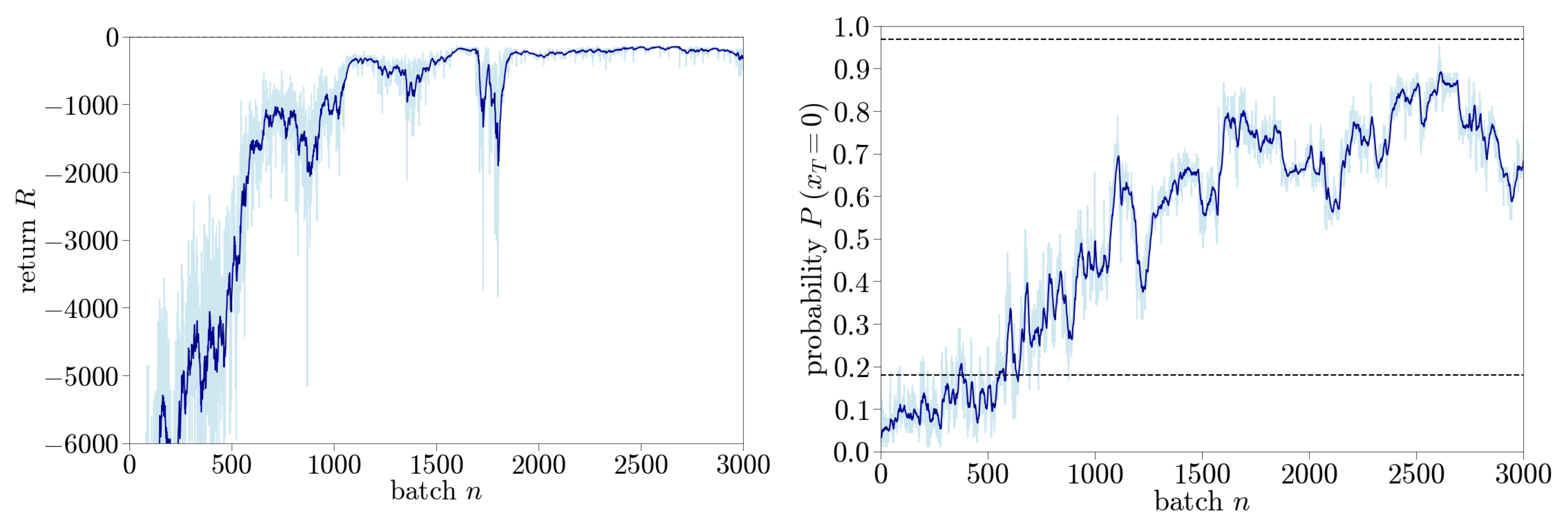} %
    \caption{Results of the two-qubit PQC-based policy-gradient RL agents trained to generate \emph{random walk bridges}  (RWBs) with trajectory length $T=200$. Left: Return $R$ as a function of batches $n$, as given in Eq.~\eqref{eq:return_RWB}, each consisting of 10 episodes (trajectories). Right: Probability of generating a rare trajectory $P(x_T = 0)$. The hyperparameters of the agent used in these plots can be found in Tab. \ref{paramsFig24} of App.\ \ref{app:hyperparameters}.}
    \label{scaling2}
\end{figure}

In Figure~\ref{scaling2} (left), we display the returns per batch (left) and the probability to generate an RWB (right) of a two-qubit policy-gradient RL agent for $T=200$. The convergence to a high return is visible in Figure~\ref{scaling2}, although the values are significantly lower due to a larger value of $s=50$. The agents are able to achieve a high probability of generating an RWB, although with a lot of fluctuation. This behavior likely originates from the sampling process. More specifically, the agent first generates ten trajectories based on a random policy, which is then optimized based on the return. At a restart, this process repeats with the updated policy. Since the probability of sampling a RWB decreases exponentially with increasing  $T$, the likelihood of obtaining it randomly, throughout the training process, becomes increasingly low. Techniques like integrating a replay buffer combined with random initializing of the policy after a certain number of batches could help to control this effect. 

\begin{figure}[h!]
    \centering
    \includegraphics[width=15cm]{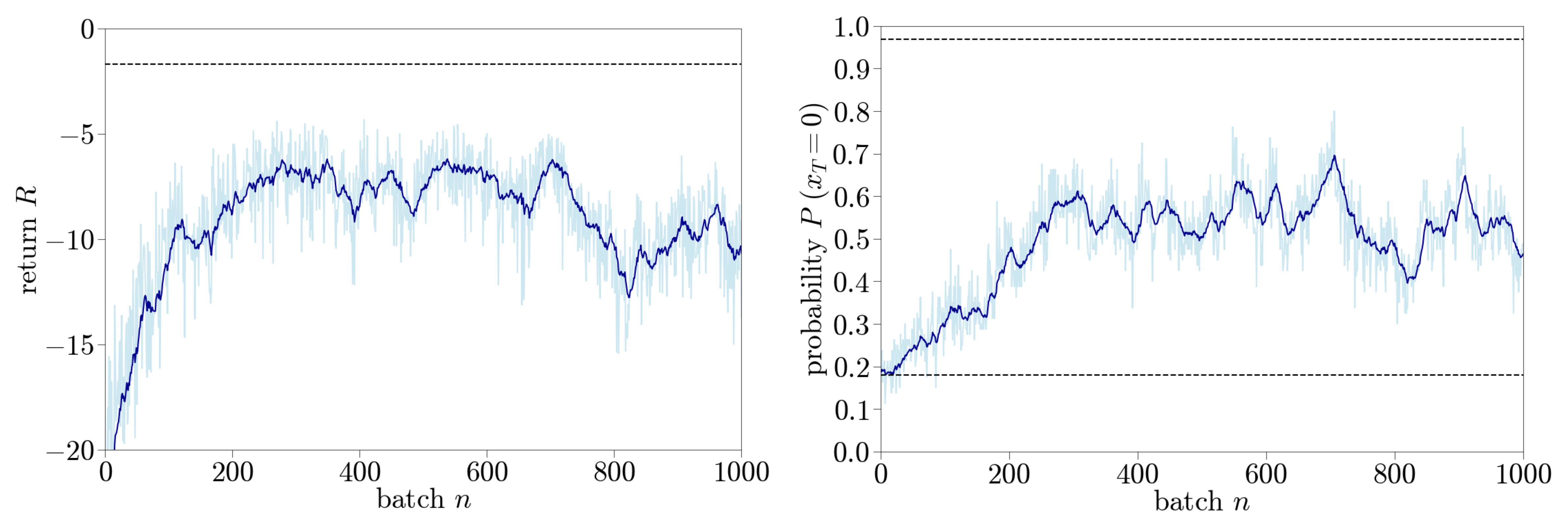} %
    \caption{Results of the eight-qubit PQC-based policy-gradient RL agents trained to generate \emph{random walk bridges}  (RWBs) with trajectory length $T=20$. Left: Return $R$ as a function of batches $n$, as given in Eq.~\eqref{eq:return_RWB}, each consisting of 10 episodes (trajectories). Right: Probability of generating a rare trajectory $P(x_T = 0)$. The hyperparameters of the agent used in these plots can be found in Tab. \ref{paramsFig8qubits} of App.\ \ref{app:hyperparameters}.}
    \label{scalingqubit}
\end{figure}

The second variant of scaling we study is that of scaling up the number of qubits, as can be seen in Fig.~\ref{scalingqubit}, where we display the returns per batch (left) and the probability to generate an RWB (right) of an eight-qubit policy-gradient RL agent for $T=20$. This PQC was build copying the circuit displayed in Fig.~\ref{CircuitReUploading} four times, thus on 8 qubits. These ``four circuits'' where then entangled with each other in the entangling layer of the hardware efficient Ansatz. The observable $Z_0 Z_1 Z_2 Z_3 Z_4 Z_5 Z_6 Z_7$ was used. In the learning phase, the agent shows a clear increase in the return and probability to generate RWBs, but convergence is not as clear as in the two-qubit case (see Section~\ref{sec:1- and 2-qubit policy gradient PQCs}). One reason for this could be an overparametrization of the simple state space, as the agent now has 48 variational parameters. This means that after obtaining an RWB through the initial sampling it could (over-)fit this single trajectory and not be incentivized to learn further. The optimization of the eight-qubit agent for further exploration is left for further work.

\section{Conclusions and outlook}
\label{sec:Outlook and discussion}

In summary, this work has been driven by two primary avenues of inquiry: Firstly, it has endeavored to heuristically probe the potential \textit{innovative application} of both established and novel \emph{quantum reinforcement learning} (QRL) methodologies in the sampling of rare trajectories. Secondly, on the basis of this successful application, this work has seeked to explore and explain \textit{novel phenomena} emerging within the realm of QRL. This has included investigating the effects of different (hyper-) parameters and architectures of the \emph{parameterized quantum circuits} (PQCs) employed by the QRL methods, and comparing the QRL performance to that of established RL methods based on \emph{neural networks} (NNs), in order to both understand QRL better and achieve quantum-inspired improvements of classical methods in the future.

In some ways, the findings presented here are reminiscent of advantages of quantum circuits 
the quantum gates of which feature complex valued entries over
discrete Markov chains in probabilistic modeling \cite{Expressive,PhysRevA.108.022411}.
Importantly, speaking with Ref.\ \cite{PRXQuantum.3.030101}, 
asymptotic quantum advantages in the sense of proven separations in sample or computational complexity may not be the only aim in the field to bring it to a next level. In the light of the observation that in meaningful learning tasks, natural data are often unstructured and it is not so clear how to relate proven separations for highly structured data with practical applications derived from unstructured data, 
such endeavours ``outside the box''
as suggested here seem worthwhile to make progress in the field of QML.

More specifically, our goal has been to approximate the reweighted dynamics from which one can efficiently sample rare trajectories of the original dynamics of a stochastic process---in this case the \emph{random walk bridges} (RWBs) of the paradigmatic one-dimensional symmetric random walk---in order to circumvent the direct computation of the exact reweighted dynamics. (This direct approach is in many cases parameter-sensitive, computationally expensive and tends to be numerically unstable, especially in cases of practical interest.) To this end, we have applied the established policy-gradient and the novel actor-critic QRL method with the aim of learning an optimal parameterized dynamics that approximates the reweighted dynamics as close as possible. The trick of RL applied to rare-trajectory sampling is that the parameterized dynamics is given by the policy of the RL agent, based on a PQC in the case of QRL, and optimization of this policy only requires the trivially computable weights of the reweighted dynamics (and not its normalization factors). Still, in order to benchmark the quality of our results we have also computed the reweighted dynamics in the tractable case of $T = 20$ time steps via the direct approach, see Section~\ref{sec:theory_methodology}.

In Section~\ref{sec:1- and 2-qubit policy gradient PQCs}, we have numerically demonstrated the surprising efficacy of training QRL agents based on PQCs, even when these PQCs are small: in this case, one- and two-qubit PQCs with three shallow data-uploading layers, optimized by policy-gradient QRL. 
We have provided a possible explanation of this success in Section~\ref{sec:1-qubit PQC} by computing the Fourier coefficients of the truncated Fourier series which represents the output of PQCs with data re-uploading and fitting these coefficients such that the parameterized dynamics approximates the reweighted one. The good fit and thus the surprising success might be explainable by the facts that the reweighted dynamics has a structure similar to a past light cone (out of which trajectories cannot reach the end point $x_T = 0$ of RWBs) and that this structure can be approximated rather well via PQC-based policies, which are functions of truncated Fourier series, in the simplest case with finitely many integer frequencies.

Furthermore, in Section~\ref{sec:Comparison of models for policy gradient and actor-critic reinforcement learning} we have compared the performance of the established \emph{policy-gradient} (PG) and that of the novel 
\emph{actor-critic} (AC) QRL algorithm, both applied to the sampling of rare trajectories. This revealed that the PG algorithm outperforms the AC one in terms of the convergence of the return, which is a measure of learning rate and success of RL agents in general, and in the case considered here it is also related to the ``closeness'' (Kullback-Leibler divergence) of the parameterized dynamics to the reweighted one. Furthermore, the ability of the PG algorithm to generate rare trajectories after training is higher. This might be related to different trade-offs between the \textit{exploration} of different policies and the \textit{exploitation} of single, relatively successful policies, arising from the different algorithms. Such differences are similar to that between classical, NN-based policy-gradient and actor-critic RL \cite{rose2021reinforcement} and thus could be hardware-agnostic. 

In Section~\ref{sec:Comparison of data re-uploading and parameter variation} we have investigated effects of the number of data-uploading layers in the PQCs on the performance. We have found that initially an increase in the number of layers and thus in the expressivity of the parameterized dynamics is beneficial, but a further increase of the number decreases the performance again, although the expressivity still rises. This effect is also present in the complementary analysis via the truncated Fourier series and could be related to a trade-off between more \textit{expressivity} and less \textit{trainability} due to increasingly complicated return/loss landscapes. Furthermore we explored the effects of deactivating specific sets of gates and further parameters of the PQCs: This has shown that (de-)activated rotation gates have the largest influence
on the learning rate and learning ability of the QRL agents and also active input scaling parameters, which lead to an ``adaptive frequency matching'', are very beneficial. Beside that we have found that (de-)activated entangling gates have little effect on the performance, as long as the input scaling parameters are active. 
Determining the effects of deactivating specific sets on the truncated Fourier is straightforward, but beyond the scope of this work. Investigating the geometry of the return/loss landscapes, depending on the number of data-uploading layers, might be interesting research for the future, also from a theory-oriented perspective. Rather regarding applications it could be interesting to investigate alternative strategies for data re-uploading and optimization of variational parameters: For example one could test strategies to utilize the optimized parameter values of QRL agents with $n$ data-uploading layers by memorizing them and initializing the parameters of QRL agents with $n + 1$ layers not randomly at the start of the learning process, but with the  values optimized for $n$ layers.

Moreover, in Section~\ref{sec:Comparison to classical neural networks} we have demonstrated numerically the enhanced performance of QRL compared to NN-based RL with comparable numbers of trainable parameters, even for one-qubit PQCs. We have analyzed the different effects of ReLU and sinusoidal activation functions in NNs, showing that RL based on NNs with sinusoidal activation function has a performance closer to that of QRL. This has been attributed to the similarity between classical Fourier encoding (equivalent to sinusoidal activation function in one layer, up to parameters) and quantum angular encoding. This suggests a promising approach for building classical surrogate models using sinusoidal activation functions. NNs with sinusoidal activation function or Fourier encoding are commonly applied to data with high-frequency variations in low-dimensional domains. Since the reweighted dynamics of the one-dimensional random walk fits into this category, this in turn seems to explain why also PQCs perform well in this context. 
Although this is not a rigorous argument, our results suggest that in this case the good problem-solution fit (as also indicated by the analysis of the truncated Fourier series) is more important for the performance than quantum resources like entanglement. This is also supported by the small effect of (de-)activating the entangling gates as discussed in the previous paragraph.
Furthermore, we have numerically investigated the effect of increasing the number of parameters in the NNs and observed a behavior different from the PQCs' behavior, which was discussed above: 
Within the tested range of numbers, NNs show performance continuously improving with increasing number of parameters. Thus, while an increase in the number of parameters similarly enhances expressivity of both PQCs and NNs, it has significantly different effects on the optimizability due to the generally good trainability of NNs.

What is more, in Figure~\ref{Fig:QPGDifProbs} of 
App.\ \ref{Appendix: Further numerics} we have shown numerically that our QRL framework is a good fit to asymmetric random walks, \textit{i.e.}, random walks with unequal probabilities to jump one step up or down. Besides that, we expect that our QRL framework can be successfully applied to the case of sampling rare trajectories (with fixed endpoints $x_T \neq 0$) other than random walk bridges. Our intuition for this case is that the reweighted dynamics should possess a structure resembling a shifted past light cone (centered at $x_T \neq 0$); in turn we expect that the parameterized dynamics should be able to approximate such a structure well. Still this remains work for the future. An investigation of both cases via the Fourier analysis is straightforward, but also beyond the scope of this work. Since our QRL framework turned out to be successful in the case of asymmetric random walks and can easily be applied to other Markov processes in principle, it might be interesting to test this application in practice, too.

Similar to many RL methods, our QRL framework is sensitive to the precise tuning the hyperparameters, whose meaning and values are summarized in App.\ \ref{app:hyperparameters}. A more detailed optimization of the hyperparameters might be important for real-world applications.
In general, there is much room for optimizing the framework for specific  applications, as the employed software and (simulated) quantum computing hardware is not specialized, but rather standard PQCs and NNs as well as standard PQ and AC algorithms were used, no memory or entropy function were utilized, etc. 
Such optimization as well as more detailed resource estimates might be necessary for more realistic benchmarking of our QRL framework against competing classical methods such as RL based on state-of-the-art NNs and exponential tilting. Only then more reliable answers can be given to questions like: How large do the PQCs of the QRL agents need to be such that they outperform RL agents based on state-of-the-art NNs?

Since the effort for a direct computation of the reweighted dynamics scales unfavorably with the number of dimensions and the size of the state space, an application of our QRL framework to random walks in more than one dimension might be interesting, also from the point of view of real-world applications.
So far, we studied mostly small instances of the paradigmatic problem to sample rare trajectories of one-dimensional random walks, \textit{i.e.}, small numbers of time steps $T$. 
On the contrary, in Section\  \ref{sec:Scaling the number of time steps} we demonstrated the scalability of our results to longer trajectories up to $T = 200$ and up to eight-qubits. In order to try to get an order-of-magnitude estimate for $T$ necessary for real-world cases, let's suppose that one wishes to assess the value of NASDAQ derivatives and that to this end one models the market fluctuations via a random walk with time steps of 1 day. Since 
there are on average 252 NASDAQ trading days per year, the time horizon of the successfully tested case of $T=200$ is more than three quarters. This estimate might be too optimistic due to the accuracy required in real-world cases, but there are also further options 
to optimize the performance of the QRL agents in these settings, \textit{e.g.}, with RL tools like experience replay or replay buffer. 
These insights invite to take an optimistic stance at the applicability of QRL for 
real-world cases. It is the hope that the present work inspires further heuristic work that explores the applicability of such approaches for practically relevant real-world cases
in the future.

\textit{The results, opinions and conclusions expressed in this publication are not necessarily those of Porsche Digital GmbH.}

\paragraph*{Acknowledgements.} We acknowledge the use of \href{https://aws.amazon.com}{AWS computing services} for this research. We also would like
to thank Maureen Krumtünger, Sofiene Jerbi, Elies Gil-Fuster, Juan Garrahan, Ryan Sweke (who has been particularly instrumental in the early stages of this project), Regina Kirschner, Marian Mularski, Mahdi Manesh, Peter Wolf, Timon Höfer, Clemens Wickboldt, Cornelius Langenbruch, Dirk Bruns, Stefan Zerweck, Mattias Ulbrich, Rolf Zöller, Ralf Knoll, and Aurelio Mir\'o for stimulating discussions
along different stages of the lifetime of this project. Furthermore, we would like to
to thank the German Federal Ministry of Education and Research (BMBF) (HYBRID, MuniQC-Atoms, QuSol),
the Federal Ministry for Economic Affairs and Climate Action (EniQmA), the QuantERA (HQCC),the Munich Quantum Valley, Berlin Quantum, the Quantum Flagship (Millenion, PasQuans2), the European Research Council (DebuQC),
and Porsche Digital 
GmbH
for their support. For the DFG (CRC 183), this has been  the result of a joint-node collaboration involving the 
two teams led by Jens Eisert and Piet Brouwer.

\bibliographystyle{alpha}  
\bibliography{references} 


\begin{appendix}

\section{Pseudocode} \label{pseudocode}

\input{pseudocode} 

\section{Further details of analysis via Fourier series}
\label{app:Fourier_analysis}


The structure of this appendix, which provides further details on the Fourier analysis of Section\  \ref{sec:1-qubit PQC}, follows the steps of that subsection:
\begin{enumerate}
    \item[(i)]
    express the expectation value ${\langle O_a\rangle}_{s,\theta}$ of the PQCs in terms of a truncated Fourier series with the input state $s = (x, t)$ as variable,

    \item[(ii)]
     determine which of its Fourier coefficients can be non-zero and are not identical to others, 

     \item[(iii)]
     fit the parameterized dynamics $P_\theta$ to the reweighted dynamics $P_W$ with these Fourier coefficients as fitting parameters, using 
     a classical optimization procedure, and

     \item[(iv)]
     investigate the results of the previous steps.
\end{enumerate}

\paragraph*{Step (i).} In Section~\ref{sec:1-qubit PQC} we have omitted to present proofs that expectation values $\langle O \rangle_{s,\theta}$ of observables $O$ of the 1- and 2-qubit PQCs considered in this work can be expressed as truncated Fourier series. For the 2-qubit PQC simply the proof in App.\ A of Ref.~\cite{schuld2020circuit} applies.
The 1-qubit case requires only a slight variation of this proof; thus we follow its lines and mainly adopt the same notation: The expectation value $\langle O \rangle_{s,\theta}$ can be written as 
\begin{align}
    \langle O \rangle_{s,\theta} = \bra{0} U^\dagger(s, \theta) O U(s, \theta) \ket{0},
    \label{eq:expectation_value_definition}
\end{align}
where the unitary transformation $U(s, \theta)$ (with $s = (x, t)$) for a PQC with $N$ data-uploading layers can be decomposed as
\begin{align}
    U(s, \theta) = V^{(N)} S(x) W^{(N)} S(t) V^{(N - 1)} \dots V^{(1)} S(x) W^{(1)} S(t) V^{(0)},
\end{align}
where in our specific case $S(\cdot) = R_X(\cdot)$, $V^{(0)}$ equals the identity, and $V^{(n)}$ ($W^{(n)}$) denotes the unitary of the second (first) variational layer in the $n$th data-uploading layer (for the sake of brevity omitting in the notation that they depend on the variational angles $\theta$), \textit{c.f.} Fig.\  \ref{Circuit1Q}, but in principle they are arbitrary. Thus we can absorb the unitaries diagonalizing $S(\cdot)$ and without loss of generality assume that they are diagonal, \textit{e.g.}, $S(x) = \text{diag} [\exp(-i \lambda_1 x), ..., \exp(-i \lambda_d x)]$ (in our case $d = 2$ and $\lambda_1 = +1/2, \lambda_2 = - 1/2$). Then the $i$th component of $U(s, \theta) \ket{0}$ can be expressed as 
\begin{align}
    \left[ U(s, \theta) \ket{0} \right]_i = &\sum_{j_1, \dots, j_N}^d \sum_{l_1, \dots, l_N}^d e^{-i (\lambda_{l_1} + \dots + \lambda_{l_N}) x} e^{-i (\lambda_{j_1} + \dots + \lambda_{j_N}) t} \nonumber\\
    &\times V^{(N)}_{i, l_N} W^{(N)}_{l_N, j_{N}} V^{(N - 1)}_{j_N, l_{N-1}} \dots V^{(1)}_{j_2, l_1} W^{(1)}_{l_1, j_1} V^{(0)}_{j_1, 1},
\end{align}
where the indices of the matrices denote their elements in the basis in which $\ket{0}$ is the first basis state
vector (for the moment omitting input scaling, which will be considered below). Then
\begin{align}
    \langle O \rangle_{s,\theta} =
    \sum_{\boldsymbol{j}, \boldsymbol{k}, \boldsymbol{l}, \boldsymbol{m}}
    a_{\boldsymbol{m}, \boldsymbol{l}; \boldsymbol{k}, \boldsymbol{j}} e^{i(\Lambda_{\boldsymbol{l}} - \Lambda_{\boldsymbol{m}}) x} e^{i(\Lambda_{\boldsymbol{k}} - \Lambda_{\boldsymbol{j}}) t}  \label{eq:ungrouped_truncated_Fourier_series_multi-index}
\end{align}
with multi-indices $\boldsymbol{j} = \{ j_1, \dots, j_N \}$ and $\boldsymbol{k}$, $\boldsymbol{l}$, and $\boldsymbol{m}$ analogously defined, $\Lambda_{\boldsymbol{j}} = \lambda_{j_1} + \dots + \lambda_{j_N}$, and $a_{\boldsymbol{m}, \boldsymbol{l}; \boldsymbol{k}, \boldsymbol{j}}$
being given by
\begin{align}
    a_{\boldsymbol{m}, \boldsymbol{l}; \boldsymbol{k}, \boldsymbol{j}} = \sum_{i, i'}  &\left( V^* \right)^{(0)}_{1, k_1} \left( W^* \right)^{(1)}_{k_1, m_1} \left( V^* \right)^{(1)}_{m_1, k_2} \dots \left( V^* \right)^{(N - 1)}_{ m_{N-1}, k_N} \left( W^* \right)^{(N)}_{k_{N}, m_N} \left( V^* \right)^{(N)}_{m_N, i} 
     \nonumber\\
    \times &O_{i, i'} V^{(N)}_{i', l_N} W^{(N)}_{l_N, j_{N}} V^{(N - 1)}_{j_N, l_{N-1}} \dots V^{(1)}_{j_2, l_1} W^{(1)}_{l_1, j_1} V^{(0)}_{j_1, 1}.
\end{align} 
Finally, Eq.\ \eqref{eq:ungrouped_truncated_Fourier_series_multi-index} is the straightforward generalization of Eq.\ (8) in Ref.~\cite{schuld2020circuit} to two-dimensional input $\boldsymbol{x}$, and thus eigenvalue differences $\Lambda_{\boldsymbol{m}} - \Lambda_{\boldsymbol{l}}$ and $\Lambda_{\boldsymbol{k}} - \Lambda_{\boldsymbol{j}}$ can be grouped to integer frequencies $n_x$ and $n_t$, respectively, and appropriate sums of $a_{\boldsymbol{m}, \boldsymbol{l}; \boldsymbol{k}, \boldsymbol{j}}$ to Fourier coefficients $c_{n_x, n_t}$ in the same way as in Ref.~\cite{schuld2020circuit}. In summary, one then arrives at 
\begin{align}
   {\langle O\rangle}_{s,\theta} = \sum_{n_x = - N_x}^{N_x} \sum_{n_t = - N_t}^{N_t} c_{n_x n_t} e^{i (n_x x + n_t t)}.
   \label{eq:truncated_Fourier_series_basis_form}
\end{align}
Equation \eqref{eq:ungrouped_truncated_Fourier_series_multi-index} also allows a straightforward generalization to the case with input scaling $x' = \arctan (x \cdot \lambda_{x}^{(n)})$, and $t' = \arctan (t \cdot \lambda_{t}^{(n)})$ for each $n$th data-uploading layer, which reads
\begin{align}
    \langle O \rangle_{s',\theta} =
    \sum_{\boldsymbol{j}, \boldsymbol{k}, \boldsymbol{l}, \boldsymbol{m}}
    a_{\boldsymbol{m}, \boldsymbol{l}; \boldsymbol{k}, \boldsymbol{j}} &\exp\left\{i \left[(\lambda_{m_1} - \lambda_{l_1}) \arctan (x \cdot \lambda_{x}^{(1)})  + \dots + (\lambda_{m_N} - \lambda_{l_N}) \arctan (x \cdot \lambda_{x}^{(N)}) \right] \right\} \nonumber\\
    \times &\exp\left\{i \left[(\lambda_{k_1} - \lambda_{j_1}) \arctan (t \cdot \lambda_{t}^{(1)})  + \dots + (\lambda_{k_N} - \lambda_{j_N}) \arctan (t \cdot \lambda_{t}^{(N)}) \right] \right\},
    \label{eq:ungrouped_truncated_Fourier_series_multi-index_adaptive_freqs}
\end{align}
where the eigenvalues $\lambda_{j_1}$, etc., are not to be confused with the input scaling parameters $\lambda_x^{(n)}$ and $\lambda_t^{(n)}$ of the $n$th layer. In some sense this can be viewed as truncated Fourier series with adaptive (non-integer) frequencies.

\paragraph*{Step (ii).} We followed two different ways to determine which of the Fourier coefficients (which in general can be non-zero complex numbers) are identically zero, real, or purely imaginary and are not identical to others: (a) by calculating them symbolically (on a classical computer; no quantum system or a simulation of one is involved) and (b) by computing the 2D discrete Fourier transform of the expectation value $\langle O \rangle_{s,\theta}$ in Eq.\ \eqref{eq:expectation_value_definition} via the fast Fourier transform (FFT) of a real-valued function for $2 N_x + 1$ and $2 N_t + 1$ equally spaced values of $x$ and $t$, respectively, (in the interval $[0, 2 \pi)$) and random variational angles $\phi$ sampled from a uniform distribution on the interval $[0,2\pi)$. (Option (b) is how in Ref.~\cite{schuld2020circuit} Schuld \textit{et al.} arrived at Fig.\  5 of that ref.)

\begin{figure}[h]
    \centering
\includegraphics[width=0.99\textwidth]{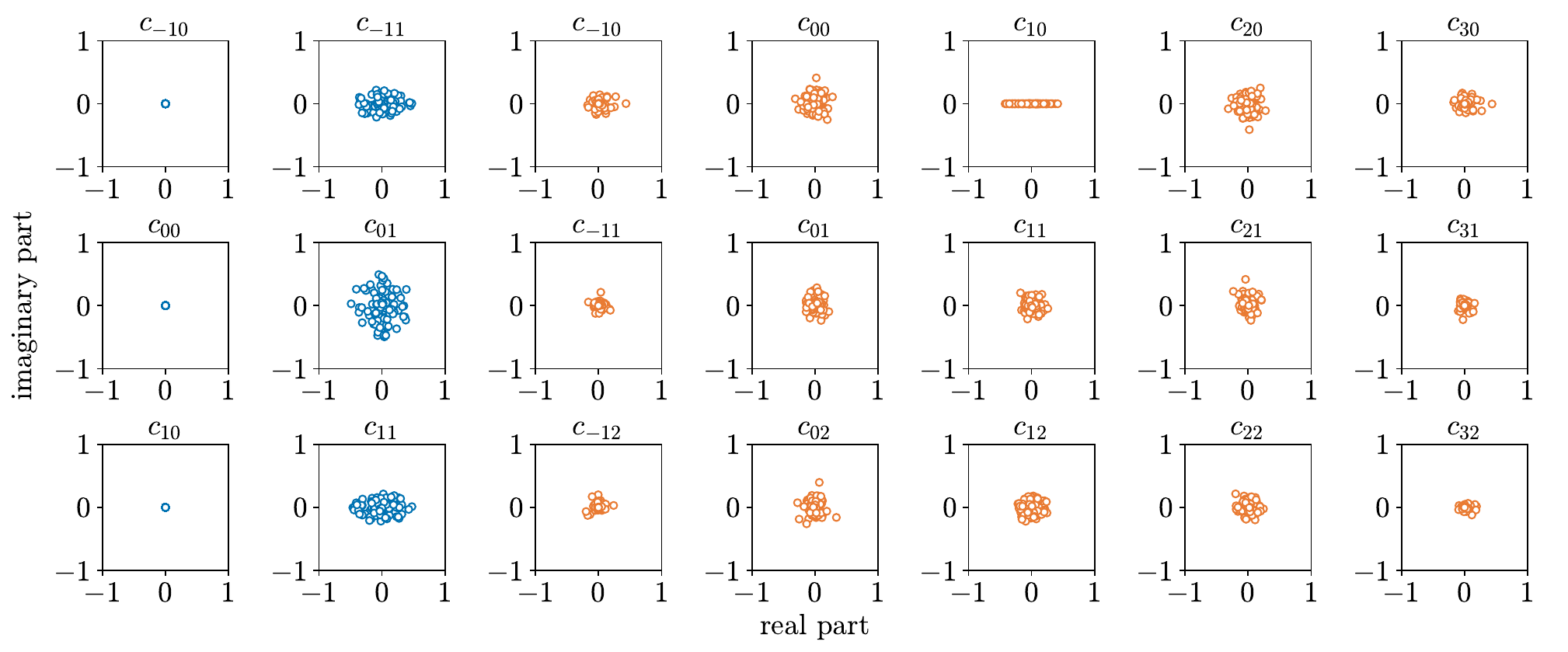}
    \caption{Fourier coefficients $c_{n_x n_t}$ of the truncated Fourier series for the expectation value $\langle Z \rangle_{s, \theta}$ of the 1-qubit PQC in Fig.\  \ref{Circuit1Q} parameterized by the variational parameters $\theta$ as function of the state $s = (x, t)$ ($n_x$ and $n_t$ denote the discrete frequencies of the variables $x$ and $t$, respectively). The horizontal (vertical) axis denotes the real (imaginary) part of the coefficients. Blue (orange): Fourier coefficients for the 1-qubit PQC with 1 (2) data-uploading layers. Each of the 100 points in each plot represents the result for a random choice of parameters $\theta$. Some of the plots which can be obtained by the relation $c_{-n_x -n_t} = c_{n_x n_t}^*$ are omitted. For further details on how the Fourier coefficients are computed see the main text of App.\ \ref{app:Fourier_analysis}, step (ii).}
\label{fig:Fourier_coeffs_random_parameters}
\end{figure}

Option (a) compared to (b) is computationally expensive and its results are unwieldy and intricate, except for the case of a single data-uploading layer only. In that case the \textit{non-zero} Fourier coefficients $c_{n_x n_t}$ of the 1-qubit PQC are given by
\begin{align}
&c_{1,-1} = c_{-1,1}^* = \frac{1}{4}  \big(\cos
   \theta_1 -\cos \theta_2 + i \sin
   \theta_1 \sin \theta_2 \big) \cos \theta_3, \nonumber\\
&c_{0, -1} = c_{0, 1}^* = \frac{i}{2}  \big (- \sin \theta_2 + i \sin \theta_1 \cos \theta_2  \big) \sin \theta_3, \nonumber\\
&c_{-1,-1} = c_{1,1}^* = \frac{1}{4}  \big(\cos
   \theta_1 +\cos \theta_2 + i \sin
   \theta_1 \sin \theta_2 \big) \cos \theta_3,
\end{align}
while for the 2-qubit PQC they read
\begin{align}
&c_{1,-1} = c_{-1,1} = c_{-1,-1} = c_{1,1} = \frac{1}{4} \cos
   \theta_1 \cos \theta_2 .
\end{align}
In contrast, option (b) does not yield such closed-form expressions, but by computing the FFT for 100 random samples of $\phi$ and plotting the results for each $c_{n_x n_t}$ in the complex plane into one figure it enables an intuitive understanding, \textit{c.f.} Fig.\  \ref{fig:Fourier_coeffs_random_parameters} (see caption for a more detailed description). These results 
suggest that for the 1-qubit PQC with 1 data-uploading layer (in red) the Fourier coefficients $c_{n_x 0}$ are identically zero (as also shown above), but none of the coefficients for that with 2 layers. Furthermore the plots indicate that the variance of the Fourier coefficients decreases for increasing frequencies $n_x$ and $n_t$. Both is also true for more than 2 layers (plots not shown here). Beside that, the plots of $c_{n_x 0}$ illustrate that the relation $c_{-n_x -n_t} = c_{n_x n_t}^*$ is always fulfilled, as required for the Fourier coefficients of a real-valued function. Thus, except for the cases without data-reuploading, the numerical evidence suggests that all $c_{n_x n_t}$ are general non-zero complex numbers and furthermore do not depend on each other except that relation.

\begin{figure}[h]
    \centering
\includegraphics[width=0.99\textwidth]{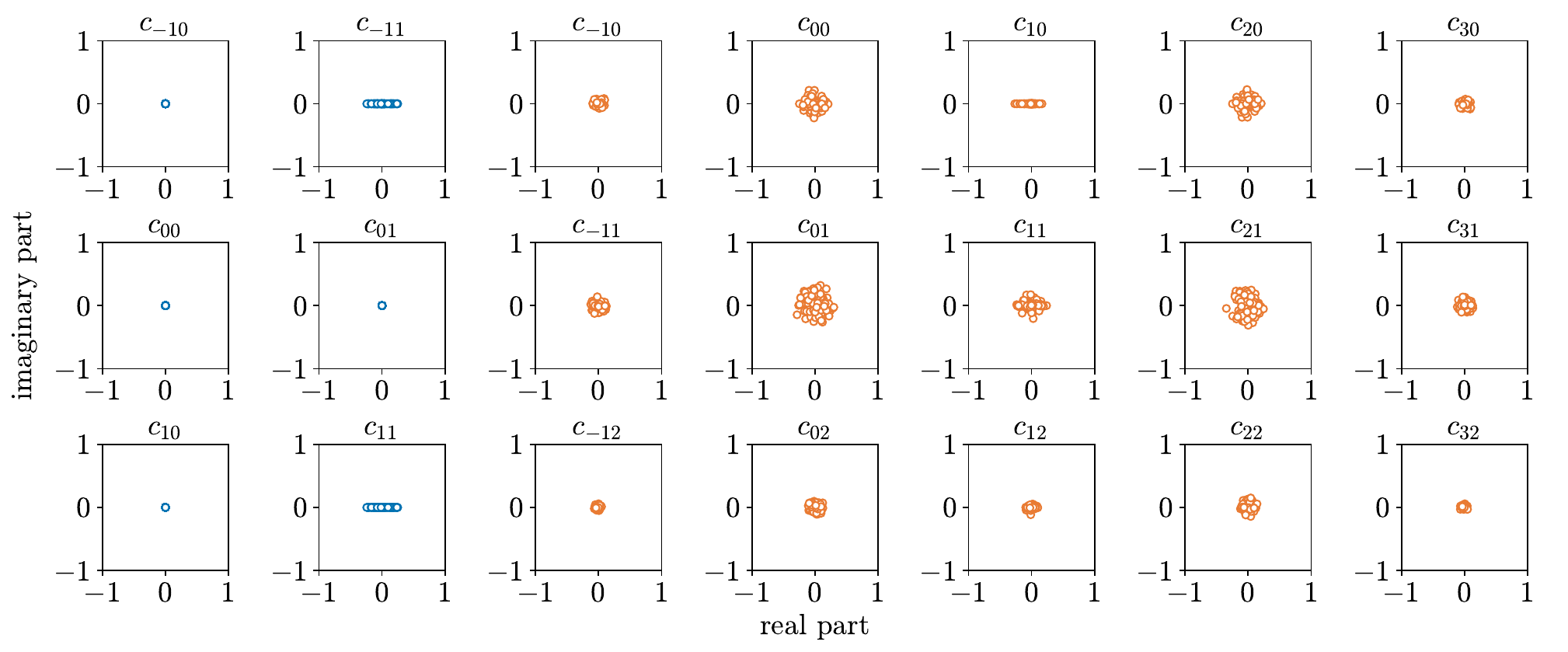}
    \caption{Fourier coefficients $c_{n_x n_t}$ like in Fig.\  \ref{fig:Fourier_coeffs_random_parameters}, but for the 2-qubit PQC in Fig.\  \ref{CircuitReUploading} ($n_x$ and $n_t$ denote the discrete frequencies of the variables $x$ and $t$, respectively).  Blue (orange): Fourier coefficients for the 2-qubit PQC with 1 (2) data-uploading layers. For further details see the caption of Fig.\  \ref{fig:Fourier_coeffs_random_parameters}.}
\label{fig:Fourier_coeffs_2_qubits_random_parameters}
\end{figure}

\paragraph*{Step (iii).} The obvious choice of fitting parameters $\alpha$ for fitting the parameterized dynamics $P_\alpha$ given by 
\begin{equation}
P_\alpha(x-1 | x, t) = \frac{1}{e^{{  \langle O\rangle}_{s,\alpha}} + 1},
\end{equation}
to the reweighted dynamics $P_W$ is $\alpha = (\lambda_x, \lambda_t, \phi, \omega')$ with $\phi$ the variational angles. Here $O = \omega' Z$ and $O = \omega' Z_0 Z_1$ in the 1- and 2-qubit case, respectively, and $s = (\arctan (x \cdot \lambda_{x})$, $\arctan (t \cdot \lambda_{t}) )$ (the connection to Eq.\ \eqref{eq:softmax-PQC_policy} is established by defining $\omega' = 2 \beta \omega$). 

But the results above suggest another choice: use the Fourier coefficients of $\langle O\rangle_{s,\alpha}$ instead of $\phi$ as fitting parameters. This choice is reasonable as the Fourier coefficients seem to be sufficiently independent of each other as functions of $\phi$, see Fig.\  \ref{fig:Fourier_coeffs_random_parameters}, except for the relation $c_{-n_x -n_t} = c_{n_x n_t}^*$ required by the truncated Fourier series representing a real-valued function (and in the 2-qubit case with 1 data-uploading layer only, see Fig.\  \ref{fig:Fourier_coeffs_2_qubits_random_parameters}, left part in red). Furthermore, this choice seems promising as it is computationally much less complex than the obvious choice, where each required function call of the loss function (to be defined below) requires additional computational effort: \textit{either} precomputing and evaluating symbolic expressions for the Fourier coefficients as functions of $\phi$ \textit{or} directly computing ${  \langle O\rangle}_{s,\alpha}$ as function of $\phi$.

In order to ensure that the relation $c_{-n_x -n_t} = c_{n_x n_t}^*$ is automatically satisfied, it is helpful to rewrite Eq.\ \eqref{eq:truncated_Fourier_series_basis_form} as
\begin{align}
    {\langle O\rangle}_{s,\theta} = \sum_{n_x = 0}^{N_x} \sum_{n_t = 0}^{N_t} a_{n_x n_t} \cos (n_x x + n_t t + \varphi_{n_x n_t}), 
\label{eq:truncated_Fourier_series_amplitudes+phases}
\end{align}
with cosine amplitudes $a_{n_x n_t}$ and phases $\varphi_{n_x n_t}$, which can be assumed to be independent of each other without violating $c_{-n_x -n_t} = c_{n_x n_t}^*$. Thus fitting parameters $\alpha = (\lambda_x, \lambda_t, \{ a_{n_x n_t} \}, \{ \varphi_{n_x n_t} \}, \omega')$ with $n_{x/t} \in \{ 0, \dots, N_{x/t} \}$ are chosen in this work. 
This choice of fitting parameters can be viewed as an equally valid choice of parameterized dynamics, different from that used for the \emph{reinforcement learning}  (RL) in the main part of this work. 

All fits in Section\  \ref{sec:1-qubit PQC} and this appendix are obtained by the method of least squares, \textit{i.e.}, by searching for values of the fitting parameters $\alpha$ such that the loss function
\begin{align}
L_{\rm MSE}(\alpha) = \sum_{x, t} \left[P_\alpha(x-1 | x, t) - P_W(x-1 | x, t) \right]^2
\label{eq:MSE-loss_function}
\end{align}
is minimized, which is the \emph{mean squared error} (MSE). To this end we use the standard SciPy function {\tt scipy.optimize.minimize} for constrained minimization with default parameters. Its algorithm requires an initial guess for the parameter to be fitted; we choose the cosine phases of the Fourier series uniformly randomly from the interval $[0,2\pi)$ and the other parameters, which are unbounded, from the standard normal distribution. We repeat this fitting procedure for  $100$ times in order to increase the chance for arriving at the \textit{global} minimum (or good local minima) of the loss function.

As mentioned in Section~\ref{sec:1-qubit PQC}, another possible choice of loss function is the empirical estimate of the \emph{Kullback-Leibler} (KL) divergence to $P_W$, also used to update the variational parameters of the RL agents, see Eqs.~\eqref{eq:gradientKL_estimate} and \eqref{update}. This loss function has the advantage that it does not require computing $P_W$, since the term containing $P_W$ is constant w.r.t. the fitting parameters $\alpha$ and such terms do \textit{not} matter for the optimization (like for the RL agents). As a consequence, Eq.\ \eqref{eq:KLdivergence} implies that the loss function is equivalent to
\begin{align}
    L_{\rm KL}(\alpha) = - \left\langle R \!\left(\omega_0^T \right) \right\rangle_{\omega_0^T\sim P_{\alpha}},
    \label{eq:KL-loss_function}
\end{align}
where $\langle \cdot \rangle_{\omega_0^T\sim P_{\alpha}}$ denotes the mean value for $N$ trajectories $\omega_0^T$ generated from $P_\alpha$, and $R \!\left(\omega_0^T \right)$ denotes the return, see Eqs.  \eqref{eq:reward_definition} and  \eqref{eq:return_definition}, an RL agent would get for generating a trajectory $\omega_0^T$ following $P_{\alpha}$ as policy. The disadvantage of this loss function is that for small values for $N$ the estimate of the KL divergence seems to be not accurate enough, yielding unreliable results, while for large $N$ loss function evaluations become computationally (too) expensive (we expect that the effort for each function evaluation scales with a factor on the order ${\cal O} (N)$).

In summary, neither the options of choosing the variational angles $\phi$ as fitting parameters nor that of choosing the KL loss function in Eq.~\eqref{eq:KL-loss_function} are eventually used in this work, due to the resulting computational complexity compared to the computationally less complex, yet useful choices of the Fourier coefficients as fitting parameters and of the MSE loss function in Eq.\ \eqref{eq:MSE-loss_function}. Still both options are implemented in the Python code published open source alongside this work, facilitating future work in this direction.

\paragraph*{Step (iv).} Tables \ref{tab:overview_num_results_step_2_ii} and \ref{tab:overview_further_num_results_step_2_ii} state all results of step (iii); as discussed above the loss after fitting is the MSE. The KL divergence can be obtained from the results via the empirical equivalent of Eq.\ \eqref{eq:KLdivergence}, \textit{i.e.},
\begin{align}
    D(P_{\theta}\Vert P_W)  &=  
    \left\langle R \!\left(\omega_0^T \right) \right\rangle_{\omega_0^T\sim P_{W}} - \left\langle R \!\left(\omega_0^T \right) \right\rangle_{\omega_0^T\sim P_{\theta}}.
\end{align}
For further details see the captions of the tables. These results form the basis of the discussion and Fig.\  \ref{fig:plot_table_results_few_qubits_cases} in Section~\ref{sec:1-qubit PQC}, step (iv).

\paragraph*{Scaling the number of qubits.}
As closing remark related to Section~\ref{sec:Scaling the number of time steps}, let us answer the question how the considerations above and in Section~\ref{sec:1-qubit PQC} need to be adapted for PQCs with larger numbers of qubits. As mentioned in Section~\ref{sec:1-qubit PQC} and illustrated by Figs.~\ref{fig:Fourier_coeffs_random_parameters} and \ref{fig:Fourier_coeffs_2_qubits_random_parameters}, the maximal number of frequencies, $N_x$ and $N_t$, of the truncated Fourier series in Eq.~\eqref{eq:truncated_Fourier_series_amplitudes+phases} increase linearly with the number of data-uploading layers. Similarly to the discussion in Sec.~III B of Ref.~\cite{schuld2020circuit}, they also increase linearly with the number of entangled copies (see Section~\ref{sec:Scaling the number of time steps}) of the considered circuits (see Figs.~\ref{Circuit1Q} and \ref{CircuitReUploading}). Thus the number of potentially non-zero Fourier coefficients is of order $O(n^d )$, where $n$ is the number of qubits and $d$ is the dimension of the input vector, here $d = 2$. That means it scales exponentially with the problem size as quantified by $d$. The only difference to scaling the number of data-uploading layers is that, due to the different unitary transformations, other Fourier coefficients which the truncated Fourier series allows to be non-zero complex numbers might be identically zero, real, or purely imaginary, and might depend on each other. Besides that, the approach described above and in Section~\ref{sec:1-qubit PQC} proceeds in the same way. 

\begin{table} [h]
    \centering
\begin{tabular}{|c|c|c||c|c|c|
} 
 \hline
 \# qubits & \# layers & \# fits & min(MSE) & $\mu$(MSE) & $\sigma$(MSE)
 \\ 
 \hline
1 & 1 & 100 & $2.76 \cdot 10^{-3}$ & $5.30 \cdot 10^{-3}$ & $6.34 \cdot 10^{-3}$ 
\\
\hline
1, 2 & 2 & 100 & $1.12 \cdot 10^{-3}$ & $3.57 \cdot 10^{-3}$ & $8.47 \cdot 10^{-3}$ 
\\
\hline
1, 2 & 3 & 100 & $0.94 \cdot 10^{-3}$ & $3.97 \cdot 10^{-3}$ & $7.89 \cdot 10^{-3}$ 
\\
\hline
1, 2 & 4 & 100 & $0.83 \cdot 10^{-3}$ & $4.19 \cdot 10^{-3}$ & $8.58 \cdot 10^{-3}$ 
\\
\hline
1, 2 & 5 & 100 & $1.08 \cdot 10^{-3}$ & $4.87 \cdot 10^{-3}$ & $6.37 \cdot 10^{-3}$ 
\\
\hline
1, 2 & 10 & 100 & $6.03 \cdot 10^{-3}$ & $23.6 \cdot 10^{-3}$ & $23.4 \cdot 10^{-3}$ 
\\
\hline
1, 2 & 15 & 100 & $ 12.9 \cdot 10^{-3}$ & $89.3 \cdot 10^{-3}$ & $77.1 \cdot 10^{-3}$ 
\\
\hline
2 & 1 & 100 & $1.49 \cdot 10^{-1}$ & $1.49 \cdot 10^{-1}$ & $\approx 0$ 
\\
\hline
\hline
$P: -$ & $-$ & $-$ & $1.49 \cdot 10^{-1}$ & $-$ & $-$
\\
\hline
$P_W: -$ & $-$ & $-$ & $0$ & $-$ & $-$ 
\\
\hline
\end{tabular}

\label{tab:Fourier_results}

\caption{Numerical results (first part) of the Fourier analysis in Section\  \ref{sec:1-qubit PQC} and App.\ \ref{app:Fourier_analysis}. ``MSE'' abbreviates ``mean squared error'' (relative to the optimal dynamics $P_W)$ of a single fit; min(MSE), $\mu$(MSE), and $\sigma$(MSE) denote the minimum, mean, and standard deviation of the MSE for the stated number of fits, respectively, and $P$ the original dynamics of the symmetric random walk. The optimal dynamics is calculated as discussed in Section\  \ref{sec:Stochastic model} with the parameter $s = 1$. 
In all cases, the trajectory length $T = 20$.}
\label{tab:overview_num_results_step_2_ii}
\end{table}

\begin{table}
    \centering
\begin{tabular}{|c|c|c||c|c|c||c|} 
 \hline
 \# qubits & \# layers & \# fits &  max($R$) & $\mu$($R$) & $\sigma$($R$)  & $P_{\text{max}(R)}(x_T = 0)$ \\ 
 \hline
1 & 1 & 100 & $-2.64$ & $-3.11$ & $1.18$ & $0.60$ \\
\hline
1, 2 & 2 & 100 & $-2.21$ & $-2.76$ & $1.65$ & $0.72$ \\
\hline
1, 2 & 3 & 100 & $-2.07$ & $-2.72$ & $1.38$ & $0.81$ \\
\hline
1, 2 & 4 & 100 & $-2.03$ & $-2.72$ & $1.55$ & $0.80$ \\
\hline
1, 2 & 5 & 100 & $-2.05$ & $-2.81$ & $1.27$ & $0.81$ \\
\hline
1, 2 & 10 & 100 & $-2.72$ & $-5.54$ & $3.87$ & $0.69$ \\
\hline
1, 2 & 15 & 100 & $-3.72$ & $-14.9$ & $12.9$ & $0.62$ \\
\hline
2 & 1 & 100 & $-19.1$ & $-20.0$ & $0.289$ & $0.17$\\
\hline
\hline
$P: -$ & $-$ & $-$ & $-20.4$ & $-$ & $-$ &  $0.18$ 
\\
\hline
$P_W: -$ & $-$ & $-$ & $-1.70$ & $-$ & $-$ &  $0.97$ 
\\
\hline
\end{tabular}

\caption{Numerical results (second part) of the Fourier analysis in Section~\ref{sec:1-qubit PQC} and App.~\ref{app:Fourier_analysis}, similar to Tab. \ref{tab:overview_num_results_step_2_ii}. The quantities max($R$), $\mu$($R$), and $\sigma$($R$) denote the maximum, mean, and standard deviation (for the stated number of fits) of the mean return $R = \left\langle R \!\left(\omega_0^T \right) \right\rangle$ (for $100.000$ trajectories $\omega_0^T$ generated based on the respective fitted policy) respectively. (Note that the values for $R$ have \textit{not} been obtained by reinforcement learning (RL), but denote the mean return an RL agent would get for generating  trajectories following the respective fitted policy.) The  probability $P_{\text{max}(R)}(x_T = 0)$ to generate a trajectory with endpoint $x_T = 0$ using the fitted policy with maximal return max$(R)$ has been estimated in a similar way, but by using the best fitted policy only.}
\label{tab:overview_further_num_results_step_2_ii}
\end{table}

\section{Further numerical considerations}
\label{Appendix: Further numerics}

\begin{figure}[h]
    \centering
    \includegraphics[width=7.5cm]{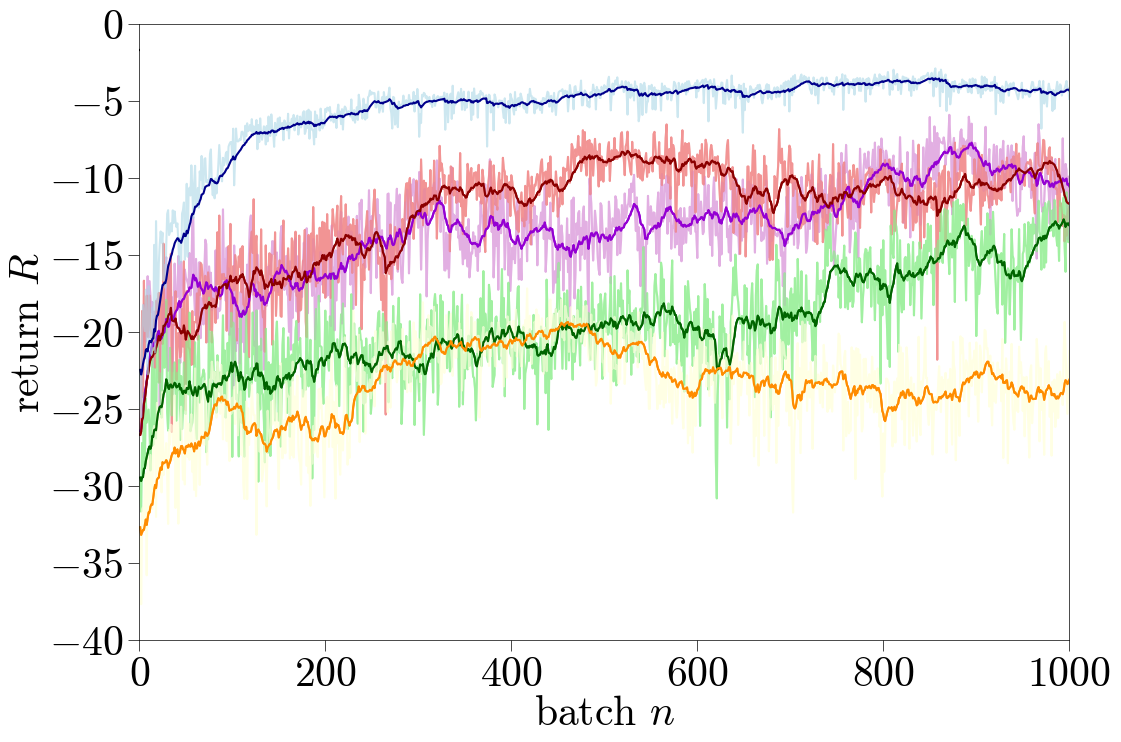} %
    \includegraphics[width=7.5cm]{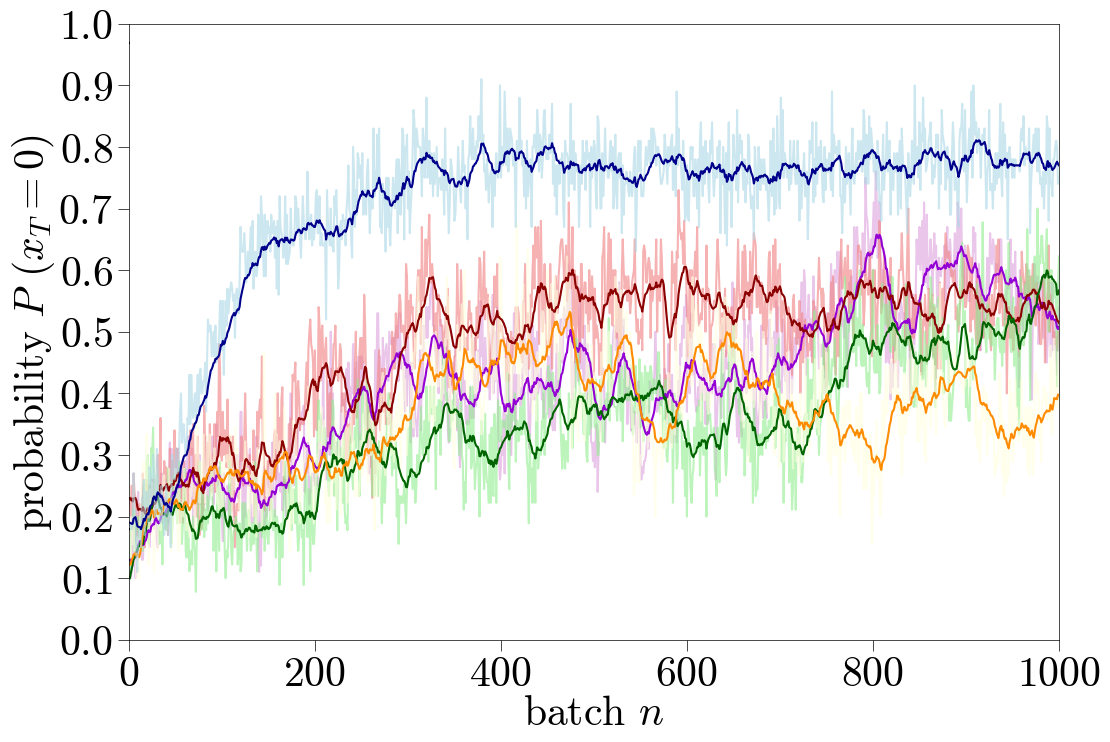} %
    \caption{Effect of different random walker jump probabilities for quantum policy gradient and $T=20$. The different colors represent different jump probabilities; blue 50/50, violet 60/40, red 70/30, green 80/20 and yellow 90/10. The lighter lines represent the average of 10 agents, each trained using the same hyperparameters. The darker lines represent the exponential moving average. Left: The average return $R$ per batch $n$. Right: Probability of generating a rare trajectory $P(x_T = 0)$. The hyperparameters of the agents used in these plots can be found in Tab. \ref{paramsFigApp1} of App.\ \ref{app:hyperparameters}.}
    \label{Fig:QPGDifProbs}
\end{figure}

In order to assess whether the success of the PQC in approximating $P_W$ is linked to the predefined equal up and down jump probabilities of the random walker, we analyzed the return the agent received over the number of batches (left) for different up and down jump probabilities, as shown in Fig.~\ref{Fig:QPGDifProbs}. The achieved return decreases as the difference in the jump probabilities increases. This behavior is expected, as the number of RWB in the sampling decreases proportionally with an increase in the disparity between up and down jump probabilities. The probability of generating a rare trajectory (right) remains high. This indicates a strong learning process, as the agent demonstrates the ability to generate a rare trajectory with high probability. Thus we conclude that this method is also applicable for random walks with different jump probabilities.

\begin{figure}[h!]
    \centering
    \includegraphics[width=7.5cm]{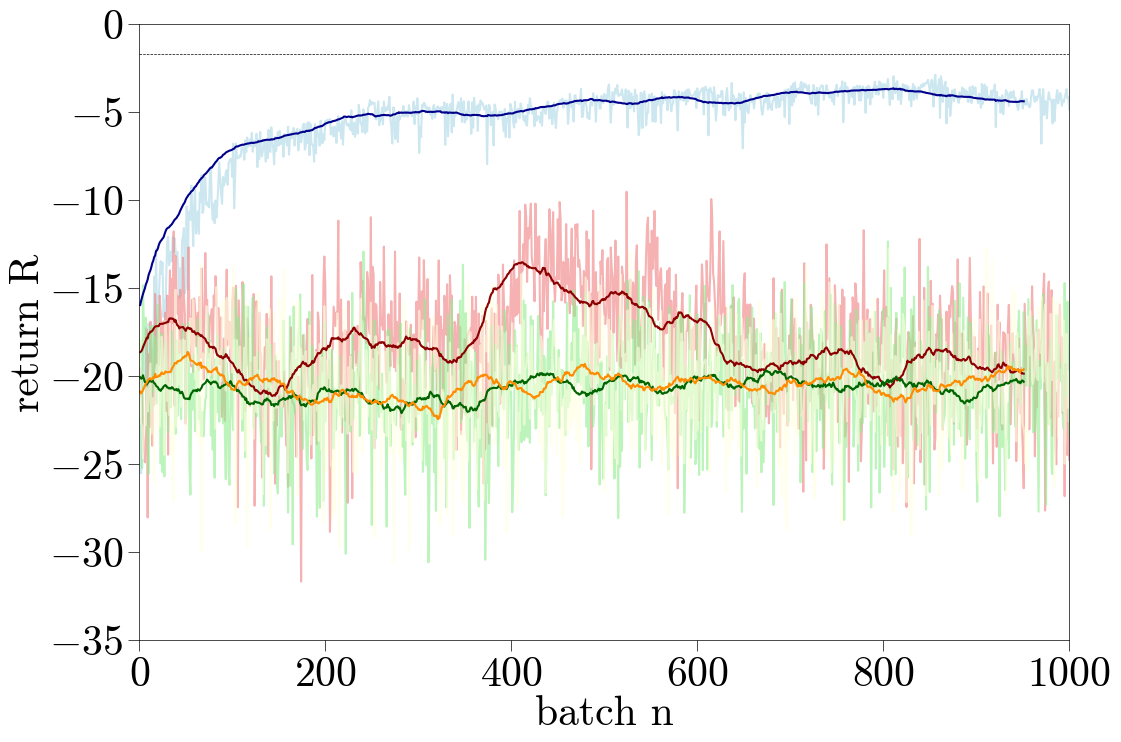} %
     \includegraphics[width=7.5cm]{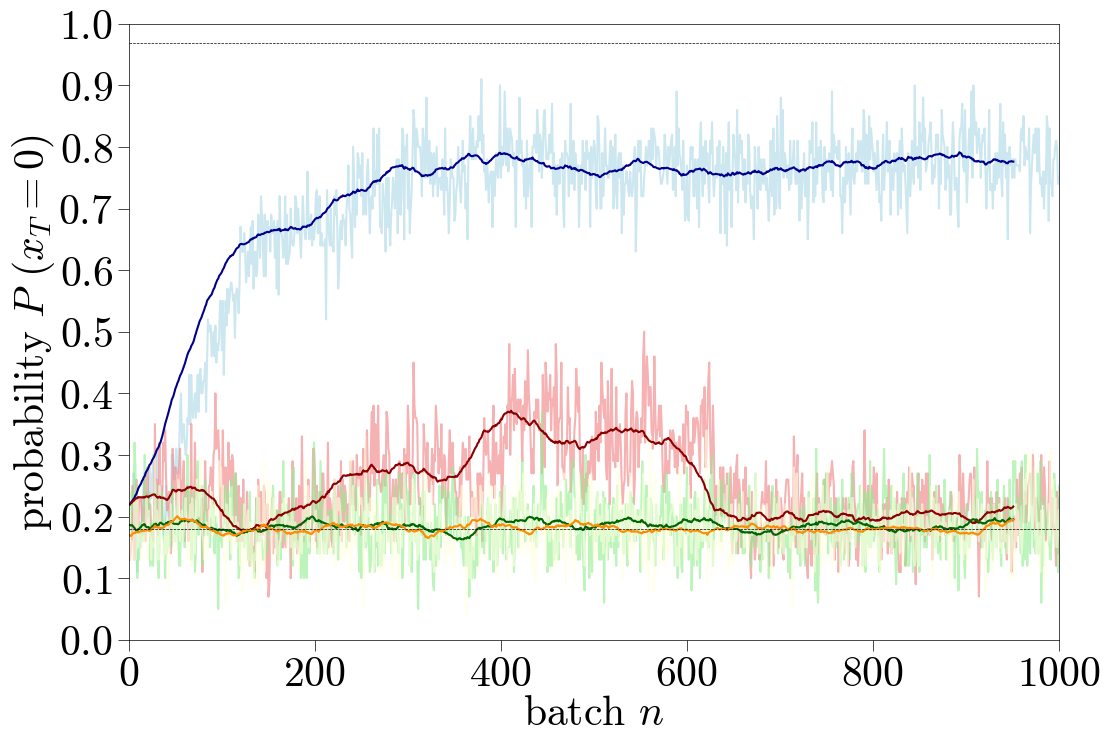}
    \caption{Effect of coherent errors in the PQCs of the \textit{policy-gradient} reinforcement-learning approach (the trajectory length $T=20$). The different colors represent  different error rates per data-encoding qubit in the circuits; blue 0\%, red 20\%, green 40\%, yellow 60\%. The lighter lines represent the average of 10 agents, each trained using the same hyperparameters. The darker lines represent the exponential moving average. Left: The average return $R$ per batch $n$. Right: Probability of generating a rare trajectory $P(x_T = 0)$. The hyperparameters of the agents used in these plots can be found in Tab. \ref{paramsFig21} of App.\ \ref{app:hyperparameters}.}
    \label{noisePG}
\end{figure}
\section{Tables of hyperparameters}
\label{app:hyperparameters}

This appendix summaries all parameters of each numerical simulation presented in this work. 

\begin{table}
    \centering
\begin{tabular}{|p{0.2\textwidth} | p{0.8\textwidth} |} 
 \hline
 Hyperparameter & Description \\
 \hline\hline
 $T$ & Length of trajectories \\
 \hline
 $s$ & Parameter in Eq.\ \eqref{eq:RWB_weight} ensuring the learning of the random walk bridge \\ 
 \hline
 $N$ & Batch size = \#trajectories generated in one batch \\
 \hline
 Noise & Decoherence noise rates\\
 \hline
 Actor layers & \# data re-uploading layers in  policy PQC \\ 
 \hline
 $\alpha_\phi$ & Learning rate for variational angles $\phi$ of policy PQC \\ 
 \hline
 $\alpha_\lambda$ & Learning rate for input scaling parameters $\lambda$ of policy PQC \\
 \hline
 $\alpha_\omega$ & Learning rate for output scaling parameters $\omega$ of policy PQC \\
 \hline
 $\beta$ &  Inverse temperature parameter of policy PQC \\
 \hline
 Critic layers & \# data re-uploading layers in  value-function PQC \\ 
 \hline
 $\alpha_\phi^{A,C}$ & Learning rate for variational angles $\phi'$ of actor (A) or critic (C) PQC\\
 \hline
 $\alpha_\lambda^{A,C}$ & Learning rate for input scaling parameters $\lambda$ of actor (A) or critic (C) PQC \\
 \hline
 $\alpha_\omega^{A,C}$ & Learning rate for output scaling parameters $\omega$ of actor (A) or critic (C) PQC \\ 
 \hline
 NN layers & \# hidden layers in neural network \\
 \hline
 NN neurons & \# neurons per hidden layer in neural network, e.g. for 2 hidden layers = (\#,\#)\\
 \hline
 NN activation & Activation function of neural network \\
 \hline
 $\beta$, $\alpha_{NN}$, & NN learning rate actor and critic\\
 \hline
\end{tabular}
    \caption{Description of hyperparameters used in this work.}
    \label{tab:description_hyperparameters}
\end{table}


\begin{table}
\centering
\begin{tabular}{|p{0.2\textwidth} |  p{0.4\textwidth} |  p{0.4\textwidth}|} 
 \hline
 Agent & PG PQC (1-qubit) & PG PQC (2-qubits)\\
 \hline \hline
 $T$ & 20 & 20 \\
 \hline
 $s$ & 1.0 & 1.0 \\ 
 \hline
 $N$ & 10 & 10 \\
 \hline
 Noise & 0 & 0 \\
 \hline
 Actor layers & 3 & 3 \\
 \hline
 $\alpha_\phi$,  $\alpha_\lambda$, $\alpha_\omega$ & 0.01, 0.05, 0.1 & 0.01, 0.05, 0.1 \\
 \hline
 $\beta$ & 1.0 & 1.0 \\ 
 \hline
\end{tabular}
    \caption{List of values for hyperparameters used for the results of Figs. \ref{Fig6}, \ref{Fig7} (right column), and \ref{fig:policy_fits_1_qubit_1_layer}.}
    \label{paramsFig6}
\end{table}

\begin{table}
\centering
\begin{tabular}{|p{0.2\textwidth} |  p{0.4\textwidth} |  p{0.4\textwidth}|} 
 \hline
 Agent & PG PQC (2-qubits) & AC PQC (2-qubits) \\
 \hline \hline
 $T$ & 20 & 20 \\
 \hline
 $s$ & 100.0 & 100.0 \\ 
 \hline
 $N$ & 10 & 10 \\
 \hline
 Noise & 0 & 0 \\
 \hline
 Actor layers & 3 & 3 \\
 \hline
  $\alpha_\phi^{A}$, $\alpha_\lambda^{A}$,
  $\alpha_\omega^{A}$ & 0.01, 0.05, 0.1 & 0.01, 0.05, 0.2 \\
 \hline
 $\beta$ & 1.0 & 1.0 \\ 
 \hline
 Critic layers & - & 3 \\
 \hline
$\alpha_\phi^{C}$, $\alpha_\lambda^{C}$, $\alpha_\omega^{C}$ & - & 0.01, 0.05, 0.7 \\
 \hline
\end{tabular}
    \caption{List of values for hyperparameters used for the results of Figs.\  \ref{fig:modelcomparison}, \ref{actorCriticPlots} (right column)}
    \label{paramsFig8}
\end{table}

\begin{table}
\centering
\begin{tabular}{|p{0.2\textwidth} |  p{0.8\textwidth}|} 
 \hline
 Agent & PG PQC (2-qubits) \\
 \hline \hline
 $T$ & 20 \\
 \hline
 $s$ & 1.0 \\ 
 \hline
 $N$ & 10\\
 \hline
 Noise & 0 \\
 \hline
 Actor layers & 1, 3, 5, 10, 15 \\
 \hline
 $\alpha_\phi$, $\alpha_\lambda$,  $\alpha_\omega$ & 0.01, 0.05, 0.1 \\
 \hline
 $\beta$ & 1.0 \\
 \hline
\end{tabular}
    \caption{List of values for hyperparameters used for the results of Fig.\  \ref{fig:layers1}.}
    \label{paramsFig11}
\end{table}

\begin{table}
\centering
\begin{tabular}{|p{0.2\textwidth} |  p{0.8\textwidth}|} 
 \hline
 Agent & PG PQC (2-qubits) \\
 \hline \hline
 $T$ & 20 \\
 \hline
 $s$ & 1.0 \\ 
 \hline
 $N$ & 10 \\
 \hline
 Noise & 0.0 \\
 \hline
 Actor layers & 3 \\
 \hline
$\alpha_\phi$, $\alpha_\lambda$, $\alpha_\omega$ & 0.01, 0.05 , 0.1 \\
 \hline
 $\beta$ & 1.0 \\
 \hline
\end{tabular}
    \caption{List of values for hyperparameters used for the results of Figs. \ref{dataReUploading} and  \ref{dataReUploading2}.}
    \label{paramsFig13}
\end{table}

\begin{table}
\centering
\begin{tabular}{|p{0.2\textwidth} |  p{0.25\textwidth} |  p{0.25\textwidth} |p{0.25\textwidth} |} 
 \hline
 Agent & PG PQC (2-qubits) & PG PQC (1-qubit) & PG NN \\
 \hline \hline
 $T$ & 20 & 20 & 20 \\
 \hline
 $s$ & 1.0 & 1.0 & 1.0 \\ 
 \hline
 $N$ & 10 & 10 & 10 \\
 \hline
 Noise & 0.0 & 0.0 & - \\
 \hline
 Actor layers & 3 & 3 & - \\
 \hline
 $\alpha_\phi$, $\alpha_\lambda$, $\alpha_\omega$ & 0.01, 0.05, 0.1 & 0.01, 0.05, 0.1& - \\
 \hline
 $\beta$ & 1.0 & 1.0 & - \\ 
 \hline
 NN layers & - & - & 2 \\
 \hline
 NN neurons & - & - & (2,2), (4,4), (5,5) \\
 \hline
 NN activation & - & - & ReLU \\
 \hline
 $\alpha_{NN}$ & - & - & 0.01 \\
 \hline
\end{tabular}
    \caption{List of values for hyperparameters used for the results of Figs.\  \ref{ReluNN}.}
    \label{paramsFig15}
\end{table}

\begin{table}
\centering
\begin{tabular}{|p{0.2\textwidth} |  p{0.25\textwidth} |  p{0.25\textwidth}| p{0.25\textwidth} |} 
 \hline
 Agent & PG PQC (2-qubits) & PG PQC (1-qubit) &PG NN \\
 \hline \hline
 $T$ & 20 & 20 & 20 \\
 \hline
 $s$ & 1.0 & 1.0 & 1.0 \\ 
 \hline
 $N$ & 10 & 10 & 10 \\
 \hline
 Noise & 0.0 & 0.0 & - \\
 \hline
 Actor layers & 3 & 3 &- \\
 \hline
 $\alpha_\phi$, $\alpha_\lambda$, $\alpha_\omega$ & 0.01, 0.05, 0.1 & 0.01, 0.05, 0.1 & - \\
 \hline
 $\beta$ & 1.0 & 1.0 & - \\ 
 \hline
 NN layers & - & - & 2 \\
 \hline
 NN neurons & - & - & (2,2), (4,4), (5,5) \\
 \hline
 NN activation & - & - & sine \\
 \hline
 $\alpha_{NN}$ & - & - & 0.01 \\
 \hline
\end{tabular}
    \caption{List of values for hyperparameters used for the results of Figs.\ \ref{SinNN}.}
    \label{paramsFig16}
\end{table}

\begin{table}
\centering
\begin{tabular}{|p{0.2\textwidth} |  p{0.8\textwidth}|}
 \hline
 Agent & PG PQC (2-qubits)\\
 \hline \hline
 $T$ & 20, 40, 60, 80, 100 \\
 \hline
 $s$ & 5.0 \\ 
 \hline
 $N$ & 10 \\
 \hline
 Noise & 0.0 \\
 \hline
 Actor layers & 3 \\
 \hline
 $\alpha_\phi$, $\alpha_\lambda$, $\alpha_\omega$ & 0.01, 0.05, 0.1 \\
 \hline
 $\beta$ & 1.0 \\ 
 \hline
\end{tabular}
    \caption{List of values for hyperparameters used for the results of Fig.\  \ref{scaling1}.}
    \label{paramsFig23}
\end{table}

\begin{table}
\centering
\begin{tabular}{|p{0.2\textwidth} |  p{0.8\textwidth}|}
 \hline
 Agent & PG PQC (2-qubits)\\
 \hline \hline
 $T$ & 200 \\
 \hline
 $s$ & 50.0 \\ 
 \hline
 $N$ & 10 \\
 \hline
 Noise & 0.0 \\
 \hline
 Actor layers & 3 \\
 \hline
 $\alpha_\phi$, $\alpha_\lambda$, $\alpha_\omega$ & 0.01, 0.05, 0.1 \\
 \hline
 $\beta$ & 1.0 \\ 
 \hline
\end{tabular}
    \caption{List of values for hyperparameters used for the results of Fig.\  \ref{scaling2}.}
    \label{paramsFig24}
\end{table}

\begin{table}
\centering
\begin{tabular}{|p{0.2\textwidth} |  p{0.8\textwidth}|}
 \hline
 Agent & PG PQC (8-qubits)\\
 \hline \hline
 $T$ & 20 \\
 \hline
 $s$ & 1.0 \\ 
 \hline
 $N$ & 10 \\
 \hline
 Noise & 0.0 \\
 \hline
 Actor layers & 3 \\
 \hline
 $\alpha_\phi$, $\alpha_\lambda$, $\alpha_\omega$ & 0.05, 0.01, 0.1 \\
 \hline
 $\beta$ & 1.0 \\ 
 \hline
\end{tabular}
    \caption{List of values for hyperparameters used for the results of Fig.\  \ref{scalingqubit}.}
    \label{paramsFig8qubits}
\end{table}

\begin{table}
\centering
\begin{tabular}{|p{0.2\textwidth} |  p{0.8\textwidth}|}
 \hline
 Agent & PG PQC (2-qubits)\\
 \hline \hline
 $T$ & 20 \\
 \hline
 $s$ & 1.0 \\ 
 \hline
 $N$ & 10 \\
 \hline
 Noise & 0.0 \\
 \hline
 Actor layers & 3 \\
 \hline
 $\alpha_\phi$, $\alpha_\lambda$, $\alpha_\omega$ & 0.01, 0.05, 0.1 \\
 \hline
 $\beta$ & 1.0 \\ 
 \hline
\end{tabular}
    \caption{List of values for hyperparameters used for the results of Fig.\  \ref{Fig:QPGDifProbs}.}
    \label{paramsFigApp1}
\end{table}
\begin{table}
\centering
\begin{tabular}{|p{0.2\textwidth} | p{0.8\textwidth}|} 
 \hline
 Agent & PG PQC (2-qubits) \\
 \hline \hline
 $T$ & 20 \\
 \hline
 $s$ & 1.0 \\ 
 \hline
 $N$ & 10 \\
 \hline
 Noise & 0.05, 0.1, 0.25, 0.5, 0.75 \\
 \hline
 Actor layers & 3 \\
 \hline
 $\alpha_\phi$,
 $\alpha_\lambda$, $\alpha_\omega$ & 0.01, 0.05, 0.1 \\
 \hline
 $\beta$ & 1.0 \\ 
 \hline
\end{tabular}
    \caption{List of values for hyperparameters used for the results of Fig.\  \ref{noisePG}.}
    \label{paramsFig21}
\end{table}

\end{appendix}

\end{document}

%% file: pseudocode.tex
    \begin{algorithm} [H]
    \caption{Quantum Policy-Gradient Reinforcement Learning }\label{alg:policy_gradient}
    \algorithmicrequire{policy PQC to approximate $\pi_{\theta}(s,a)$}\vskip 3pt
    
    \algorithmicparam{trajectory length $T$, batch size $N$, number of batches $B$ }

    \algorithmichyperparam{ weighting parameter $s$, inverse-temperature parameter $\beta$, learning rates $\alpha_\lambda$, $\alpha_\phi$, $\alpha_\omega$}\vskip 3pt
    
    \algorithmicinit{trainable parameters $\theta$ randomly following a uniform distribution between $0$ and $2\pi$ }\vskip 3pt
    \algorithmicensure{optimized PQC for $\pi_{\theta}(s,a)$}
    \begin{algorithmic}[1]

        \FOR{batch $n \in\{1,2, ..., B\}$}
        \FOR{episode $i \in\{1, 2, ..., N\}$}
        
        \FOR{time $t \in\{0, 1, ..., T-1\}$}

        \STATE $\pi (s_t , a) \gets \frac{e^{\beta \expval{O_a}_{s_t, \theta}}}{\sum_a^\prime e^{\beta \expval{O_{a^\prime}}_{s_t, \theta}}}$
        \STATE choose $a$ according $\pi_\theta (s_t , a)$
        \STATE $s_{t+1} \gets (x_t + a_t, t+1)$

        \ENDFOR

        \STATE $L_{P} \gets 0$
        
        \FOR{$t \in\{0, 1, ..., T-1\}$}


        \STATE $L_{P} \gets L_{P} + R_{t-1} \nabla_{\theta} P_{\theta}^{(t)} (x_t | x_{t-1})$
        
        \ENDFOR

        \FOR{$j \in \{ \lambda, \phi, \omega \}$}
        \STATE $j^{\theta_A} \gets j^{\theta_A} + \alpha_j^{\theta_A} L_{P}$
        \ENDFOR
        
        \ENDFOR

        \ENDFOR
        
    \end{algorithmic}
\end{algorithm}
    \begin{algorithm} [H]
    \caption{Quantum Actor-Critic Reinforcement Learning}\label{alg:actor_critic}
    \algorithmicrequire{policy PQC to approximate $\pi_{\theta_A}(s,a)$; PQC for value-function approximation $V_{P_{\theta_C}}(x_t,t)$; trajectory length $T$, batch size $N$, number of batches $B$, weighting parameter $s$, inverse-temperature parameter $\beta$} \vskip 3pt
    
    \algorithmicparam{learning rates $\alpha_\lambda^{\theta_A}$, $\alpha_\phi^{\theta_A}$, $\alpha_\omega^{\theta_A}$, $\alpha_\lambda^{\theta_C}$, $\alpha_\phi^{\theta_C}$, $\alpha_\omega^{\theta_C}$} \vskip 3pt
    
    \algorithmicinit{trainable parameters $\theta_A$ and $\theta_C$ randomly following a uniform distribution between $0$ and $2\pi$ } \vskip 3pt
    
    \algorithmicensure{optimized PQCs for $\pi_{\theta_A}(s,a)$ and $V_{P_{\theta_C}}(x_t,t)$}
    \begin{algorithmic}[1]

        \FOR{batch $n \in\{1,2, ..., B\}$}
        \FOR{episode $i \in\{1, 2, ..., N\}$}
        
        \FOR{time $t \in\{0, 1, ..., T-1\}$}

        \STATE $\pi (s_t , a) \gets \frac{e^{\beta \expval{O_a}_{s_t, \theta_A}}}{\sum_a^\prime e^{\beta \expval{O_{a^\prime}}_{s_t, \theta_A}}}$
        \STATE choose $a_t$ according $\pi (s_t , a)$
        \STATE $s_{t+1} \gets (x_t + a_t, t+1)$

        \ENDFOR

        \STATE $L_{P} \gets 0$
        \STATE $L_{V} \gets 0$
        
        \FOR{$t \in\{0, 1, ..., T-1\}$}

        \STATE $V_{P_{\theta_C}}(x_t, t) \gets \expval{O}_{s_t,\theta_C}$
        \STATE $V_{P_{\theta_C}}(x_{t+1}, t+1) \gets \expval{O}_{s_{t+1},\theta_C}$

        \STATE $\delta_{TD} \gets V_{P_{\theta_C}}(x_{t+1}, t+1) + r(x_{t+1},x_{t},t) - V_{P_{\theta_C}}(x_t, t)$
        \STATE $L_{P} \gets L_{P} + \delta_{TD} \nabla_{\theta_A} P_{\theta_A} (x_{t+1} | x_{t}, t+1)$
        \STATE $L_{V} \gets L_{V} + \delta_{TD} \nabla_{\theta_C} V_{P_{\theta_C}}(x_t, t)$
        
        \ENDFOR

        \FOR{$j \in \{ \lambda, \phi, \omega \}$}
        \STATE $j^{\theta_A} \gets j^{\theta_A} + \alpha_j^{\theta_A} L_{P}$
        \STATE $j^{\theta_C} \gets j^{\theta_C} + \alpha_j^{\theta_C} L_{V}$
        \ENDFOR
        
        \ENDFOR

        \ENDFOR
        
    \end{algorithmic}
\end{algorithm}